\documentclass[12pt]{article}

\PassOptionsToPackage{nosort
}{cite}
\usepackage{chetnew}
\usepackage{xparse}
\usepackage{amssymb, amsmath}
\usepackage{xcolor}
\usepackage{tikz}
\usepackage{cite}
\usepackage[linktocpage]{hyperref}
\usepackage{mathtools}
\usepackage{physics}
\usepackage{float}
\usepackage{arydshln}

\usetikzlibrary{decorations.pathmorphing}
\usetikzlibrary{patterns} 
\usepackage{subcaption}

\DeclareFontShape{OT1}{cmr}{mx}{n}%
    {<->cmr10}{}
\renewcommand{\tilde}{\widetilde}

\allowdisplaybreaks

\preprint{
NITEP 287}

\title{On phases of 4d $Spin/SO$ Yang-Mills theories}

\author{Ryo Hamachika$^{1\spadesuit}$ and Takahiro
Nishinaka$^{1,2,3\diamondsuit}$}
\affiliation{\smallskip  $^1$Department of Physics, Graduate School of Science\\
Osaka Metropolitan University, Osaka 558-8585, Japan\\
\bigskip
$^2$Nambu Yoichiro Institute of Theoretical and Experimental Physics (NITEP)\\
Osaka Metropolitan University, Osaka 558-8585, Japan\\
\bigskip
$^3$Osaka Central Advanced Mathematical Institute (OCAMI)\\
Osaka Metropolitan University, Osaka 558-8585, Japan
\emails{$^{\spadesuit}$sf22775d@st.omu.ac.jp, 
$^{\diamondsuit}$nishinaka@omu.ac.jp}
}

\abstract{
We study the low-temperature phase diagrams of $Spin(N)/SO(N)_\pm$ Yang-Mills
theories as functions of the $\theta$-parameter. In particular, for
every even integer $N\geq 4$, we
find a phase diagram consistent with the center/CP mixed
anomaly, via the technique of mixed gauge theory. The resulting phase
diagrams are in perfect agreement with the phase diagrams of
$Spin(N)/SO(N)_\pm$ super Yang-Mills theory softly broken by a small
gaugino mass. In the case of $Spin(4k)$ Yang-Mills for an integer $k\geq
2$,  only two phases
appear in the phase diagram, while in the $Spin(4)$ case the phase
diagram contains four SPT phases protected by the
$\mathbb{Z}_2\times \mathbb{Z}_2$ center symmetry.
In the case of $Spin(4k+2)$ Yang-Mills for a positive integer $k$, the
phase diagram contains four SPT phases protected by the $\mathbb{Z}_4$
center symmetry.
}

\begin{document}

\maketitle
\setcounter{tocdepth}{2}
\toc

\section{Introduction}

% It is widely believed that the Yang-Mills theory is confined at zero
% temperature. One consistency check of this is given by studying a
% supersymmetric Yang-Mills (SYM) theory with
% a soft supersymmetry (SUSY) breaking mass term. For instance, in the
% case of
% $SU(N)$ SYM, there are $N$
% degenerate vacua; at one vacuum the monopole
% condensation occurs while at the others dyons condensate. Since a particle with non-vanishing magnetic charge
% condensates at these vacua, the color confinement occurs via the dual
% Meissner effect. When a small gaugino mass $m$ is turned on, the
% supersymmetry is softly broken, giving rise to a unique vacuum in the
% infrared. If this situation holds in the limit of $m\to\infty$,
% one can confirm that the (non-supersymmetric) $SU(N)$ Yang-Mills theory
% has a unique confining (or oblique confining) vacuum.

It has recently been pointed out that the mixed 't Hooft anomaly between the
CP and center symmetries sheds light on the vacuum structure
of non-supersymmetric Yang-Mills theories \cite{Gaiotto:2017yup}.
In particular, for
the $SU(2)$ gauge group, a phase diagram of the
non-supersymmetric Yang-Mills theory was read off from
the CP/center mixed anomaly, as a function of the $\theta$-parameter.
This result turned out to be in perfect agreement with the phase diagram
of the super Yang-Mills (SYM) with a small SUSY-breaking mass $m$ turned
on. Since the non-supersymmetric Yang-Mills can be regarded as the
$m\to \infty$ limit, this suggests that the phase diagram is
independent of  $m$.
A crucial step in this analysis is to consider the {\it mixed gauge
theory,} which makes the mixed center/CP anomaly transparent.
One can generalize this analysis to the $SU(N)$ gauge group \cite{Gaiotto:2017yup, Tachikawa:2017gyf}.
For recent applications of similar discrete anomalies to phase
structures of 4d gauge theories,
see
\cite{Tanizaki:2017bam, Shimizu:2017asf, Gaiotto:2017tne,
Tanizaki:2017mtm, Aitken:2018kky,  Cordova:2018acb, Tanizaki:2018wtg, 
Yamaguchi:2018xse, Anber:2018xek, Wan:2018djl,
Yonekura:2019vyz,  Karasik:2019bxn, Poppitz:2019fnp, Cordova:2019bsd,Wang:2019obe,
Chen:2020syd, Honda:2020txe,
Bolognesi:2020mpe,Sulejmanpasic:2020zfs,Furusawa:2020qdz, 
Bolognesi:2021yni, 
Kaidi:2021xfk,
Bolognesi:2022beq, 
Morikawa:2022liz,
Kobayashi:2023ajk,Nguyen:2023fun,Hayashi:2023wwi,
Hayashi:2024qkm, Bolognesi:2024wje, Hayashi:2024gxv, Brennan:2026ira, Bolognesi:2026eso
}.

Given the above progress, it is important to apply the technique of
CP/center mixed anomaly to other gauge groups and see
how the technique of the mixed gauge theory works in a more general setup. In
particular, it is interesting to study
the generalization to $Spin/SO$ Yang-Mills theory and read off the corresponding
phase diagram  from the center/CP mixed anomaly as a function of $\theta$. While such a
generalization was already done for the
$Spin(3)/SO(3)$ Yang-Mills theory in
\cite{Gaiotto:2017yup},
 the situation for general
$Spin/SO$ gauge groups is still to be understood.

In this paper, we apply the technique of the center/CP mixed anomaly and
mixed gauge theory to the
$Spin(N)/SO(N)$ Yang-Mills theory for all even $N$ larger than $3$. We
particularly identify a
phase diagram consistent with the mixed anomaly by studying the mixed
gauge theories constructed by compactifying these $SO/Spin$
Yang-Mills theories on $S^1$ and gauging a 3d one-form
symmetry. We find that the $\mathbb{D}_8$ symmetry that plays an
important role in the $SU(2)$ case
is replaced by $(\mathbb{Z}_2)^4\rtimes
\mathbb{Z}_2$ or $(\mathbb{Z}_4)^2\rtimes \mathbb{Z}_2$, depending on the
value of $N$. The $Spin(N)/SO(N)$ Yang-Mills theory for odd $N$ has no
center/CP anomaly, and therefore the same technique is not available.

One can regard these Yang-Mills theories as 
the $m\to \infty$ limit of the softly broken $Spin/SO$ SYM with a gaugino mass
$m$. It is therefore interesting to compare the above results with the phase
diagrams of the softly-broken SYM for a small $|m|$, which can be read
off in a perturbative analysis. We perform this comparison to
find that the above results are in perfect agreement
with the phase diagram of SYM for a small $|m|$, which suggests that the
phase diagram is independent of $m$.

Among these $Spin/SO$ gauge groups, $Spin(4)/SO(4)$ is special. Since $Spin(4) \simeq SU(2)\times SU(2)$, the
$Spin(4)/SO(4)$ Yang-Mills theory has two independent
$\theta$-parameters, in contrast to the other $Spin/SO$
cases. Moreover, since the $\mathbb{Z}_2$ quotient giving rise to $SO(4)$ from $Spin(4)$ is associated
with the
diagonal sub-group of $\mathbb{Z}_2\times \mathbb{Z}_2\subset
SU(2)\times SU(2)$, the vacuum structure of $SO(4)$ Yang-Mills theory
depends non-trivially on both of these two $\theta$-parameters.
As a result, the $SO(4)_\pm$ Yang-Mills theories are related by shifts of the
 $\theta$-parameters as
\begin{equation}
 SO(4)_\pm^{\theta_1+2\pi,\theta_2} =
  SO(4)_{\pm}^{\theta_1,\theta_2+2\pi} =SO(4)_{\mp}^{\theta_1,\theta_2}
  = SO(4)_{\mp}^{\theta_1+2\pi,\theta_2+2\pi}~.
\label{eq:SO4_periodicity}
\end{equation}
This is totally different from the $Spin(N)/SO(N)$ cases for $N>4$, where $SO_+$
 and $SO_-$ theories are not related by any shift of $\theta$.

 Given the above special features of $Spin(4)/SO(4)$, we start our analysis with the
$Spin(4)/SO(4)$ case in Sec.~\ref{sec:Spin4}, after reviewing the $SU(2)$
case in Sec.~\ref{sec:SU2}. In particular, we identify a phase diagram
of $Spin(4)/SO(4)_\pm$ Yang-Mills theories from their center/CP mixed
anomaly, using the technique of mixed gauge theory.
This result is shown in Fig.~\ref{fig:phasetheta1and2-0} and
Table~\ref{table:phase_Spin4}. Its consistency with the soft
SUSY-breaking of $Spin(4)/SO(4)_\pm$ SYM is also discussed.
 We then generalize these analyses  to the other $Spin(\text{even})/SO(\text{even})$ gauge
 groups in Sec.~\ref{sec:Spin4k}
and Sec~\ref{sec:Spin4k+2}. A phase diagram read off from the center/CP anomaly for
$Spin(4k)/SO(4k)_\pm$ Yang-Mills is shown in
Fig.~\ref{fig:Spin4k_phases} and
Table~\ref{table:Spin4k_phases}, and a similar phase diagram
for $Spin(4k+2)/SO(4k+2)_\pm$ Yang-Mills is shown in
Fig.~\ref{fig:Spin4k+2_phases} and Table~\ref{table:Spin4k+2_phases}.
They all turn out to be in perfect agreement with the result of the soft SUSY breaking.
While the technique of center/CP anomaly is not available for
$Spin(2k+1)/SO(2k+1)_\pm$ Yang-Mills for $k\geq 2$, the soft SUSY breaking
 analysis is still possible for them, which we describe in
 Sec.~\ref{sec:Spin2k+1} for completeness. In Sec.~\ref{sec:summary}, we
 conclude and discuss future directions. In appendix \ref{app:line}, we
 summarize our notation for the charges of line operators.

\section{Review of $SU(2)$ case}
\label{sec:SU2}

Here we briefly review, following \cite{Gaiotto:2017yup}, how the phase diagram of $SU(2)$
Yang-Mills theory can be read off from the CP/center mixed anomaly via
the technique of mixed gauge theory.

\subsection{CP/center mixed anomaly}
The action of the $SU(2)$ Yang-Mills theory
is 
\begin{align}
  S = \int
\Tr(-\frac{1}{4g} F \wedge \ast F + \frac{i \theta}{8 \pi^2} F \wedge
 F)~, 
\label{eq:action}
\end{align}
where $F$ is the $SU(2)$ gauge field strength, 
and $g$ is the Yang-Mills coupling.
In the absence of the background gauge field of the center symmetry, the
parameter $\theta$ has periodicity $\theta \sim \theta + 2\pi$
since the instanton number
$\frac{1}{8\pi^2}\int \text{Tr}(F\wedge F)$ is always an integer. This
implies that, at least classically, the theory at $\theta = 0,\pi$ is
CP-invariant. Indeed, the CP transformation 
\begin{equation}
  CP : \theta \to -\theta
\end{equation}
preserves $\theta =0$, and also preserves $\theta = \pi$ up to the
$2\pi$-shift of the $\theta$-parameter.

The $SU(2)$ Yang-Mills theory has a 1-form $\mathbb{Z}_2$ global
symmetry called the center symmetry. 
It keeps all local operators
invariant but acts non-trivially on line operators such as the Wilson
line. For the $SU(2)$ gauge group, the only genuine line operator is the
Wilson line $W$ in the fundamental representation, which transforms as
\begin{equation}
 W \to -W
\end{equation}
under the $\mathbb{Z}_2$ center symmetry.

One can gauge the above center symmetry by coupling the theory
with a two-form $\mathbb{Z}_2$ gauge field \cite{Kapustin:2014gua, Gaiotto:2014kfa}. This
gauging adds an extra interaction term to the action \eqref{eq:action},
which breaks the invariance under $\theta \to \theta + 2\pi$ 
\cite{Gaiotto:2017yup}. This implies that the CP symmetry at $\theta =
\pi$ is
explicitly broken when the center symmetry is gauged, signaling
the presence of a CP/center mixed anomaly.

To be more specific, when the theory is coupled to a background gauge
field $B$ for the $\mathbb{Z}_2$ center symmetry, the $2\pi$-shift of
the $\theta$ parameter induces the following change of the action
\begin{align}
    \Delta S = \frac{i\pi}{2} \int_{\mathcal{M}}
\mathfrak{P}(B)~, \qquad B \in 
Z^2(\mathcal{M}, \mathbb{Z}_2)~,
\label{eq:4danomaly}
\end{align}
where $\mathcal{M}$ is the spacetime manifold, and $B$ is a two-cocycle
describing the background $\mathbb{Z}_2$ gauge field. The cohomology class
$\mathfrak{P}(B)$ is the Pontryagin
square of $B$, which is an element of
$H^{4}(\mathcal{M},\mathbb{Z}_4)$.\footnote{For the
definition of the Pontryagin square, see for instance
\cite{Hatcher_Chap5} and \cite{Kapustin:2013qsa, Gaiotto:2014kfa, Benini:2018reh}.}
Its integral $\int_{\mathcal{M}} \mathfrak{P}(B)$
can be an odd or even integer
when $\mathcal{M}$ is an arbitrary
four-manifold, but is always even 
when $\mathcal{M}$ is a spin four-manifold. In the rest of this paper, we
assume that $\mathcal{M}$ is a spin manifold, and therefore
$\int_{\mathcal{M}} \frac{\mathfrak{P}(B)}{2} \in \mathbb{Z}$. In that case, 
we have 
\begin{equation}
 \Delta S \in \pi i \, \mathbb{Z}~.
\label{eq:DeltaS2}
\end{equation}
Since \eqref{eq:DeltaS2} is not generally an integer multiple of $2\pi$,
we see that the invariance under $\theta \to \theta +2\pi$ can be broken by
coupling the theory with a two-form gauge field $B$. This  
is the mixed CP/center anomaly.

One can interpret \eqref{eq:DeltaS2} as a modified periodicity of the
$\theta$-parameter. Indeed it 
implies that, in the presence of 
the background gauge field for the center symmetry,
the periodicity of $\theta$
is replaced by
\begin{equation}
 \theta \sim \theta + 4\pi~.
\end{equation}
This change of periodicity plays an important role in the study of the
phase diagram as reviewed below.

\subsection{Mixed $SU(2)$ gauge theory}

In the rest of this section, we review how the CP/center mixed
anomaly 
can be
used to read off the phase structure of the $SU(2)$ Yang-Mills theory. A
key ingredient is what is called the {\it mixed gauge theory} \cite{Gaiotto:2017yup}.

To describe it,
we first compactify the
four-dimensional $SU(2)$ Yang-Mills theory on $S^1$ with the
circumference $\beta$.
The $\mathbb{Z}_2$ center symmetry in four dimensions now
gives rise to a $\mathbb{Z}_2$ 1-form symmetry and a $\mathbb{Z}_2$
0-form symmetry in three dimensions.
When $\theta=n\pi$ for $n\in\mathbb{Z}$, the theory also has an extra
$\mathbb{Z}_2$ 0-form symmetry
corresponding to the 4d CP symmetry. 
We want to identify a vacuum structure that is consistent with these
global symmetries.

To that end, we consider gauging the 3d $\mathbb{Z}_2$ 1-form
symmetry, 
which gives rise to a $\mathbb{Z}_2$
0-form global symmetry.\footnote{In $d$-dimensions, gauging a discrete $p$-form
global symmetry gives rise to a $(d-p-2)$-form global symmetry.} In other words, we convert the 3d 1-form symmetry
into a 3d 0-form symmetry. Since one can always recover the former
symmetry by gauging the latter back, no information is lost by
this conversion. The resulting 3d theory is called the {\it mixed
$SU(2)$ gauge theory}. Since the global symmetries are now
all 0-form symmetries, one can analyze the vacuum structure of this
theory by the
standard method of quantum field theory.

\begin{table}
 \centering
\begin{tabular}{c||c|c|c}
 $\theta$ modulo $4\pi$ & $\theta \neq n\pi$ for any $n\in \mathbb{Z}$ & $\theta=0,2\pi$ & $\theta  =\pi,3\pi$ \\ 
\hline
global symmetry &  $\mathbb{Z}_2^{(c)}\times \mathbb{Z}_2^{(h)}$ & $\mathbb{Z}_2^{(c)}\times
	 \mathbb{Z}_2^{(h)}\times \mathbb{Z}_2^{(CP)}$ &
	    $\mathbb{D}_8$ \\
\end{tabular}
\caption{The global symmetry of the 3d mixed $SU(2)$
 gauge theory. These are all 0-form symmetries.}
\label{table:globalsym_SU2}
\end{table}

The global symmetry of the mixed $SU(2)$ theory is shown in
Table~\ref{table:globalsym_SU2}.
When $\theta\neq n\pi$ for any $n\in\mathbb{Z}$, the mixed gauge theory
has only a $\mathbb{Z}_2\times \mathbb{Z}_2$ 0-form symmetry.
Here, one $\mathbb{Z}_2$ 0-form symmetry arises from the 4d center symmetry in
the $S^1$-compactification. We denote it by $\mathbb{Z}_2^{(c)}$ and
its generator by $c$. The
other $\mathbb{Z}_2$ 0-form symmetry arises when
gauging the 3d $\mathbb{Z}_2$ 1-form symmetry. We denote it 
by $\mathbb{Z}_2^{(h)}$ and its generator by $h$.

When $\theta = n\pi$ for an integer $n$, the mixed theory has an extra $\mathbb{Z}_2$ 0-form
symmetry arising from the 4d CP symmetry. We denote it by
$\mathbb{Z}_2^{(CP)}$ and its generator by $CP$. 
When $n$
is even, the  generators $c, h$ and $CP$
commute with each other, and therefore the full global symmetry group is $\mathbb{Z}_2^{(c)}\times \mathbb{Z}_2^{(h)}\times \mathbb{Z}_2^{(CP)}$.
However, when $n$ is odd,
$c$ and
$CP$ do {\it not} commute. Indeed, the 4d CP/center mixed anomaly \eqref{eq:4danomaly}
implies
the following relations among $c,h$ and $CP$ (at
$\theta = n\pi$ for odd $n$):
\begin{align}
CP\cdot c\cdot CP = c\cdot h~,\qquad CP\cdot h = h\cdot CP~,\qquad
 c\cdot h=h\cdot c~,\qquad c^2 = h^2=CP^2=1~.
  \label{eq:relation_generators_su(2)}
\end{align}
This means  that the full global symmetry group for odd $n$ is 
$\mathbb{D}_8$.

\subsection{Phase diagram of mixed $SU(2)$ gauge theory}

One can identify the vacuum structure of the mixed $SU(2)$ gauge theory
so that it correctly gives a representation of the global symmetry
group shown in Table~\ref{table:globalsym_SU2}.

To see this, let us first consider 4d line operators wrapped on the
$S^1$. From the 3d view point, they are local operators with
non-vanishing electro-magnetic charge $(\lambda_e,\lambda_m)\in
\mathbb{Z}^2$. 
We take the 't Hooft
line with $(\lambda_e,\lambda_m) = (0,1)$ and the dyonic line with
$(\lambda_e,\lambda_m) = (1,1)$ as the basic line operators. We denote the
't Hooft line wrapped on $S^1$ by $T$, and the dyonic line wrapped on
$S^1$ by $D$. 
The action of  $\mathbb{Z}_2^{(c)}\times \mathbb{Z}_2^{(h)}$ on
these local operators is expressed as
\begin{align}
c &: D\rightarrow -D~,
\label{eq:c}
\\[1mm]
h &: T\rightarrow -T~,\quad D\rightarrow -D~,
\label{eq:h}
\end{align}
which can be read off from the fact that $\mathbb{Z}_2^{(c)}$ 
and $\mathbb{Z}_2^{(h)}$ are respectively related to the electric and
magnetic 1-form symmetries of the 4d theory.
The action of $\mathbb{Z}_2^{(CP)}$ (which is present only at $\theta =
n\pi$ for $n\in\mathbb{Z}$) is more non-trivial. When $\theta = n\pi$
for even $n$, it just flips the sign of the electric charge, and
therefore preserves $(T,D)$.\footnote{To be precise, the $CP$ maps $D$ to a line
with charge $(\lambda_e,\lambda_m) = (-1,1)$. However, this is
equivalent to $(\lambda_e,\lambda_m)=(1,1)$ up to the 
charge of the adjoint representation. 
Since the electro-magnetic charges of line operators are
classified up to shifts by the root lattice, we see the line with
$(\lambda_e,\lambda_m)=(-1,1)$ is equivalent to $D$.} However, when $\theta =
n\pi$ for odd $n$, the CP 
also involves $\theta \to \theta + 2\pi$. Since this 
induces a Witten effect \cite{Witten:1979ey}, the action of $\mathbb{Z}_2^{(CP)}$ is
now expressed as
\begin{equation}
 CP: T \leftrightarrow D \qquad (\text{when}\;\;\theta = n\pi \;\;
  \text{for odd} \;\; n)~.
\end{equation}

\begin{figure}[h]
    \centering
    \begin{tikzpicture}[scale=.7]
        \draw[thick, ->] (-2,0) -- (12,0) node [right] {$\theta$};
        \path[pattern=dots, pattern color=red] (-0.7,0) rectangle +(5.4,3);
        \path[pattern=dots, pattern color=blue] (5.3,0) rectangle +(5.4,3);
        \path[pattern=dots, pattern color=red] (11.2,0) rectangle +(0.8,3);
        \path[pattern=dots, pattern color=blue] (-2,0) rectangle +(0.8,3);
        \node[fill=white] at (2, 1.5) {region 1};
        \node[fill=white] at (8, 1.5) {region 2};
        \draw decorate[decoration={name=zigzag,
                  amplitude=2mm,segment length=4mm}]
                      { (5,0) -- (5,3) };
        \draw decorate[decoration={name=zigzag,
                  amplitude=2mm,segment length=4mm}]
                      { (-1,0) -- (-1,3) };
        \draw decorate[decoration={name=zigzag,
                  amplitude=2mm,segment length=4mm}]
                      { (11,0) -- (11,3) };
        \draw[thick] (8,-0.1) -- (8,0.1);
        \node at (2,0) [below] {$0$};
        \node at (5,0) [below] {$\pi$};
        \node at (8,0) [below] {$2 \pi$};
        \draw[thick] (-1,-0.1) -- (-1,0.1);
        \draw[thick] (11,-0.1) -- (11,0.1);
        \node at (-1,0) [below] {$-\pi$};
        \node at (11,0) [below] {$3\pi$};
        \draw[thick] (2,-0.1) -- (2,0.1);
    \end{tikzpicture}
    \caption{Region 1 is $\{\theta \in\mathbb{R}\,|\, -\pi +4\pi n < \theta <\pi +
 4\pi n$ for $n\in\mathbb{Z}\}$ and region 2 is
 $\{\theta\in\mathbb{R}\,|\,\pi+4\pi n < \theta <3\pi + 4\pi n$ for
 $n\in\mathbb{Z}\}$. There is a phase transition at $\theta = \pi +
 2\pi n$ for $n\in\mathbb{Z}$.}
    \label{fig:SU2_phases}
\end{figure}
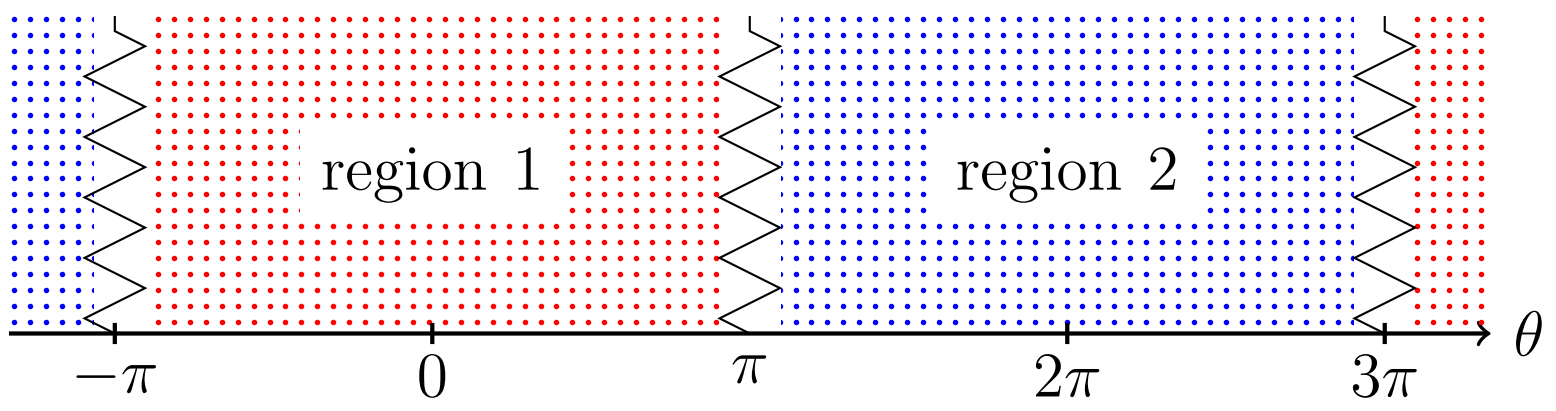

\begin{figure}
  \centering
  \begin{tikzpicture}
      \draw[thick] (1,2.5) -- (3,2.5) node [below] {$T$};
      \draw[thick] (2,1.5) -- (2,3.5) node [above] {$D$};
      \fill (1.5,2.5) circle (2pt);
      \fill (2.5,2.5) circle (2pt);
      \node[below] at (2.1,1.3) {region 1};
      \draw[thick] (9,2.5) -- (11,2.5) node [below] {$T$};
      \draw[thick] (10,1.5) -- (10,3.5) node [above] {$D$};
      \fill (10,3) circle (2pt);
      \fill (10,2) circle (2pt);
      \node[below] at (10.1,1.3) {region 2};
      \draw[thick] (5,2.5) -- (7,2.5) node [below] {$T$};
      \draw[thick] (6,1.5) -- (6,3.5) node [above] {$D$};
      \fill (5.5,2.5) circle (2pt);
      \fill (6.5,2.5) circle (2pt);
      \fill (6,3) circle (2pt);
      \fill (6,2) circle (2pt);
      \node[below] at (6.1,1.3) {\Large $\substack{\theta = \pi,\, 3\pi
   \\[1mm] (\text{modulo }4\pi)}$};
  \end{tikzpicture}
  \caption{The vacuum structure of the 3d mixed $SU(2)$ gauge theory below
 the deconfining temperature. Regions 1 and 2 are specified in
 Fig.~\ref{fig:SU2_phases}. Each
 dot stands for a single vacuum, and the axes represent the VEVs of $T$
 and $D$.}
  \label{fig:vacua_mixed_SU2}
\end{figure}
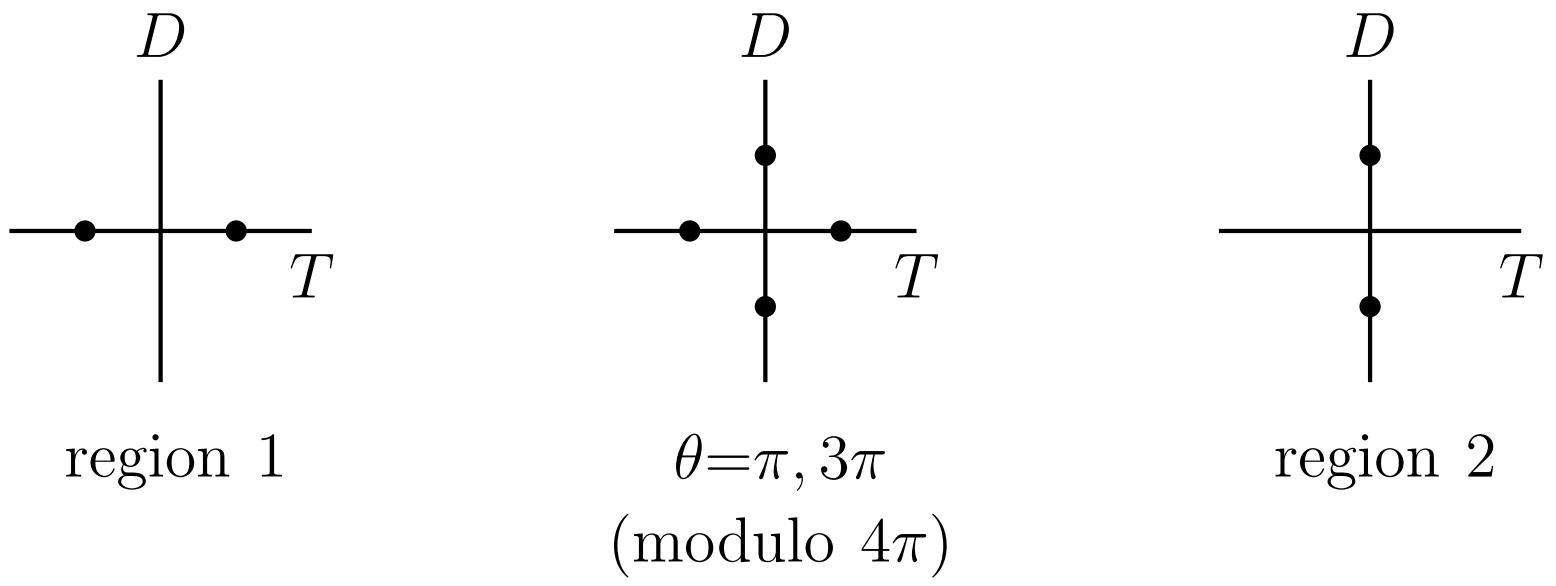

A vacuum structure consistent with the above actions of $c,h$ and $CP$ 
was identified in \cite{Gaiotto:2017yup}. In Fig.~\ref{fig:SU2_phases} and Fig.~\ref{fig:vacua_mixed_SU2},
we reproduce the vacuum structure and phase diagram for temperature lower than the deconfining temperature.\footnote{For
 the complete vacuum structure including the high temperature region,
 see Fig.~2 of \cite{Gaiotto:2017yup}. 
 For an early discussion on the implication of the axial
 anomaly for QCD at finite temperature, see
 \cite{Itoyama:1982up}.}
There are two vacua for generic values of $\theta$, but four vacua when
$\theta  = n\pi$ for odd $n$. 

Note that the above is not the unique possible phase diagram consistent
with the center/CP anomaly. In particular, it is based on an assumption that
the $SU(2)$ Yang-Mills is gapped for all values of $\theta$. For an
alternative scenario without this assumption, see Sec.~7 of \cite{Gaiotto:2017yup}.
We here follow \cite{Gaiotto:2017yup} and assume that
(at zero temperature)
the $SU(2)$ Yang-Mills theory is gapped for all values of $\theta$ and
has a trivial non-degenerate vacuum at $\theta=0$. 
Then the above phase diagram is the unique one consistent with the
global symmetry.

\subsection{Phase diagram of $SU(2)$ and $SO(3)_\pm$ Yang-Mills theories}

If the above phase diagram of the mixed gauge theory is correct,
one can now read off the phase diagram of the $SU(2)/SO(3)_\pm$ Yang-Mills
theories by discrete gaugings.
Indeed, these
Yang-Mills theories are obtained by the following
$\mathbb{Z}_2$-gaugings from the mixed $SU(2)$ theory:
\begin{align}
    SU(2) &= \big(\text{mixed } SU(2)\big)\,\big/\, \mathbb{Z}_2^{(h)}~,
\label{eq:SU2}
\\
    SO(3)_{+} &= \big(\text{mixed } SU(2)\big)\,\big/\, \mathbb{Z}_2^{(c)}~,
\label{eq:SO3+}
\\
    SO(3)_{-} &= \big(\text{mixed }SU(2)\big)\,\big/\, \mathbb{Z}_2^{(c\cdot h)}~,
  \label{eq:SO3-}
\end{align}
where a quotient means a discrete gauging. 
Here, the relation
\eqref{eq:SU2} follows from the definition of the mixed gauge
theory. The next one \eqref{eq:SO3+} follows from the fact that $SO(3)$
gauge theory is obtained from $SU(2)$ gauge theory by gauging its
$\mathbb{Z}_2$ center symmetry. From the 3d viewpoint, this means to
gauge the 3d 0-form $\mathbb{Z}_2^{(c)}$ as well as the 3d 1-form
$\mathbb{Z}_2$ symmetry of the compactified theory. Since the latter was already gauged in
constructing the mixed gauge theory, we only need to gauge
$\mathbb{Z}_2^{(c)}$ of the mixed theory
to obtain an $SO(3)$ Yang-Mills theory. 
The
difference between $SO(3)_+$ and $SO(3)_-$ is the spectrum of genuine line
operators, which is translated into the difference between
\eqref{eq:SO3+} and \eqref{eq:SO3-}.

Let us now look at the $SU(2)$ theory and analyze its phase diagram
using \eqref{eq:SU2}. Note that, since $\mathbb{Z}_2^{(h)}$ acts on $T$ and $D$ as in
Eq~\eqref{eq:h}, neither $T$ nor $D$ is invariant under
$\mathbb{Z}_2^{(h)}$. The gauge-invariant operator is instead the Wilson line
\begin{equation}
 W =TD~.
\end{equation}
The vacuum structure is read off
from 
Fig.~\ref{fig:vacua_mixed_SU2} by identifying two vacua related by
$\mathbb{Z}_2^{(h)}$ as a single vacuum. We see that only a single vacuum with $\langle W \rangle
=0$ exists both in
regions 1 and 2 of Fig.~\ref{fig:SU2_phases},
which is in the confining
phase in region 1, 
and in an oblique confining phase in region 2. 
These two phases are distinguished only by the $\mathbb{Z}_2$ center
symmetry.
Note that, 
when $\theta/\pi$ is an odd integer, these two vacua exist
simultaneously. 
When a background gauge field of the center symmetry is turned on,
these two vacua can be distinguished and therefore the periodicity of
$\theta$ is $\theta \sim \theta + 4\pi$. When the background gauge field
is turned off, one cannot distinguish the two vacua in regions 1 and 2,
and therefore the periodicity is  $\theta \sim \theta +
2\pi$. 
This result is summarized in the first line of Table~\ref{table:phases_SU2}.

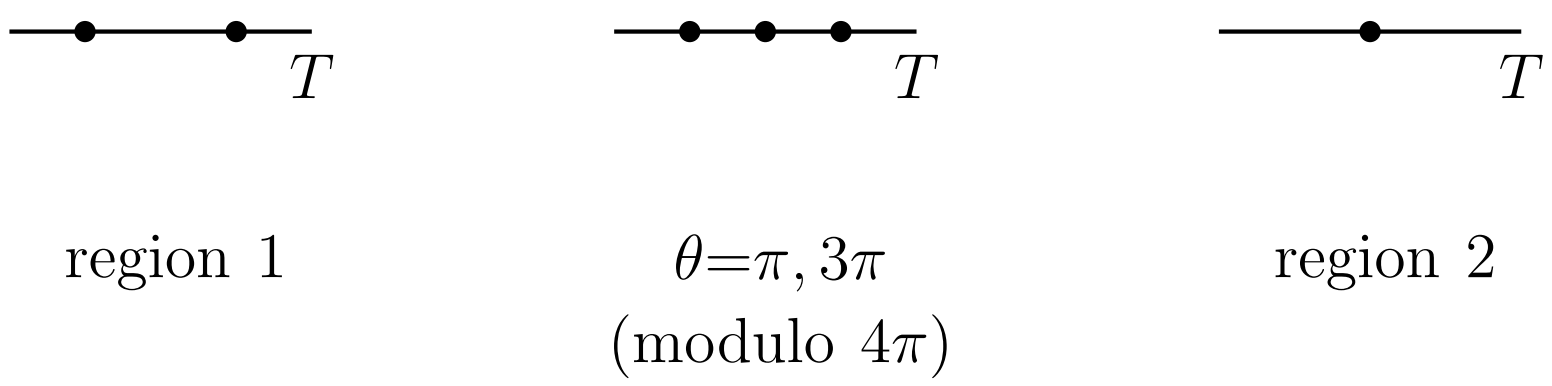
\begin{figure}
  \centering
  \begin{tikzpicture}
      \draw[thick] (1,2.5) -- (3,2.5) node [below] {$T$};
      \fill (1.5,2.5) circle (2pt);
      \fill (2.5,2.5) circle (2pt);
      \node[below] at (2.1,1.3) {region 1};
      \draw[thick] (9,2.5) -- (11,2.5) node [below] {$T$};
      \fill (10,2.5) circle (2pt);
      \node[below] at (10.1,1.3) {region 2};

      \draw[thick] (5,2.5) -- (7,2.5) node [below] {$T$};
      \fill (5.5,2.5) circle (2pt);
      \fill (6.5,2.5) circle (2pt);
      \fill (6,2.5) circle (2pt);
      \node[below] at (6.1,1.3) {\Large $\substack{\theta = \pi,\, 3\pi
   \\[1mm] (\text{modulo }4\pi)}$};
  \end{tikzpicture}
  \caption{The vacuum structure of the $SO(3)_+$ Yang-Mills theory compactified
 on $S^1$. The two vacua in region 1 corresponds to two different
 holonomies of the unbroken $\mathbb{Z}_2$ gauge symmetry along the
 $S^1$. When going back to the original 4d theory, they are mapped to a
 single vacuum supporting the $\mathbb{Z}_2$ TQFT.}
  \label{fig:vacua_SO3+}
\end{figure}

One can similarly read off the phase diagrams of the $SO(3)_+$ Yang-Mills
theory using \eqref{eq:SO3+}. Since $\mathbb{Z}_2^{(c)}$ acts on $T$ and $D$ as in
\eqref{eq:c}, the gauge-invariant operator
is now $T$ instead of $W$. The vacua of $SO(3)_+$ Yang-Mills theory is obtained
by taking the $\mathbb{Z}_2^{(c)}$-quotient of the vacua shown in
Fig.~\ref{fig:vacua_mixed_SU2}. We see that there are two vacua
characterized by $\langle T\rangle \neq 0$ in region 1, but there is only
one vacuum with $\langle T\rangle =0$ in region 2
(Fig.~\ref{fig:vacua_SO3+}). Note that, in region 1, the
$\mathbb{Z}_2^{(c)}$ gauge symmetry does not couple to anything, and
therefore there is an unbroken $\mathbb{Z}_2$ gauge theory in the
infrared.\footnote{There are two vacua supporting this $\mathbb{Z}_2$
TQFT in three dimensions. In the original 4d $SO(3)$ theory, there is
only a single vacuum supporting the $\mathbb{Z}_2$ TQFT, which gives
rise to two distinct vacua upon compactification. These two vacua are
specified by the $\mathbb{Z}_2$ holonomy along the compactified direction. } In contrast, in region 2, the $\mathbb{Z}_2$ gauge symmetry is
broken and therefore the single vacuum is in the trivially gapped phase.
These results
are summarized in the second line of Table~\ref{table:phases_SU2}.

Finally, the phase digram of the
$SO(3)_-$ Yang-Mills theory is obtained exactly in the same way. The result is
precisely such that 
\begin{equation}
 SO(3)_\pm^{\theta + 2\pi}  = SO(3)_\mp^{\theta}~,
\end{equation}
as expected.
Namely, the roles of regions 1 and 2 are exchanged compared to
$SO(3)_+$, as shown in the third line of Table~\ref{table:phases_SU2}.

\begin{table}
 \centering
\begin{tabular}{c||c|c} 
gauge group $\backslash$ regions in Fig.~\ref{fig:SU2_phases} &
     region 1 & region 2  \\
\hline\hline
$SU(2)$ & trivial & SPT  \\
\hline
$SO(3)_+$ & $\mathbb{Z}_2$ TQFT & trivial  \\
\hline
$SO(3)_-$ & trivial & $\mathbb{Z}_2$ TQFT  \\
\end{tabular}
\caption{The IR TQFT for various vacua of 4d $SU(2)/SO(3)_\pm$ Yang-Mills theory}
\label{table:phases_SU2}
\end{table}

\section{$Spin(4)/SO(4)_\pm$ Yang-Mills theory}
\label{sec:Spin4}

In this section, 
we apply the technique of the mixed center/CP anomaly to the
$Spin(4)/SO(4)_{\pm}$ Yang-Mills theories and identify their
phase diagrams. Since $Spin(4)\simeq SU(2)\times SU(2)$, the relevant
center/CP anomaly can be read off from that of the $SU(2)$ Yang-Mills
theory. 

We first consider the mixed $Spin(4)$ gauge theory and identify
its phase diagram, from which we read off the phase
diagram of the $Spin(4)/SO(4)_\pm$ Yang-Mills theories by discrete gaugings. We also show that the resulting phase
diagrams are in perfect agreement with what are suggested by the softly
broken $\mathcal{N}=1$ SYM.

\subsection{$Spin(4)$ gauge theory}

We begin by considering $Spin(4)\simeq SU(2) \times SU(2)$ Yang-Mills
theory on $\mathbb{R}^4$.
Its action is 
\begin{align}
  S = \int 
\sum_{k=1}^2 \Tr(-\frac{1}{4g_k} F^{(k)} \wedge \ast F^{(k)} + \frac{i \theta_k}{8 \pi^2} F^{(k)} \wedge F^{(k)} 
)~, 
\end{align}
where $F^{(1)}$ and $F^{(2)}$
are the field strengths of two $SU(2)$ gauge
fields, and $g_1$ and $g_2$
are Yang-Mills couplings.
The parameters $\theta_k \; (k = 1, 2)$ are periodic 
$\theta_k \sim \theta_k + 2\pi$
unless a background gauge field of the $\mathbb{Z}_2\times
\mathbb{Z}_2$ center symmetry is turned on.
The CP transformation flips the signs of these
$\theta$-parameters simultaneously as
\begin{equation}
  CP : \theta_k \to -\theta_k \quad (k=1,2)~,
\end{equation}
and therefore the CP is a symmetry only at $(\theta_1,\theta_2) = (n_1\pi,\,n_2\pi)$
for $n_1,n_2\in \mathbb{N}$.

When $(\theta_1,\theta_2) = (n_1\pi,n_2\pi)$ for $n_k\in\mathbb{N}$ and
at least one of $n_k$ is odd, the
theory has a mixed center/CP anomaly.
 Specifically, in the
presence of the background gauge field $(B_1,B_2)$ of the
$\mathbb{Z}_2\times \mathbb{Z}_2$ center symmetry, shifting the
$\theta$-parameters as $\theta_k \to \theta_k+2\pi n_k$
induces an extra term
\begin{equation}
 \Delta S = \frac{i\pi}{2} \sum_{k=1}^2 n_k \int_{\mathcal{M}} \mathfrak{P}(B_k)
\label{eq:DeltaSk}
\end{equation}
 in the action.
When $\mathcal{M}$ is a spin manifold, the integrals
 $\int_{\mathcal{M}}\mathfrak{P}(B_k)$ are even integers and
 therefore \eqref{eq:DeltaSk} is expressed as
\begin{equation}
 \Delta S = i\pi \sum_{k=1}^2 \ell_k
\end{equation}
for some $\ell_k\in \mathbb{Z}$. 
If $n_1$ or $n_2$ is odd, the sum $\sum_{k=1}^2 \ell_k$ can
be odd depending on the values of $B_k$. In that case, the shift
$\theta_k\to \theta_k+2\pi n_k$ does not preserve the theory.
This means that the CP at $(n_1\pi, n_2\pi)$ is broken in the presence
of non-vanishing $B_k$, signaling the the mixed
center/CP anomaly at these values of $\theta_k$.
This also implies that, when generic background gauge
fields $(B_1,B_2)$ are turned on, the periodicity of $\theta_k$
is modified to be
\begin{equation}
 \theta_k  \sim \theta_k + 4\pi \quad (k=1,2)~.
\end{equation}

\subsection{3d mixed $Spin(4)$ theory and its phase diagram}

When compactifying the 4d $Spin(4)$ theory
on $S^1$, the $\mathbb{Z}_2\times \mathbb{Z}_2$
center symmetry splits into a 0-form and a 1-form $\mathbb{Z}_2\times
\mathbb{Z}_2$ symmetry in three dimensions. By gauging the 3d 1-form
$\mathbb{Z}_2\times \mathbb{Z}_2$ symmetry, we obtain the mixed
$Spin(4)$ gauge theory.

The mixed $Spin(4)$ theory is a 3d theory with at least a 0-form
$(\mathbb{Z}_2)^4$ symmetry.
We denote by $\mathbb{Z}_2^{(c_1)} \times \mathbb{Z}_2^{(c_2)}$ the
3d 0-form symmetry arising from the 4d center symmetry upon 
compactification, and by
$\mathbb{Z}_2^{(h_1)}\times \mathbb{Z}_2^{(h_2)}$ the 0-form symmetry
obtained by gauging the 3d 1-form symmetry. All these discrete symmetries act on the
space of 3d local operators arising from 4d line operators wrapping
on $S^1$. We denote 't Hooft lines wrapped
on $S^1$ by $T_1$ and $T_2$ where the sub-script represents under which of the
two $SU(2)$ factors in $Spin(4)\simeq SU(2)_1\times SU(2)_2$ they are charged magnetically.
Similarly, we denote by $D_k$ dyonic lines wrapped on $S^1$.
 Then the action of
$\mathbb{Z}_2^{(c_1)}\times \mathbb{Z}_2^{(h_1)}\times
\mathbb{Z}_2^{(c_2)}\times \mathbb{Z}_2^{(h_2)}$ on them is expressed as
\begin{equation}
 c_k:D_k\to -D_k~,\qquad h_k:T_k\to -T_k,\;D_k\to -D_k~,
\label{eq:TkDk}
\end{equation}
where $c_1,\,c_2,\,h_1$ and $h_2$ are the generators of
$\mathbb{Z}_2^{(c_1)},\,\mathbb{Z}_2^{(c_2)},\,\mathbb{Z}_2^{(h_1)}$ and
$\mathbb{Z}_2^{(h_2)}$, respectively.

The full global symmetry of the mixed $Spin(4)$ theory depends on the values of
$\theta_1$ and $\theta_2$. When $\theta_k/\pi\not \in \mathbb{Z}$ for
some $k$,
 the
4d classical Lagrangian is not CP-invariant, and therefore the above
$(\mathbb{Z}_2)^4$ symmetry is the full global symmetry of the mixed
$Spin(4)$ theory.
When $\theta_k/\pi \in \mathbb{Z}$ for both $k=1$ and $k=2$, the
mixed theory has an additional $\mathbb{Z}_2$ symmetry corresponding to the 4d CP
symmetry.
We denote its generator  by $CP$. When $\theta_k/\pi$ is an odd integer,
the $CP$ exchanges $T_k$ and $D_k$, i.e.,
\begin{equation}
 T_k \leftrightarrow D_k~.
\end{equation}
In the presence of $\mathbb{Z}_2^{(CP)}$, the full global symmetry is a 
$\mathbb{Z}_2^{(CP)}$-extension of $\mathbb{Z}_2^{(c_1)}\times
\mathbb{Z}_2^{(c_2)}\times \mathbb{Z}_2^{(h_1)}\times \mathbb{Z}_2^{(h_2)}$, as shown in
Table~\ref{table:globalsym_SO4}.
In particular, when both $\theta_1/\pi$ and
$\theta_2/\pi$ are odd integers, the full global symmetry is
$(\mathbb{Z}_2)^4\rtimes \mathbb{Z}_2$ such that
\begin{equation}
 CP\cdot c_k \cdot CP = c_k\cdot h_k~,\qquad CP\cdot h_k = h_k\cdot
 CP~,\qquad c_k\cdot h_\ell = h_\ell\cdot c_k~,\qquad c_k^2 = h_k^2 = CP^2={\bf 1}~.
\label{eq:globalsym_SO(4)_atpi}
\end{equation}
Here, the first two equations show that the semi-direct product is
associated with the most non-trivial homomorphism from $\mathbb{Z}_2^{(CP)}$ to $\text{Aut}\,(\mathbb{Z}_2)^4$.\footnote{There
are two non-trivial semi-direct products of the form $(\mathbb{Z}_2)^4
\rtimes \mathbb{Z}_2$ up to isomorphisms, specified by a homomorphism
$\varphi:\mathbb{Z}_2\to\text{Aut}((\mathbb{Z}_2)^4) \simeq
GL(4,\mathbb{Z}_2)$. When $\varphi(1)-{\bf 1}$ is a rank-one matrix, the corresponding semi-direct product is
isomorphic to $\mathbb{D}_8\times (\mathbb{Z}_2)^2$. The semi-direct
product discussed around \eqref{eq:globalsym_SO(4)_atpi} is the one
corresponding to the case in which 
$\text{rank}(\varphi(1)-{\bf 1})=2$.}

\begin{table}
 \centering
\begin{tabular}{c||c|c|c}
 & $\theta_1/\pi\not\in\mathbb{Z}$ &
	 $\theta_1/\pi$ is an even integer & $\theta_1/\pi$ is an odd integer \\ 
\hline\hline
$\theta_2/\pi\not\in \mathbb{Z}$  &  $(\mathbb{Z}_2)^4$ & $(\mathbb{Z}_2)^4$ & $(\mathbb{Z}_2)^4$ \\
\hline
$\theta_2/\pi$ is an even integer & $(\mathbb{Z}_2)^4$ & $(\mathbb{Z}_2)^5$ & $(\mathbb{Z}_2)^2 \times
	     \mathbb{D}_8$\\
\hline
$\theta_2/\pi$ is an odd integer & $(\mathbb{Z}_2)^4$ & $\mathbb{D}_8\times (\mathbb{Z}_2)^2$ &
	    $(\mathbb{Z}_2)^4\rtimes \mathbb{Z}_2$  \\
\end{tabular}
\caption{The global symmetry of the 3d mixed $SU(2)\times SU(2)$
 theory. These are all 0-form symmetries.}
\label{table:globalsym_SO4}
\end{table}

\begin{figure}[h]
  \centering
  \begin{tikzpicture}[scale=.7]
      \draw[thick, ->] (-1,0) -- (11,0) node [right] {$\theta_1$};
      \draw[thick, ->] (0,-1) -- (0,11) node [above] {$\theta_2$};

      \draw[thick] (10,-0.1) -- (10,0.1);
      \node at (0,0) [anchor=north east] {$0$};
      \node[below] at (2.5,0) {$\pi$};

      \node[below] at (7.5,0) {$3 \pi$};
      \node[below] at (10,0) {$4 \pi$};
      \node[left] at (0,2.5) {$\pi$};

      \node[left] at (0,7.5)  {$3 \pi$};
      \node[left] at (0,10) {$4 \pi$};

      \draw[thick, dashed] (0,2.5) -- (10,2.5);

      \draw[thick, dashed] (0,7.5) -- (10,7.5);
      \draw[thick] (0,10) -- (10,10);
      \draw[thick, dashed] (2.5,0) -- (2.5,10);

      \draw[thick, dashed] (7.5,0) -- (7.5,10);
      \draw[thick] (10,0) -- (10,10);

      \path[pattern=horizontal lines, pattern color=red] (0,0) rectangle +(2.5,2.5);
      \path[pattern=vertical lines, pattern color=red] (2.5,2.5) rectangle +(5,5);
      \path[pattern=horizontal lines, pattern color=blue] (2.5,0) rectangle +(5,2.5);
      \path[pattern=vertical lines, pattern color=blue] (0,2.5) rectangle +(2.5,5);
      \path[pattern=horizontal lines, pattern color=red] (7.5,7.5) rectangle +(2.5,2.5);
      \path[pattern=horizontal lines, pattern color=red] (7.5,0) rectangle +(2.5,2.5);
      \path[pattern=horizontal lines, pattern color=red] (0,7.5) rectangle +(2.5,2.5);
      \path[pattern=horizontal lines, pattern color=blue] (2.5,7.5) rectangle +(5,2.5);
      \path[pattern=vertical lines, pattern color=blue] (7.5,2.5) rectangle +(2.5,5);

      \node[fill=white] at (1.25,1.25) {\small region 1};
      \node[fill=white] at (5,5) {\small region 2};
      \node[fill=white] at (5,1.25) {\small region 3};
      \node[fill=white] at (1.25,5) {\small region 4};
  \end{tikzpicture}
  \caption{The phase diagram of the mixed $Spin(4)$ theory is periodic
 under $\theta_k\to \theta_k+4\pi$. We show a
 single fundamental domain above.
The ordinary $Spin(4)/SO(4)$ Yang-Mills theories have the same
 periodicity in the presence of the background gauge field of the 1-form
 symmetry.   The IR theories present in the four
 regions are summarized in Table~\ref{table:phase_Spin4}.}
  \label{fig:phasetheta1and2-0}
\end{figure}
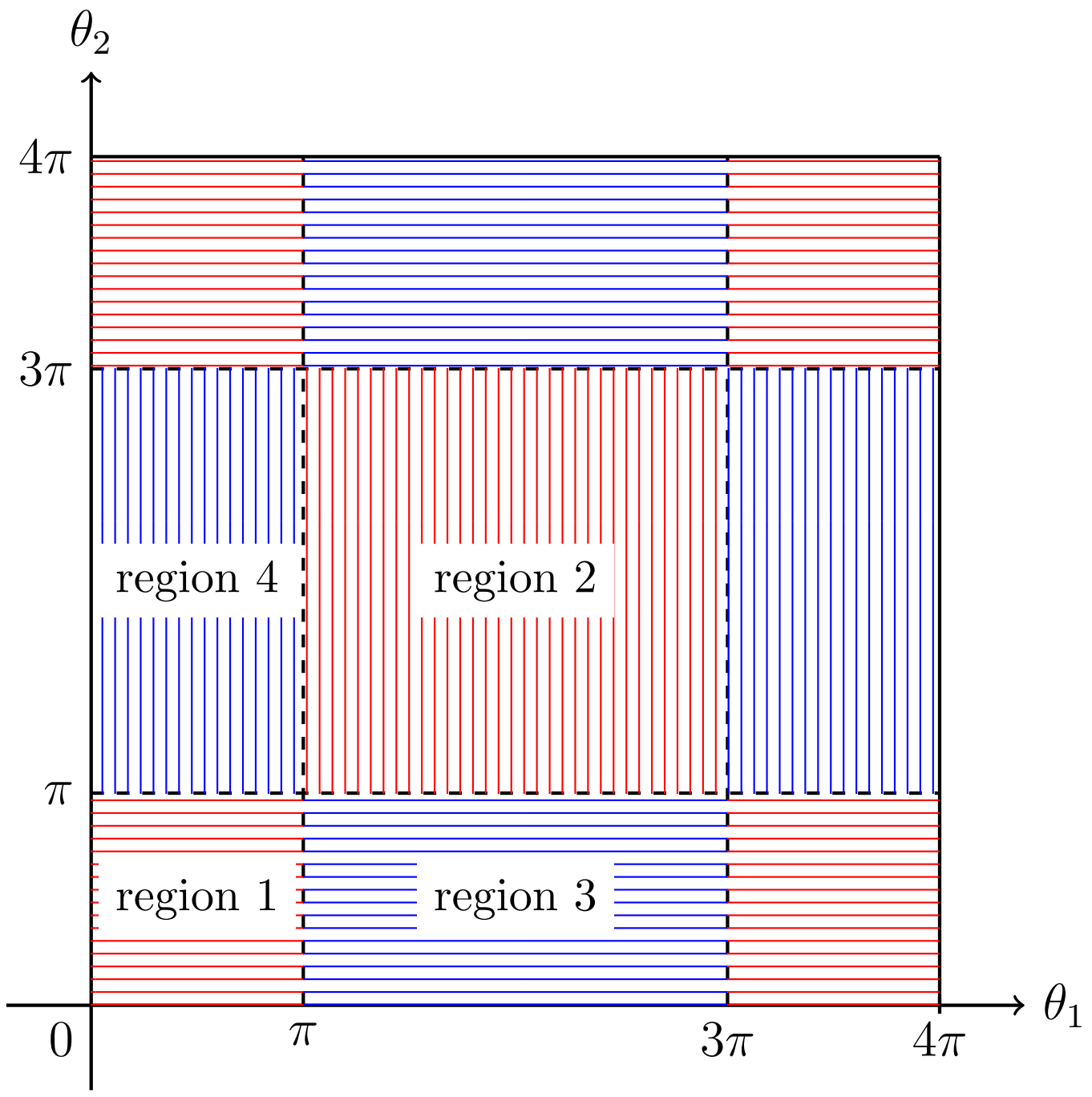

As in the case of the $SU(2)$ Yang-Mills theory reviewed in Sec.~\ref{sec:SU2}, one
can find a phase
diagram of the mixed $Spin(4)$ theory that is consistent with the global
symmetry groups. 
We assume that at zero temperature the (ordinary)
$Spin(4)$ Yang-Mills theory is gapped for all values of $\theta_k$ and has a
trivial non-degenerate vacuum at $\theta_k = 0$, which is the same
assumption as in most sections of \cite{Gaiotto:2017yup}.
Then the phase diagram of the {\it mixed} $Spin(4)$ theory at zero
temperature is described
as in 
Fig.~\ref{fig:phasetheta1and2-0} and Table~\ref{table:vacua0}.

The four phases of the mixed $Spin(4)$ theory shown in Fig.~\ref{fig:phasetheta1and2-0} are characterized
 by the
 VEVs of $T_k$ and $D_k$, as described in Table.~\ref{table:vacua0}. For instance, in region 1,
 we have $\langle D_k\rangle = 0$ and $\langle T_k\rangle \neq 0$ for
 both $k=1,2$. Depending on the sign of $\langle T_1\rangle$ and
 $\langle T_2\rangle$, there are four different vacua in region
 1. 
This will turn out to be necessary and sufficient for the ordinary
 $Spin(4)$ Yang-Mills theory to have a unique gapped confining (or
 oblique confining) vacuum for generic values of $\theta_k$.

\begin{table}
 \centering
\begin{tabular}{c||c|c|c|c}
 & region 1& region 2&  region 3& region 4 \\ 
\hline\hline
$T_1$ & $\pm a_1$ & $0$ & $0$ & $\pm a_1$  \\
$D_1$ & $0$ & $\pm b_1$ & $\pm b_1$ & $0$ \\
\hdashline
$T_2$ & $\pm a_2$ & $0$ & $\pm a_2$ & $0$ \\
$D_2$ & $0$ & $\pm b_2$ & $0$ & $\pm b_2$ \\
\end{tabular}
\bigskip
\caption{The VEVs of $T_k$ and $D_k$ in various regions in
 Fig.~\ref{fig:phasetheta1and2-0}, where $a_k$ and $b_k$ are
 non-vanishing values. The signs are {\it not} correlated. There are four vacua in each region.}
\label{table:vacua0}
\end{table}

\begin{table}[h]
\centering
\bigskip
\begin{tabular}{c||c|c|c|c}
regions in Fig.~\ref{fig:phasetheta1and2-0} &region 1 & region 2 & region 3 & region 4 \\
\hline
unbroken symmetry & $\mathbb{Z}_2^{(c_1)}\times \mathbb{Z}_2^{(c_2)}$ &
	 $\mathbb{Z}_2^{(c_1h_1)}\times \mathbb{Z}_2^{(c_2h_2)}$ &
	     $\mathbb{Z}_2^{(c_1h_1)}\times \mathbb{Z}_2^{(c_2)}$ &
		 $\mathbb{Z}_2^{(c_1)}\times \mathbb{Z}_2^{(c_2h_2)}$
\end{tabular}
\caption{The unbroken global symmetry of the {\it mixed} $Spin(4)$ gauge
 theory in various regions
 in Fig.~\ref{fig:phasetheta1and2-0}. Here, $\mathbb{Z}_2^{(g)}$ stands for a
 $\mathbb{Z}_2$ generated by $g$.}
\label{table:phases}
\end{table}

The above four phases are also characterized by which sub-group of
$\mathbb{Z}_2^{(c_1)}\times \mathbb{Z}_2^{(h_1)}\times
\mathbb{Z}_2^{(c_2)}\times \mathbb{Z}_2^{(h_2)}$ is unbroken, as shown in Table~\ref{table:phases}.
From this, we see that the phase transitions of the mixed $Spin(4)$ theory (that occur when crossing
dashed lines in Fig.~\ref{fig:phasetheta1and2-0}) are all conventional
phase transitions rather than SPT phase transitions.

Along each dashed line in Fig.~\ref{fig:phasetheta1and2-0}, two regions are
attached to each other. When the $\theta$-parameters are on such a line,
the vacua present in these two regions exist simultaneously, leading to
eight vacua.
The most non-trivial situation occurs where two dashed lines
intersect, i.e., when both $\theta_1/\pi$ and $\theta_2/\pi$ are odd
integers. In this case, the mixed $Spin(4)$ theory has sixteen vacua
such that
\begin{equation}
 (T_1,D_1;T_2,D_2) \;=\; (\pm a_1,0; \pm a_2,0),\; (0,\pm a_1;0,\pm a_2),\;(\pm
  a_1,0;0,\pm a_2),\;(0,\pm a_1;\pm a_2,0),
\label{eq:D8}
\end{equation}
where $a_1$ and $a_2$ are a positive real number determined by the temperature. 
Note that these sixteen vacua belong to two orbits of
$(\mathbb{Z}_2)^4\rtimes \mathbb{Z}_2$.

\subsection{Phase diagram of $Spin(4)$ Yang-Mills theory}
\label{subsec:Spin4}

\begin{table}
 \centering
\begin{tabular}{c||c|c|c|c} 
gauge group $\backslash$ regions in Fig.~\ref{fig:phasetheta1and2-0} &
     region 1 & region 2 & region 3 & region 4 \\
\hline\hline
$Spin(4)$ & trivial & SPT$_{1,1}$ & SPT$_{1,0}$ & SPT$_{0,1}$ \\
\hline
$SO(4)_+$ & $\mathbb{Z}_2$ TQFT & $\mathbb{Z}_2$ TQFT & trivial & SPT \\
\hline
$SO(4)_-$ & trivial & SPT & $\mathbb{Z}_2$
		 TQFT & $\mathbb{Z}_2$ TQFT   \\
\end{tabular}
\caption{The IR theories present in the four regions in
 Fig.~\ref{fig:phasetheta1and2-0} for 4d
 $Spin(4)/SO(4)_\pm$ Yang-Mills theory. In the first line, ``SPT$_{n_1,n_2}$'' stands for the SPT
 phase protected by the center $\mathbb{Z}_2\times \mathbb{Z}_2$
 symmetry of the $Spin(4)$ Yang-Mills theory, and ``trivial'' stands for the trivial TQFT,
 i.e., SPT$_{0,0}$. In the second and third lines, the SPT is regarded as
 the SPT phase protected by the electric $\mathbb{Z}_2$ and the magnetic
 $\mathbb{Z}_2$ 1-form symmetries of
 $SO(4)$ Yang-Mills theory.}
\label{table:phase_Spin4}
\end{table}

The $Spin(4)$ Yang-Mills theory is obtained from the mixed $Spin(4)$
theory by gauging
its $\mathbb{Z}_2^{(h_1)}\times \mathbb{Z}_2^{(h_2)}$ symmetry, i.e.,
\begin{equation}
 Spin(4) = \big(\text{mixed }Spin(4)\big) \Big / \big(\mathbb{Z}_2^{(h_1)}\times \mathbb{Z}_2^{(h_2)}\big)~.
\end{equation}
The action of $\mathbb{Z}_2^{(h_1)}\times \mathbb{Z}_2^{(h_2)}$ on $T_k$ and $D_k$ are shown in
\eqref{eq:TkDk}. At generic points in regions $1\sim 4$, gauging
$\mathbb{Z}_2^{(h_1)}\times \mathbb{Z}_2^{(h_2)}$ reduces four
vacua into a single vacuum. 
Along the dashed line in Fig.~\ref{fig:phasetheta1and2-0}, eight vacua
of the mixed gauge theory are reduced to two vacua. This
can be seen by analyzing how $\mathbb{Z}_2^{(h_1)}\times \mathbb{Z}_2^{(h_2)}$
acts on these eight vacua. Finally, at intersections of two dashed
lines, the sixteen vacua of the mixed $Spin(4)$ theory reduce to four
vacua in the (ordinary) $Spin(4)$ Yang-Mills theory.

The order parameters of the $Spin(4)$ Yang-Mills theory are
the Wilson lines
\begin{equation}
 W_k \equiv T_kD_k
\end{equation}
for $k=1,2$. The above various vacua all have $\langle
W_k\rangle = 0$ for $k=1,2$, and therefore they are in the confining or
an oblique confining phase.
 In the convention that region 1 corresponds to the confining
phase (trivial phase), the other three regions correspond to oblique
confining phases, i.e., SPT phases
protected by the $\mathbb{Z}_2\times \mathbb{Z}_2$ center symmetry \cite{Tachikawa:2016xvs}. This
result is summarized in the first line of Table~\ref{table:phase_Spin4}.
Thus, the conventional phase transitions of the mixed $Spin(4)$ theory
are now all mapped to SPT phase transitions of the (ordinary) $Spin(4)$
Yang-Mills theory.

\subsection{Phase diagram of $SO(4)_\pm$ Yang-Mills theory}
\label{subsec:SO4}

Let us now turn to the $SO(4)_\pm$ Yang-Mills theory. These theories are
obtained from the mixed $Spin(4)$ theory by gauging the
$\mathbb{Z}_2^{(c_1\cdot c_2)}\times \mathbb{Z}_2^{(h_1\cdot h_2)}$ or
$\mathbb{Z}_2^{(c_1\cdot c_2\cdot h_1)}\times \mathbb{Z}_2^{(h_1\cdot h_2)}$
symmetry, i.e.,\footnote{Note here that $c_1,\, c_2,\,h_1$ and $h_2$ commute with each other, and
are squared to be the identity. Therefore, one can take for instance
$c_1\cdot c_2\cdot h_2$ instead of $c_1\cdot c_2\cdot h_1$ in the second
line without changing the result.}
\begin{align}
 SO(4)_+ &= \big(\text{mixed } Spin(4)\big) \Big/ \big(\mathbb{Z}_2^{(c_1\cdot
 c_2)}\times \mathbb{Z}_2^{(h_1\cdot h_2)}\big)~,
\\
SO(4)_- &= \big(\text{mixed }Spin(4)\big) \Big/\big(\mathbb{Z}_2^{(c_1\cdot
 c_2\cdot h_1)}\times \mathbb{Z}_2^{(h_1\cdot h_2)}\big)~.
\end{align}
From the viewpoint of the $Spin(4)$ Yang-Mills theory, these discrete
gauging correspond to gauging the diagonal $\mathbb{Z}_2$ sub-group of
the $\mathbb{Z}_2\times \mathbb{Z}_2$ center symmetry. As a result, the
$SO(4)_\pm$ Yang-Mills theory has an electric 1-form $\mathbb{Z}_2$
symmetry
and a magnetic 1-form
$\mathbb{Z}_2$ symmetry. 

\subsubsection{$SO(4)_+$}
\label{subsubsec:SO4+}

Let us first study the phase diagram of the $SO(4)_+$ theory. Since $T_k$ and $D_k$
transform under $c_1\cdot c_2$ and
$h_1\cdot h_2$ as
\begin{align}
 c_1\cdot c_2&:\quad  D_1 \to -D_1~,\quad D_2\to -D_2~,
\label{eq:c1c2}
\\
h_1\cdot h_2&: \quad T_1\to -T_1~,\quad T_2\to-T_2~,\quad D_1\to -D_1~,\quad
 D_2\to -D_2~,
\label{eq:h1h2}
\end{align}
gauging $\mathbb{Z}_2^{(c_1\cdot c_2)}\times
\mathbb{Z}_2^{(h_1\cdot h_2)}$ 
 removes 
$T_k$ and $D_k$ from the
spectrum, leaving the gauge-invariant combinations
\begin{equation}
 T_1T_2~,\qquad D_1D_2~.
\label{eq:TTDD}
\end{equation}
The VEVs of these two gauge-invariant operators are the order parameters
of the $SO(4)_+$ Yang-Mills theory. In the case $\langle T_1T_2\rangle \neq 0$ the magnetic 1-form $\mathbb{Z}_2$ is
broken, and in the case $\langle D_1D_2\rangle \neq 0$  the
electric and magnetic $\mathbb{Z}_2$ are both broken.

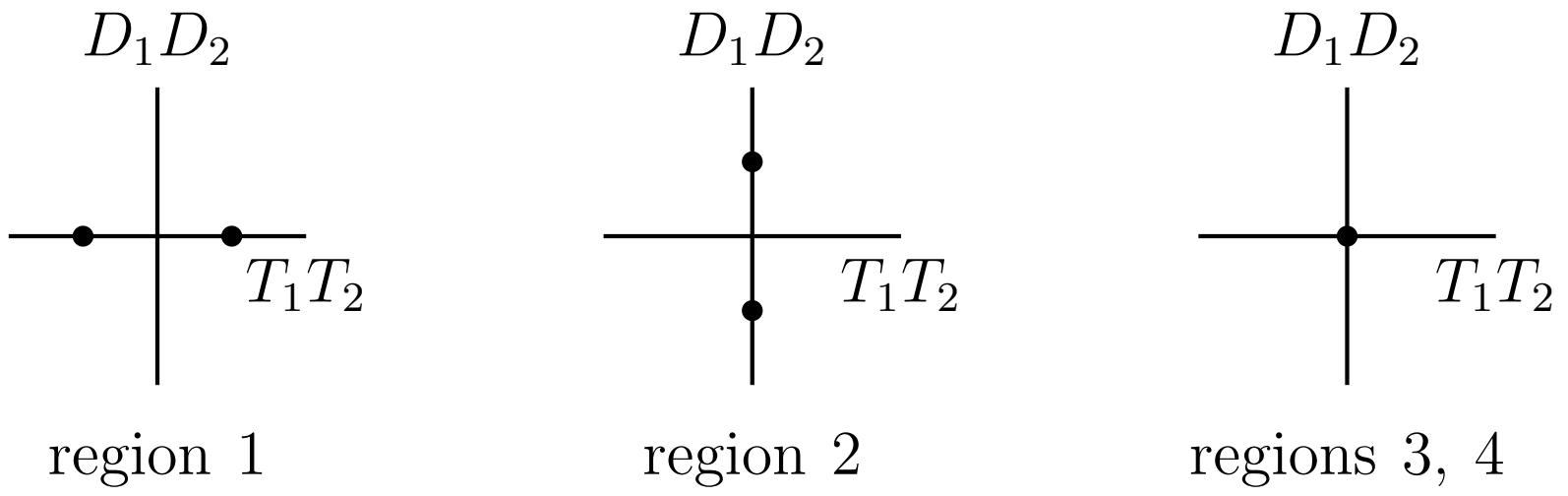
\begin{figure}
  \centering
  \begin{tikzpicture}

      \draw[thick] (1,2.5) -- (3,2.5) node [below] {$T_1 T_2$};
      \draw[thick] (2,1.5) -- (2,3.5) node [above] {$D_1 D_2$};

      \fill (1.5,2.5) circle (2pt);
      \fill (2.5,2.5) circle (2pt);

      \draw[thick] (5,2.5) -- (7,2.5) node [below] {$T_1 T_2$};
      \draw[thick] (6,1.5) -- (6,3.5) node [above] {$D_1 D_2$};

      \fill (6,2) circle (2pt);
      \fill (6,3) circle (2pt);

      \draw[thick] (9,2.5) -- (11,2.5) node [below] {$T_1 T_2$};
      \draw[thick] (10,1.5) -- (10,3.5) node [above] {$D_1 D_2$};

      \fill (10,2.5) circle (2pt);

      \node at (2,1) {region 1};
      \node at (6,1) {region 2};
      \node at (10,1) {regions 3, 4};
  \end{tikzpicture}
  \caption{The vacuum structure of $SO(4)_+$ Yang-Mills theory compactified on
 $S^1$ for generic
 values of the $\theta$-parameters. Two vacua in each of regions 1 and 2
 are associated with different $\mathbb{Z}_2$ holonomies along
 $S^1$. When going back to four dimensions, they are mapped to a single vacuum that supports a $\mathbb{Z}_2$ TQFT.}
  \label{fig:vacua1}
\end{figure}

When the
$\theta$-parameters are in region 3 or 4 of
Fig.~\ref{fig:phasetheta1and2-0}, the $\mathbb{Z}_2^{(c_1\cdot c_2)}\times
\mathbb{Z}_2^{(h_1\cdot h_2)}$ freely acts on the four vacua of the
mixed $Spin(4)$ theory, and therefore gauging it leads to
a single vacuum of the $SO(4)_+$ theory, as shown in the
rightmost picture of Fig.~\ref{fig:vacua1}. These are in the trivial and
SPT phases protected by the 1-form symmetry.  On the other hand, when the
$\theta$-parameters are in region 1 or 2, there exists a $\mathbb{Z}_2$
sub-group of $\mathbb{Z}_2^{(c_1\cdot c_2)}\times
\mathbb{Z}_2^{(h_1\cdot h_2)}$ that trivially acts on the four vacua. In this case, gauging
$\mathbb{Z}_2^{(c_1\cdot c_2)}\times \mathbb{Z}_2^{(h_1\cdot h_2)}$
gives rise to two
degenerate vacua accompanied with an unbroken
$\mathbb{Z}_2$ TQFT (the left and middle pictures of
Fig.~\ref{fig:vacua1}). These results are summarized in the second line of Table~\ref{table:phase_Spin4}.

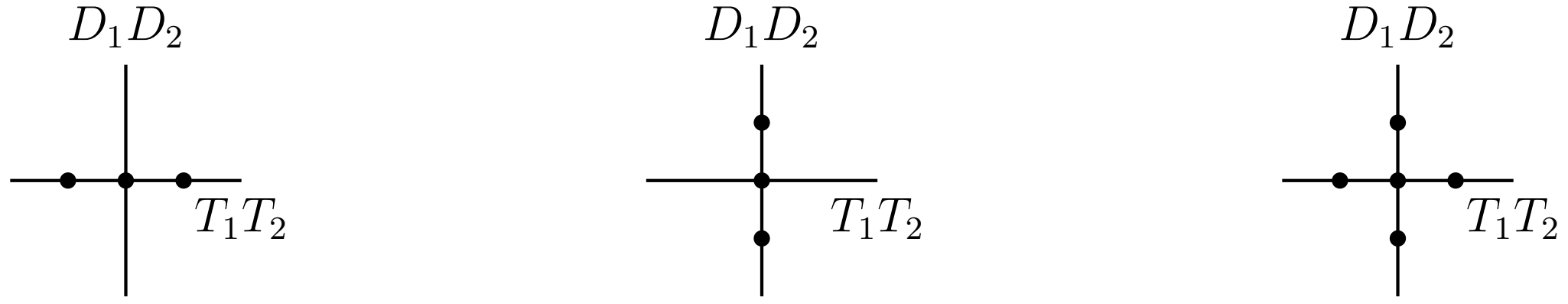
\begin{figure}
  \centering
\bigskip
  \begin{tikzpicture}
      \draw[thick] (1,2.5) -- (3,2.5) node [below] {$T_1 T_2$};
      \draw[thick] (2,1.5) -- (2,3.5) node [above] {$D_1 D_2$};

      \fill (1.5,2.5) circle (2pt);
      \fill (2.5,2.5) circle (2pt);
      \fill (2,2.5) circle (2pt);

      \draw[thick] (5+1.5,2.5) -- (7+1.5,2.5) node [below] {$T_1 T_2$};
      \draw[thick] (6+1.5,1.5) -- (6+1.5,3.5) node [above] {$D_1 D_2$};

      \fill (6+1.5,2) circle (2pt);
      \fill (6+1.5,3) circle (2pt);
      \fill (6+1.5,2.5) circle (2pt);

      \draw[thick] (9+3,2.5) -- (11+3,2.5) node [below] {$T_1 T_2$};
      \draw[thick] (10+3,1.5) -- (10+3,3.5) node [above] {$D_1 D_2$};
      \fill (10+3,3) circle (2pt);
      \fill (10+3,2.5) circle (2pt);
      \fill (10+3,2) circle (2pt);
      \fill (10+3+.5,2.5) circle (2pt);
      \fill (10+3-.5,2.5) circle (2pt);
      \node at (2-.3,1) {between regions 1 and 3};
      \node at (2-.3,.5) {or between regions 1 and 4};
      \node at (7.5,1) {between regions 2 and 3};
      \node at (7.5,.5) {or between regions 2 and 4};
      \node at (13+.5,1) {intersections of the four regions};
  \end{tikzpicture}
  \caption{The vacuum structure of $SO(4)_+$ theory compactified on
 $S^1$ when the
 $\theta$-parameters are at boundaries of different regions. 
There is a single vacuum at the origin $\langle T_1T_2\rangle =
 \langle D_1D_2\rangle = 0$ in the left two pictures, while there are
 two degenerate vacua at the origin in the rightmost picture. Every pair
 of vacua
 connected by $T_1T_2 \to -T_1T_2$ and $D_1D_2\to -D_1D_2$ is
 associated with different $\mathbb{Z}_2$ holonomies along the $S^1$,
 and therefore corresponds to a single vacuum in the original 4d
 theory.}
  \label{fig:vacua1-1}
\end{figure}

When the $\theta$-parameters are on a dashed line in Fig.~\ref{fig:phasetheta1and2-0},
the $SO(4)_+$ Yang-Mills theory has
three degenerate vacua as shown in the left two pictures of
Fig.~\ref{fig:vacua1-1}; two of the three are  accompanied by an unbroken
$\mathbb{Z}_2$ TQFT and the remaining one is in the confining/oblique
confining phase. When the $\theta$-parameters are at an intersection of
dashed lines,
the $SO(4)_+$ Yang-Mills theory has six vacua as shown in the
rightmost picture of Fig.~\ref{fig:vacua1-1}, where two vacua are
located at
$\langle T_1T_2\rangle = \langle D_1D_2\rangle = 0$.

\subsubsection{$SO(4)_-$}

The phase diagram of the $SO(4)_-$ theory is obtained similarly. We now gauge the
$\mathbb{Z}_2^{(c_1\cdot c_2\cdot h_1)}\times \mathbb{Z}_2^{(h_1\cdot
h_2)}$ symmetry of the mixed $Spin(4)$ theory, whose action is
characterized by
\begin{align}
 c_1\cdot c_2\cdot h_1&: \quad T_1 \to -T_1~,\quad D_2\to -D_2~,
\label{eq:cch}\\
h_1\cdot h_2&: \quad T_1\to -T_1~,\quad T_2\to-T_2~,\quad D_1 \to
 -D_1~,\quad D_2\to -D_2~.
\label{eq:hh2}
\end{align}
Since this can be simply obtained from \eqref{eq:c1c2} and \eqref{eq:h1h2} by $T_1
\leftrightarrow D_1$, the phase diagram of the $SO(4)_-$ Yang-Mills
theory is simply obtained by exchanging the roles of $T_1$ and $D_1$ in
our discussion of $SO(4)_+$.
In particular, the gauge-invariant order parameters are now the VEVs of
\begin{equation}
 T_1 D_2~,\qquad D_1T_2~.
\end{equation}
The resulting phase structure is summarized in the third
line of Table~\ref{table:phase_Spin4}. We see that these results are
consistent with the relation \eqref{eq:SO4_periodicity}.

\subsection{Comparison to soft SUSY breaking}

In this sub-section, we study the phase diagrams of supersymmetric
$Spin(4)/SO(4)_\pm$ Yang-Mills theories with a small gaugino mass $m$
turned on in the small $|m|$ approximation. Since the non-supersymmetric Yang-Mills is regarded as the
$m\to\infty$ limit of this theory, the phase diagram in this
approximation does not necessarily coincide with the diagrams we
obtained above.
However, we will find the two diagrams in
perfect agreement with each other, which suggests that the phase diagram
is independent of $m$. A similar analysis for the $SU(N)$ Yang-Mills
theory was performed in \cite{Gaiotto:2017yup}.

\subsubsection{$\mathcal{N}=1$ $Spin(4)$ SYM}

Let us start with the $\mathcal{N}=1$ $Spin(4)$ SYM, which has four
supersymmetric vacua  labeled by two
parameters $\epsilon_1$ and $\epsilon_2$ taking values in $\{+1,-1\}$
\cite{Intriligator:1995id, Intriligator:1995au,
Aharony:2013hda}. These four vacua
are confining and oblique confining vacua, protected by the
$\mathbb{Z}_2\times \mathbb{Z}_2$ center
symmetry. We take the convention that the vacuum for $\epsilon_1 = \epsilon_2 = 1$
corresponds to the confining phase (i.e., the trivial phase), and the
other vacua correspond to oblique confining phases.

When we introduce a soft SUSY breaking mass term, the degeneracy of
the four vacua is resolved. To see this, let us introduce a small
gaugino mass term
 \begin{equation}
\delta \mathcal{L} = m\Big(\text{Tr}\left(\eta_\alpha\eta^\alpha\right) 
		       + \text{Tr}\left(\xi_\alpha\xi^\alpha\right)\Big)
		       \;\; + \;\; \text{c.c.}~,
\end{equation}
where $(\eta_\alpha, \xi_\alpha)$ are gaugino fields in the adjoint
 representation of $Spin(4)\simeq SU(2)\times SU(2)$. The gaugino condensation is
 characterized by
\begin{equation}
 \langle \text{Tr}(\eta_\alpha \eta^\alpha)\rangle =
  \epsilon_1 \kappa_1\Lambda_1^3~,\qquad \langle \text{Tr}(\xi_\alpha \xi^\alpha)\rangle =
  \epsilon_2\kappa_2\Lambda_2^3~,
\label{eq:GC}
\end{equation}
where $\kappa_1$ and $\kappa_2$ are non-vanishing complex constants, and
$\Lambda_1$ and $\Lambda_2$ are two dynamical scales so that
$\Lambda_k^6 = |\Lambda_k|^6e^{i\theta_k}$. We take the origins of
$\theta_k$ so that $m\kappa_1$ and $m\kappa_2$ are positive real numbers.

When $|m|$ is sufficiently small, the effective potential is evaluated as $V = -m(\langle \text{Tr}(\eta_\alpha\eta^\alpha)\rangle 
		       + \langle
		       \text{Tr}(\xi_\alpha\xi^\alpha)\rangle)
		       +  \text{c.c.}$
Using \eqref{eq:GC} and $\Lambda_k^6 = |\Lambda_k|^6 e^{i\theta_k}$, one
		       can rewrite it as 
\begin{equation}
 V = - 2\sum_{k=1}^2\text{Re}\left( \epsilon_k
						      |m\kappa_k\Lambda_k^3|e^{\frac{i\theta_k}{2}}\right)
 = -2\sum_{k=1}^2\epsilon_k |m\kappa_k\Lambda_k^3|\cos \frac{\theta_k}{2} ~,
\label{eq:soft-mass}
\end{equation}
which is periodic under $\theta_k
\to\theta_k + 4\pi$ for each of $k=1,2$. 
This effective potential  implies that four vacua
corresponding to $(\epsilon_1,\epsilon_2) = (+1,+1)$, $(+1,-1)$, $(-1,+1)$ and $(-1,-1)$
now have different vacuum energies.
For
generic values of the $\theta$-parameters,
there is only one true vacuum with the minimum energy.\footnote{To be specific, when $\pi <
\theta_1,\theta_2 < 3\pi$ modulo $4\pi$, the
vacuum corresponding to $\epsilon_1=\epsilon_2 = -1$ has the minimum energy. When $\pi
<\theta_1<3\pi$ with $0\leq \theta_2 < \pi$ or $3\pi < \theta_2 < 4\pi$, the vacuum
corresponding to $\epsilon_1 = -\epsilon_2 = -1$ has the minimum energy. When
$0\leq \theta_1<\pi$ or $3\pi <\theta_1<4\pi$ with $\pi <\theta_2<3\pi$,
the vacuum corresponding to $\epsilon_1 = -\epsilon_2= 1$ has the minimum
energy. Finally, when both of $\theta_1$ and $\theta_2$ take values in
$[0,\pi)\cup (3\pi,4\pi)$, the vacuum with $\epsilon_1 = \epsilon_2 =1$
has the minimum energy.}

There are special values of the $\theta$-parameters which give rise to
degenerate vacua.
For instance, when $\theta_1/\pi$ is an odd integer but $\theta_2/\pi$ is not, the effective
potential \eqref{eq:soft-mass} reduces to
\begin{equation}
 V = -2\epsilon_2|m\kappa_2\Lambda_2^3|\cos\frac{\theta_2}{2}~,
\label{eq:soft-mass2}
\end{equation}
which implies two degenerate vacua corresponding to $\epsilon_1=
\pm1$.\footnote{The other parameter $\epsilon_2$ is fixed to $+1$ or
$-1$, depending on the value of $\theta_2$.} The same situation occurs
when $\theta_2/\pi$ is an odd integer but $\theta_1/\pi$ is not. The most
non-trivial situation occurs when both $\theta_1/\pi$ and $\theta_2/\pi$
are odd integers, in which case the effective
potential vanishes:
\begin{equation}
V = 0~.
\end{equation}
Then the four vacua corresponding to $(\epsilon_1,\epsilon_2) =
(1,1),(1,-1),(-1,1)$ and $(-1,-1)$ are all degenerate even after a small
gaugino mass $m$ is turned on.

The above results in the small $|m|$ approximation are in perfect agreement with the
phase diagram of the $Spin(4)$ Yang-Mills theory that we have identified
using the center/CP anomaly in
Sec.~\ref{subsec:Spin4}. 

\subsubsection{$\mathcal{N}=1$ $SO(4)_\pm$ SYM}

We now move on to the soft SUSY breaking of the $SO(4)_\pm$ SYM. In this
case, the supersymmetric theory still has four vacua corresponding to
$\epsilon_1,\epsilon_2=\pm 1$, but the IR degrees of freedom
present in these vacua are different from the $Spin(4)$ case. Indeed,
for the $SO(4)_-$ SYM, the vacua for $\epsilon_1=\epsilon_2 = \pm 1$ are confining and
oblique confining phases, while those for $\epsilon_1 = -\epsilon_2 =
\pm 1$ support a $\mathbb{Z}_2$ TQFT. For the $SO(4)_+$ SYM,
the roles of $\epsilon_1=\epsilon_2$ and $\epsilon_1=-\epsilon_2$ are
exchanged. 

This means that, when
$\theta_1,\theta_2\in (\pi,3\pi)$ or $\theta_1,\theta_2\in
[0,\pi)\cup(3\pi, 4\pi)$, the $SO(4)_-$ theory is in the confining or
oblique confining phase. On the other hand, when one of $\theta_1$ and
$\theta_2$ is in $(\pi,3\pi)$ and the other is in $[0,\pi)\cup
(3\pi,4\pi)$, the IR degrees of freedom in the $SO(4)_-$ Yang-Mills
theory is a $\mathbb{Z}_2$ TQFT. The result for $SO(4)_+$ is obtained
similarly, which will turn out to be consistent with the relation
$SO(4)_+^{\theta_1,\theta_2} = SO(4)_-^{\theta_1+2\pi,\theta_2}$.

One finds that the above results for a small gaugino mass $m$ are
precisely identical to the phase diagrams of $SO(4)_\pm$ Yang-Mills
theories obtained via the center/CP anomaly in
Sec.~\ref{subsec:SO4}.

\section{$Spin(4k)/SO(4k)$ Yang-Mills theory}
\label{sec:Spin4k}

In this section, we generalize our discussions in Sec.~\ref{sec:Spin4}
to $Spin(4k)/SO(4k)$ Yang-Mills theories for $k> 1$. An important
difference from $Spin(4)/SO(4)_\pm$ is that there is only a single $\theta$-parameter. Furthermore, the center/CP anomaly is slightly different
between odd and even $k$.

\subsection{Center/CP anomaly and mixed gauge theory}

\subsubsection{Center/CP anomaly}

When $k$ is odd, the mixed center/CP
anomaly for the $Spin(4k)$ Yang-Mills theory is almost identical
to that of the $Spin(4)$ Yang-Mills \cite{Witten:2000nv, Aharony:2013hda,
Cordova:2019uob, Chen:2020syd}. Namely, when the $\theta$-parameter
is shifted by $2\pi$, the action receives
\begin{equation}
 \Delta S = \frac{i\pi}{2}\int_{\mathcal{M}}\left(\mathfrak{P}(B_1) +
					     \mathfrak{P}(B_2)\right)~,
\label{eq:anomaly_Spin4k}
\end{equation}
where $(B_1,B_2)\in Z^2(\mathcal{M},\mathbb{Z}_2\times \mathbb{Z}_2)$
stands for the background gauge fields for the $\mathbb{Z}_2\times
\mathbb{Z}_2$ center symmetry. 
Since
this is precisely the same as the $Spin(4)/SO(4)_\pm$ case
with the constraint
\begin{equation}
 \theta_1 = \theta_2 = \theta~,
\end{equation}
it is straightforward to read off the phase diagrams of the
$Spin(4k)/SO(4k)_\pm$ Yang-Mills theory for odd $k$ from the results of
Sec.~\ref{sec:Spin4}. In particular, the periodicity of $\theta$ for generic values of $(B_1,B_2)$ is $\theta\sim \theta + 4\pi$.

When $k$ is even, the above anomaly \eqref{eq:anomaly_Spin4k}
is modified to be \cite{Witten:2000nv, Aharony:2013hda, Cordova:2019uob,
Chen:2020syd}
\begin{equation}
 \Delta S = \pi i \int_{\mathcal{M}} B_1\cup B_2~.
\label{eq:anomaly_Spin4k2}
\end{equation}
Since $\int_{\mathcal{M}}B_1\cup B_2\in\mathbb{Z}_2$, we find again that the
periodicity under a general background field is $\theta \sim
\theta + 4\pi$.

Below, we consider the mixed $Spin(4k)$ theory separately for odd and even $k$.

\subsubsection{Mixed $Spin(4k)$ theory for odd $k$}

Let us first consider the case of odd $k$, and study the mixed
$Spin(4k)$ theory. Namely, we
compactify the $Spin(4k)$ Yang-Mills on $S^1$ and gauge the 3d 1-form
$\mathbb{Z}_2\times \mathbb{Z}_2$ symmetry. 
The global symmetry of this theory can be read off
from the diagonal elements of Table~\ref{table:globalsym_SO4}.

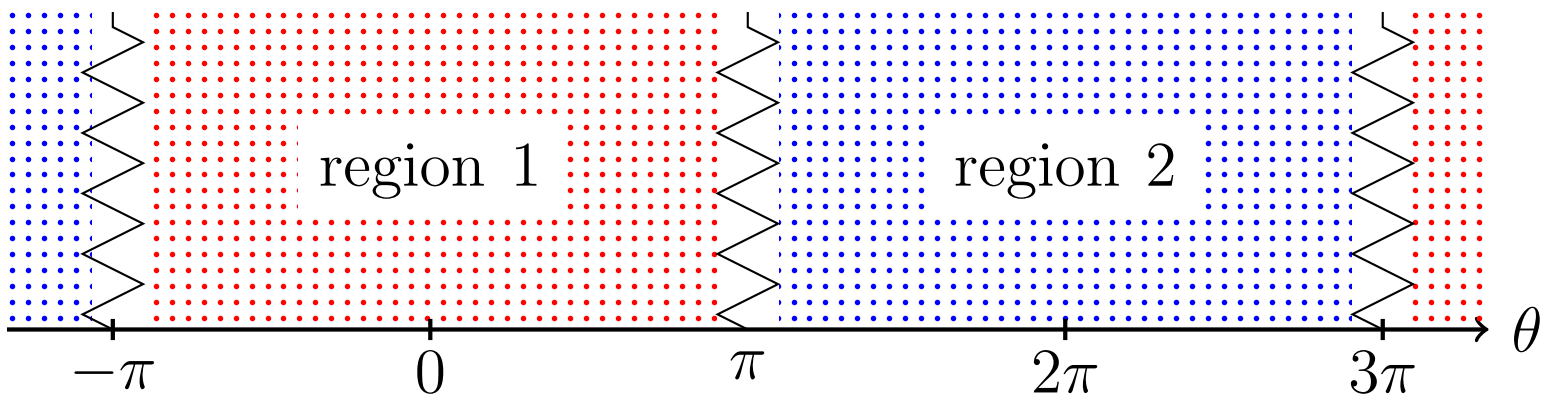
\begin{figure}
    \centering
    \begin{tikzpicture}[scale=.7]
        \draw[thick, ->] (-2,0) -- (12,0) node [right] {$\theta$};
        \path[pattern=dots, pattern color=red] (-0.7,0) rectangle +(5.4,3);
        \path[pattern=dots, pattern color=blue] (5.3,0) rectangle +(5.4,3);
        \path[pattern=dots, pattern color=red] (11.2,0) rectangle +(0.8,3);
        \path[pattern=dots, pattern color=blue] (-2,0) rectangle +(0.8,3);
        \node[fill=white] at (2, 1.5) {{region 1}};
        \node[fill=white] at (8, 1.5) {{region 2}};
        \draw decorate[decoration={name=zigzag,
                  amplitude=2mm,segment length=4mm}]
                      { (5,0) -- (5,3) };
        \draw decorate[decoration={name=zigzag,
                  amplitude=2mm,segment length=4mm}]
                      { (-1,0) -- (-1,3) };
        \draw decorate[decoration={name=zigzag,
                  amplitude=2mm,segment length=4mm}]
                      { (11,0) -- (11,3) };
        \draw[thick] (8,-0.1) -- (8,0.1);
        \node at (2,0) [below] {$0$};
        \node at (5,0) [below] {$\pi$};
        \node at (8,0) [below] {$2 \pi$};
        \draw[thick] (-1,-0.1) -- (-1,0.1);
        \draw[thick] (11,-0.1) -- (11,0.1);
        \node at (-1,0) [below] {$-\pi$};
        \node at (11,0) [below] {$3\pi$};
        \draw[thick] (2,-0.1) -- (2,0.1);
    \end{tikzpicture}
    \caption{There are two different phases of $Spin(4k)/SO(4k)_\pm$
 theories for $k\geq 2$, depending on the value of $\theta$.}
    \label{fig:Spin4k_phases}
\end{figure}

The 4d center/CP anomaly \eqref{eq:anomaly_Spin4k} implies that
the periodicity of the $\theta$-parameter for the mixed $Spin(4k)$
theory is  $\theta \sim \theta + 4\pi$.
From the global symmetry group, one heuristically finds
a vacuum structure shown in Fig.~\ref{fig:Spin4k_phases} and
Table~\ref{table:mixed_Spin4k}, where $T_1,T_2,D_1$ and $D_2$ are line
operators labeled respectively by $(0,0;1,0),\,(0,0;0,1),\,(1,0;1,0)$,
$(0,1;0,1)\in\mathbb{Z}_2^4$ (See appendix \ref{app:line} for our
notation of charges of line operators).  Note that this vacuum structure is
consistent with the fact that the Witten effect  $(T_1,T_2) \to
(D_1,D_2)$ under $\theta\to \theta+2\pi$. When the value of $\theta$ is generic, the
mixed $Spin(4k)$ theory has four vacua, while it has eight vacua when
$\theta/\pi$ is an odd integer. All vacua now belong to a single
orbit of $(\mathbb{Z}_2)^4\rtimes \mathbb{Z}_2$
in
contrast to 
the $Spin(4)$ case. 

\begin{table}[h]
 \centering
\begin{tabular}{c||c|c}
& region 1& region 2  \\ 
\hline\hline
$T_1$ & $\pm a_1$ & $0$   \\
$D_1$ & $0$ & $\pm b_1$  \\
\hdashline
$T_2$ & $\pm a_2$ & $0$  \\
$D_2$ & $0$ & $\pm b_2$  \\
\end{tabular}
\bigskip
\caption{The VEVs of $T_i$ and $D_i$ in various regions of
 Fig.~\ref{fig:Spin4k_phases}, where $a_i$ and $b_i$ are
 non-vanishing values. The signs are {\it not} correlated. There are
 four vacua in each region. The $T_1$ and $T_2$ are associated with the
 lines labeled by $(0,0;1,0),\,(0,0;0,1) \in \mathbb{Z}_4^2$, respectively. The $D_1$
 and $D_2$ are respectively associated with $(1,0;1,0),\,(0,1;0,1)$ for
 odd $k$, and with $(0,1;1,0),\,(1,0;0,1)$ for even $k$. }
\label{table:mixed_Spin4k}
\end{table}

\subsubsection{Mixed $Spin(4k)$ theory for even $k$}

When $k$ is even, the mixed center/CP anomaly is different from the
$Spin(4)$ case, and therefore the corresponding mixed gauge theory has a
different global symmetry.

Indeed, replacing \eqref{eq:anomaly_Spin4k} with
\eqref{eq:anomaly_Spin4k2} implies that the relations
\eqref{eq:globalsym_SO(4)_atpi} for the $0$-form
global symmetry 
at $\theta = \pi$ are now replaced with
\begin{align}
 CP\cdot c_1\cdot CP &= c_1\cdot h_2~,\qquad CP\cdot c_2 \cdot CP =
 c_2\cdot h_1~,\qquad CP\cdot h_k = h_k \cdot CP~,
\label{eq:second-semi1}
\\
c_k\cdot h_\ell &= h_\ell \cdot c_k
 ~,\qquad c_k^2 = h_k^2 = CP^2 = {\bf 1}~.
\label{eq:second-semi2}
\end{align}
This shows that the global symmetry at $\theta = \pi$ is still 
expressed as $(\mathbb{Z}_2)^4 \rtimes \mathbb{Z}_2$, but the relevant
semi-direct product is different from the one for odd $k$. 

Note that, due to \eqref{eq:anomaly_Spin4k2}, the Witten effect now
implies $(T_1,T_2) \longrightarrow (T_1W_2, T_2 W_1)$ under $\theta \to
\theta + 2\pi$, where $W_1$ and $W_2$ are Wilson lines labeled
respectively by $(1,0;0,0)\in\mathbb{Z}_2^4$ and
$(0,1;0,0)\in\mathbb{Z}_2^4$. This implies that the vacuum structure
shown in Table~\ref{table:mixed_Spin4k} is still valid for even $k$ if
$D_1$ and $D_2$ are understood as $D_1 = T_1W_2$ and $D_2=T_2W_1$.

\subsection{Phase diagrams of $Spin(4k)/SO(4k)_\pm$ theories}

In the rest of this section, we use the notation that $(D_1,D_2) = (T_1W_1,T_2W_2)$ for odd $k$
and $(D_1,D_2) = (T_1W_2,T_2W_1)$ for even $k$, so that we can treat odd
and even
$k$ in the same way.
Below, we read off the
phase diagrams of the $Spin(4k)/SO(4k)_\pm$ theories from the above
results for the mixed $Spin(4k)$ theory, via the relations
\begin{align}
Spin(4k) &= \big(\text{mixed } Spin(4k)\big) \Big/ \big(\mathbb{Z}_2^{(h_1)}\times \mathbb{Z}_2^{(h_2)}\big)~,
\\
 SO(4k)_+ &= \big(\text{mixed } Spin(4k)\big) \Big/ \big(\mathbb{Z}_2^{(c_1\cdot
 c_2)}\times \mathbb{Z}_2^{(h_1\cdot h_2)}\big)~,
\label{eq:SO4k+}
\\
SO(4k)_- &= \big(\text{mixed }Spin(4k)\big) \Big/\big(\mathbb{Z}_2^{(c_1\cdot
 c_2\cdot h_1)}\times \mathbb{Z}_2^{(h_1\cdot h_2)}\big)~.
\label{eq:SO4k-}
\end{align}

Let us start with the $Spin(4k)$ Yang-Mills, which is obtained by gauging the $\mathbb{Z}_2^{(h_1)}\times
\mathbb{Z}_2^{(h_2)}$ symmetry of the mixed $Spin(4k)$ theory. Here,
$\mathbb{Z}_2^{(h_1)}\times \mathbb{Z}_2^{(h_2)}$ is generated by $h_1$
and $h_2$, whose actions on
$T_k$ and $D_k$ are shown in \eqref{eq:TkDk}. We see that both in
regions 1 and 2 there exists a single vacuum corresponding to $\langle
W_k\rangle = 0$.  In the convention that region 1 corresponds
to the trivial phase, region 2 corresponds to an SPT phase
protected by the center symmetry of $Spin(4k)$.
 These results are summarized in the first line
of Table~\ref{table:Spin4k_phases}.

\begin{figure}
  \centering
  \begin{tikzpicture}

      \draw[thick] (1,2.5) -- (3,2.5) node [below] {$T_1 T_2$};
      \draw[thick] (2,1.5) -- (2,3.5) node [above] {$D_1 D_2$};

      \fill (1.5,2.5) circle (2pt);
      \fill (2.5,2.5) circle (2pt);

      \draw[thick] (6,2.5) -- (8,2.5) node [below] {$T_1 T_2$};
      \draw[thick] (7,1.5) -- (7,3.5) node [above] {$D_1 D_2$};

      \fill (7,2) circle (2pt);
      \fill (7,3) circle (2pt);

      \draw[thick] (11,2.5) -- (13,2.5) node [below] {$T_1 T_2$};
      \draw[thick] (12,1.5) -- (12,3.5) node [above] {$D_1 D_2$};
      \fill (12.5,2.5) circle (2pt);
      \fill (11.5,2.5) circle (2pt);
      \fill (12,2) circle (2pt);
      \fill (12,3) circle (2pt);
      \node at (2,1) {region 1};
      \node at (7,1) {region 2};
      \node at (12,1) {$\theta = (2k+1)\pi$ for $k\in \mathbb{Z}$};
  \end{tikzpicture}
  \caption{The vacuum structure of $SO(4k)_+$ Yang-Mills theory
 compactified on $S^1$ for $k\geq
 1$. Each vacuum is accompanied with a $\mathbb{Z}_2$ TQFT. When going
 back to four dimensions, every pair of vacua connected by $T_1T_2 \to
 -T_1T_2$ and $D_1D_2\to -D_1D_2$ is mapped to a single vacuum.}
  \label{fig:SO4k+}
\end{figure}
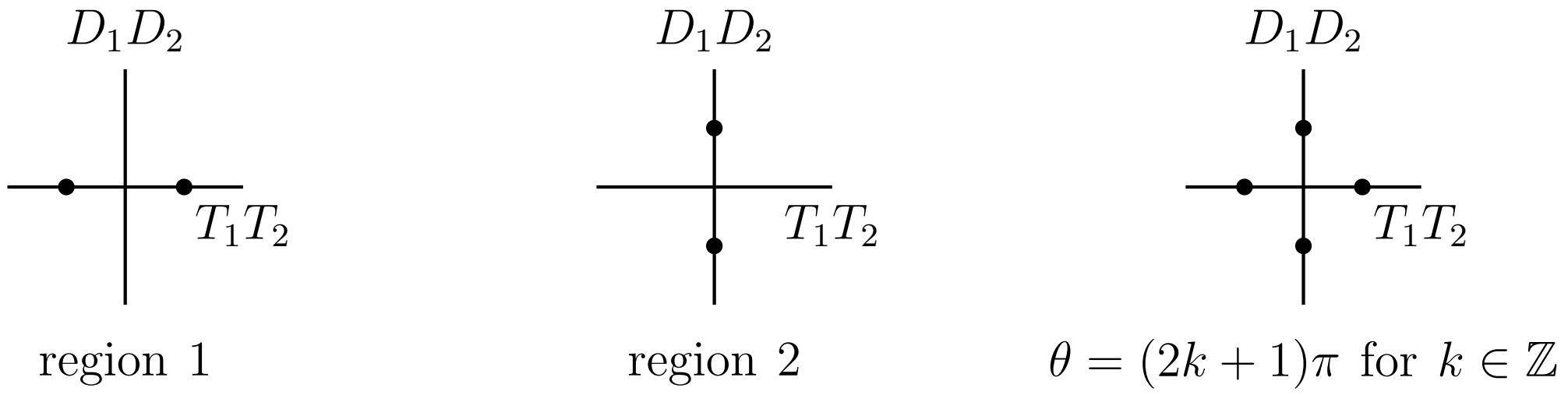

Let us next consider the $SO(4k)_+$ theory using \eqref{eq:SO4k+}.
 Since $\mathbb{Z}_2^{(c_1\cdot c_2)}\times \mathbb{Z}_2^{(h_1\cdot
h_2)}$ acts on $T_k$ and $D_k$ as in \eqref{eq:c1c2} and
\eqref{eq:h1h2}, we see that the resulting vacuum structure is as shown in Fig.~\ref{fig:SO4k+}. For generic values of
$\theta$, there are two vacua accompanied with an unbroken
$\mathbb{Z}_2$ gauge theory. When $\theta/\pi$ is an odd integer, the
$SO(4k)_+$  theory has four degenerate vacua. These results are
summarized in the second line of Table~\ref{table:Spin4k_phases}.

One can analyze the $SO(4k)_-$ theory in the same way.
The $\mathbb{Z}_2^{(c_1\cdot c_2\cdot h_1)}\times
\mathbb{Z}_2^{(h_1\cdot h_2)}$ acts on $T_k$ and $D_k$ as in
\eqref{eq:cch} and \eqref{eq:hh2}, leaving only one
vacuum for generic values of $\theta$. This single vacuum is
characterized by $\langle T_1D_2\rangle = \langle D_1T_2\rangle =
0$.
When the vacuum for region 1 is called the trivial vacuum, region 2
is in the non-trivial SPT phase protected by the 1-form symmetry. When $\theta/\pi$ is an odd integer,
 the $SO(4k)_-$ theory has two degenerate vacua. These results are
 shown in the third line of Table~\ref{table:Spin4k_phases}.

Note that, in contrast to the case of
$SO(4)_\pm$, the IR physics of the $SO(4k)_+$ theory and that of the
$SO(4k)_-$ theory are
completely different. This is possible since for $k> 1$ one has
\begin{equation}
 SO(4k)_\pm^{\theta + 2\pi} = SO(4k)_\pm^{\theta}~,
\end{equation}
and therefore these two theories are
not related by any shift of the $\theta$-parameter.

\begin{table}
 \centering
\begin{tabular}{c||c|c|c} 
gauge group $\backslash$ regions in Fig.~\ref{fig:Spin4k_phases} &
     region 1  & region 2 & $\theta = (2k+1)\pi$ for $k\in \mathbb{Z}$ \\
\hline\hline
$Spin(4k)$ & trivial & SPT & trivial \& SPT \\
\hline
$SO(4k)_+$ & $\mathbb{Z}_2$ TQFT & $\mathbb{Z}_2$ TQFT &
	     two vacua supporting $\mathbb{Z}_2$ TQFTs\\
\hline
$SO(4k)_-$ & trivial & SPT & trivial \& SPT\\
\end{tabular}
\caption{The IR theory present in the two regions in
 Fig.~\ref{fig:Spin4k_phases} for 4d 
 $Spin(4k)/SO(4k)_\pm$ Yang-Mills with $k\geq 2$.}
\label{table:Spin4k_phases}
\end{table}

\subsection{Comparison to soft SUSY breaking}
\label{subsec:SUSY_Spin4k}

Here we compare the above phase diagrams 
with those of supersymmetric $Spin(4k)/SO(4k)_\pm$
Yang-Mills theories with a small gaugino mass turned on.  We will see that they are in perfect agreement with
each other.

Let us start with the $\mathcal{N}=1$ $Spin(4k)$ SYM theory. For $k\geq
2$, the theory has $(4k-2)$ SUSY vacua
\cite{Intriligator:1995id, Intriligator:1995au, Aharony:2013hda}. In terms of the gaugino bilinear operator
$\lambda_\alpha\lambda^\alpha$, these $(4k-2)$ vacua are characterized
by
\begin{equation}
 \langle \text{Tr}(\lambda_\alpha\lambda^\alpha)\rangle = \kappa
  \Lambda^{3}e^{-\frac{2\pi i }{4k-2}n}~,\quad \qquad n=0,1,2,\cdots, 4k-3~.
\end{equation}
where $\kappa$ is a constant, and $\Lambda$ is the dynamical
scale such that $\Lambda^{3(4k-2)} = |\Lambda^{3(4k-2)}|e^{i\theta}$. 
Note that shifting $\theta$ by $2\pi$ shifts $n$ by one. 
Since the
center of the gauge group is $(\mathbb{Z}_2)^2$, line
operators are characterized by $(z_1,z_2;w_1,w_2) \in
(\mathbb{Z}_2)^4$, where $(z_1,z_2)$
corresponds to the electric charges of the line (up to root lattice of the
gauge group), while $(w_1,w_2)$ corresponds to the magnetic charges.\footnote{See appendix \ref{app:line} for our
notation for the charges of line operators.} The
$Spin(4k)$ SYM contains $(1,0;0,0),\,(0,1;0,0)$ and $(1,1;0,0)$ as
genuine line operators. 
At vacua for even $n$, purely magnetic monopoles are condensed, while at the other vacua dyons are
condensed.\footnote{The charges of these dyons are such that the lines
of charges $(1,0;1,0)$ and
$(0,1;0,1)$ have a perimeter law for odd $k$, and those of $(0,1;1,0)$
and $(1,0;0,1)$ have a perimeter law for even $k$.} From this we see that, at any vacuum, all the
genuine line operators have an area law, since every genuine line
operator has a non-vanishing Dirac pairing with at least one condensed particle.\footnote{
The
Dirac's electro-magnetic paring of $(z_1,z_2;w_1,w_2)$ and $(\tilde
z_1,\tilde z_2;\tilde w_1,\tilde w_2)$ is given by $z_1 \tilde w_1 -w_1
\tilde z_1 +z_2\tilde w_2 - w_2\tilde z_2$ for any integer $k$.
This is a different notation from the one used in  \cite[Sec.~5.2]{Aharony:2013hda}
when $k$ is even.} Therefore, all the
$(4k-2)$ vacua are in the confining or an oblique confining phase. In
the convention that the vacua for even $n$ are in the confining phase, those for odd $n$ are in an oblique confining phase. This oblique
confining phase is an SPT phase protected by 
the $\mathbb{Z}_2\times \mathbb{Z}_2$ center symmetry.

Turning on a soft SUSY breaking term $\delta \mathcal{L} = m\,
\text{Tr}(\lambda_\alpha \lambda^\alpha) + \text{c.c.}$, we obtain the effective
potential\footnote{We again choose the origin of $\theta$ so that
$\kappa m$ is a positive real number.} 
\begin{equation}
 V = -2|m\kappa \Lambda^3| \cos\left( \frac{\theta - 2\pi n}{4k-2}\right)~,
\end{equation}
which resolves the above $(4k-2)$ degeneracy. Specifically, the vacuum
for $n=\ell$ has the minimum energy when $(2\ell-1)\pi < \theta <(2\ell
+ 1)\pi$. When $\theta/\pi$ is an odd integer, two vacua are degenerate. Although there are $(4k-2)$ different values of $n$, one
can only distinguish vacua for even $n$ from those for odd $n$.
 Thus, we observe only two different phases in the infrared. The vacua for even $n$
are in the trivial phase, and those for odd $n$ are in an SPT phase protected by
the center $\mathbb{Z}_2\times \mathbb{Z}_2$ symmetry. This result is in
perfect agreement with the first line of
Table~\ref{table:Spin4k_phases} read off from the center/CP anomaly of
non-supersymmetric $Spin(4k)$ Yang-Mills theory.

The generalization of the above discussion to $SO(4k)_\pm$ theories is
straightforward. The only difference is that each vacuum is associated
with a different
TQFT due to different spectra of line
operators. Let us first consider the $SO(4k)_+$ Yang-Mills theory, in which genuine
line operators have charge $(z_1,z_2;w_1,w_2) = (1,1;0,0),\, (0,0;1,1)$
and $(1,1;1,1)$ \cite{Aharony:2013hda}. Note that, while the line for $(1,1;0,0)$
has an area law, either $(0,0;1,1)$ or $(1,1;1,1)$ has a perimeter law
since its charge is proportional to that of the condensed
particles up to a shift by the root lattice.
This implies that the IR TQFT is a $\mathbb{Z}_2$ TQFT at all
vacua. Note that the $\mathbb{Z}_2$ TQFT for $-\pi <
\theta<\pi$ and that for $\pi < \theta <3\pi$ are different, which
simultaneously exist when $\theta/\pi$ is an odd integer.
This result is in perfect agreement with the second line of
Table~\ref{table:Spin4k_phases}. The case of $SO(4k)_-$ can be analyzed
in the same way, whose genuine line operators are those of charge
$(1,1;0,0),\,(1,0;1,1)$ and $(0,1;1,1)$. Since all these lines always have an area law, every vacuum is in the trivial phase or in
the SPT phase protected by the 1-form symmetry.
This
result is
identical to the third line of Table~\ref{table:Spin4k_phases}.

\section{$Spin(4k+2)/SO(4k+2)$ Yang-Mills theory}
\label{sec:Spin4k+2}

In this section, we study the phase diagrams of the $Spin(4k+2)/SO(4k+2)_\pm$ Yang-Mills
theories for a positive integer $k$. Since the center of $Spin(4k+2)$ is
$\mathbb{Z}_4$ instead of $\mathbb{Z}_2\times \mathbb{Z}_2$, we will find essentially
different phase diagrams from what we have obtained in the previous
sections.

In
Sec.~\ref{subsec:Spin4k+2_anomaly}, we will study it using the mixed
center/CP anomaly of the theory. In Sec.~\ref{subsec:Spin4k+2_SUSY}, we
will show that the result of Sec.~\ref{subsec:Spin4k+2_anomaly} is in
agreement with the result suggested by the soft SUSY breaking of
$\mathcal{N}=1$ SYM.

\subsection{Phase digram from the center/CP anomaly}
\label{subsec:Spin4k+2_anomaly}

The mixed center/CP anomaly for $Spin(4k+2)$ theory is expressed as \cite{Aharony:2013hda,
Witten:2000nv, Cordova:2019uob, Chen:2020syd}
\begin{equation}
 \Delta S = \pm \frac{i\pi}{4} \int_{\mathcal{M}}\mathfrak{P}(B)~,
\label{eq:Spin4k+2_anomaly}
\end{equation}
where $\Delta S$ is the change of the action under the shift $\theta \to
\theta +2\pi$ of the $\theta$-parameter, and $B\in
Z^2(\mathcal{M},\mathbb{Z}_4)$ is the background gauge field for the
center $\mathbb{Z}_4$ symmetry of the $Spin(4k+2)$ Yang-Mills theory. The sign
on the right-hand side depends on whether $k$ is even or odd.
The above expression for $\Delta S$ implies that the periodicity is 
\begin{equation}
 \theta \sim \theta + 8\pi~,
\end{equation}
in the presence of generic $B\neq 0$.

For later use, let us briefly discuss the line operators of the $Spin(4k+2)$
Yang-Mills theory. Since
the center of $Spin(4k+2)$ is $\mathbb{Z}_4$, 4d line operators are
labeled by $(z;w)\in \mathbb{Z}_4\times \mathbb{Z}_4$ (See appendix \ref{app:line} for our notation of the charges
of line operators). One can take the
one
corresponding to $(z;w) = (0;1)$ and the one corresponding to $(z;w) =
(1;0)$ as primitive line operators. In the $Spin(4k+2)$ theory,
the $(0;1)$ line is not a genuine line operator (and therefore attached
to a surface), but will play an important role in
the study of the mixed gauge theory below.

\subsubsection{Mixed $Spin(4k+2)$ gauge theory and its phase diagram}

Compactifying the 4d $Spin(4k+2)$ Yang-Mills on $S^1$, we obtain a 1-form
$\mathbb{Z}_4$ symmetry and a 0-form $\mathbb{Z}_4$ symmetry in three
dimensions. We call this $0$-form $\mathbb{Z}_4$ symmetry
$\mathbb{Z}_4^{(c)}$ and denote its generator by $c$. 
We also gauge the 3d 1-form $\mathbb{Z}_4$ symmetry to obtain another
0-form $\mathbb{Z}_4$ symmetry, which we call $\mathbb{Z}_4^{(h)}$ and
whose generator we denote by $h$. The resulting 3d theory with the
0-form $\mathbb{Z}_4^{(c)}\times \mathbb{Z}_4^{(h)}$ symmetry is called the
mixed $Spin(4k+2)$ theory.

For generic values of $\theta$, the 4d CP symmetry is explicitly broken
by the $\theta$-term, and therefore the 3d mixed $Spin(4k+2)$ theory has
only $\mathbb{Z}_4^{(c)}\times \mathbb{Z}_4^{(h)}$ as a global
symmetry. However, when $\theta/\pi$ is an integer, the 4d CP
symmetry gives rise to an extra $\mathbb{Z}_2$ 0-form symmetry of the
mixed $Spin(4k+2)$ theory, which we denote by $\mathbb{Z}_2^{(CP)}$ and whose generator
we denote by $CP$. The full global symmetry group is an extension of $\mathbb{Z}_4^{(c)}\times \mathbb{Z}_4^{(h)}$ by $\mathbb{Z}_2^{(CP)}$.

Specifically, 
when $\theta = \ell \pi + 4\pi
n$ for $\ell = 0,1,2$ or $3$ and
$n\in \mathbb{Z}$, the $CP$ involves a shift $\theta \to \theta
+2\pi \ell$. In this case, the 4d mixed anomaly \eqref{eq:Spin4k+2_anomaly}
implies the following relations:
\begin{align}
 CP\cdot c\cdot CP &= c^{-1}\cdot h^{\pm \ell}~,
\label{eq:CP1}
\\
CP \cdot h &= h\cdot CP~,
\label{eq:CP2}
\\
c\cdot h &= h\cdot c~,
\\
c^4 &= h^4 = CP^2 = {\bf 1}~,
\label{eq:CP4}
\end{align}
where the power of $h$ on the right-hand side of \eqref{eq:CP1} is
$+\ell$ for odd $k$ and $-\ell$ for even
$k$.
Note that these relations are slightly different from
\eqref{eq:relation_generators_su(2)} for $SU(2)$ gauge group.\footnote{The reason that $c^{-1}$ appears on the RHS of
\eqref{eq:CP1} is that $CP$ flips the electric charge and
therefore maps the generator of the electric symmetry $c$ to its
inverse. This subtlety does not matter in the cases of $SU(2)$ and
$Spin(4k)$ gauge groups, since in these cases the center of the gauge
group is $\mathbb{Z}_2$ or $\mathbb{Z}_2\times \mathbb{Z}_2$ and
therefore $c$ and its inverse are identical.}  Indeed, this global
symmetry group is identified
as $(\mathbb{Z}_4^{(c)} \times \mathbb{Z}_4^{(h)})\rtimes_{\varphi_\ell}
\mathbb{Z}_2^{(CP)}$ associated with a group homomorphism
\begin{equation}
 \varphi_\ell: \mathbb{Z}_2 \to \text{Aut}\,(\mathbb{Z}_4)^2
\end{equation} 
 specified by \eqref{eq:CP1} and \eqref{eq:CP2}. 
The full global symmetry of the
mixed $Spin(4k+2)$ theory is then identified as shown in
Table~\ref{table:globalsym_Spin4k+2}. 

\begin{table}
 \centering
\begin{tabular}{c||c|c}
the value of $\theta$ &  $\theta\neq \ell \pi$ for any $\ell\in \mathbb{Z}$ & $\theta = \ell \pi + 4\pi n$ for $\ell=0,1,2,3$ and $n\in\mathbb{Z}$\\ 
\hline
 global symmetry &  $(\mathbb{Z}_4)^2$ & $(\mathbb{Z}_4)^2\rtimes_{\varphi_\ell} \mathbb{Z}_2$\\
\end{tabular}
\caption{The global symmetry of the 3d mixed $Spin(4k+2)$
 theory for a positive integer $k$. These are all 0-form
 symmetries. The $\varphi_\ell$ is a group homomorphism $\varphi_\ell: \mathbb{Z}_2
 \to \text{Aut}(\mathbb{Z}_4)^2$ specified by \eqref{eq:CP1} and
 \eqref{eq:CP2}, where $CP$ is the generator of $\mathbb{Z}_2$ and $h$ and $c$ are generators of $(\mathbb{Z}_4)^2$.}
\label{table:globalsym_Spin4k+2}
\end{table}

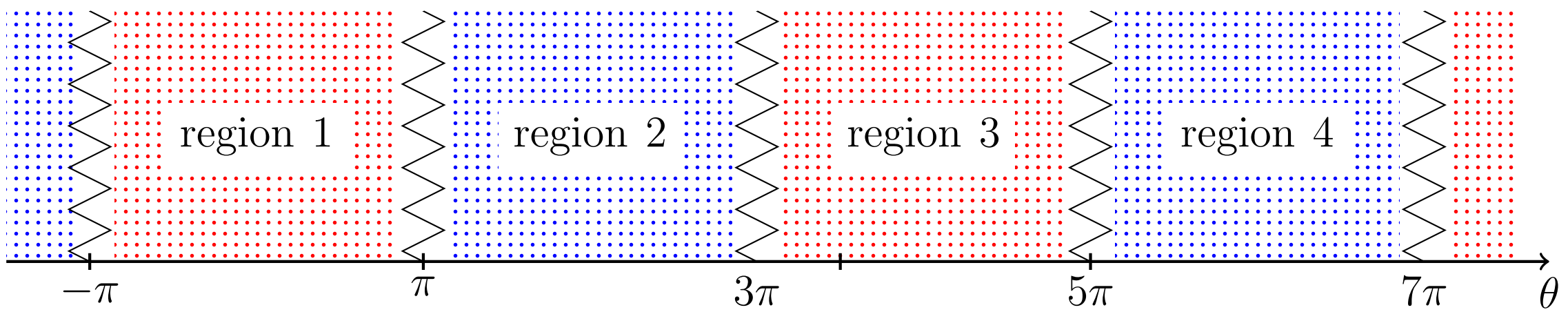
\begin{figure}
    \centering
    \begin{tikzpicture}[scale=.8]
        \draw[thick, ->] (-2,0) -- (16.5,0) node [below] {$\theta$};
        \path[pattern=dots, pattern color=red] (-0.7,0) rectangle +(3.4,3);
        \path[pattern=dots, pattern color=blue] (3.3,0) rectangle +(3.4,3);
        \path[pattern=dots, pattern color=red] (7.3,0) rectangle +(3.4,3);
        \path[pattern=dots, pattern color=blue] (-2,0) rectangle
     +(0.8,3);
        \path[pattern=dots, pattern color=blue] (11.3,0) rectangle
     +(3.4,3);
        \path[pattern=dots, pattern color=red] (15.3,0) rectangle +(.8,3);
        \node[fill=white] at (1, 1.5) {{region 1}};
        \node[fill=white] at (5, 1.5) {{region 2}};
        \node[fill=white] at (9, 1.5) {{region 3}};
        \node[fill=white] at (13, 1.5) {{region 4}};
        \draw decorate[decoration={name=zigzag,
                  amplitude=2mm,segment length=4mm}]
                      { (3,0) -- (3,3) };
        \draw decorate[decoration={name=zigzag,
                  amplitude=2mm,segment length=4mm}]
                      { (-1,0) -- (-1,3) };
        \draw decorate[decoration={name=zigzag,
                  amplitude=2mm,segment length=4mm}]
                      { (7,0) -- (7,3) };
        \draw decorate[decoration={name=zigzag,
                  amplitude=2mm,segment length=4mm}]
                      { (11,0) -- (11,3) };
        \draw decorate[decoration={name=zigzag,
                  amplitude=2mm,segment length=4mm}]
                      { (15,0) -- (15,3) };
        \draw[thick] (8,-0.1) -- (8,0.1);
        \node at (3,0) [below] {$\pi$};
        \node at (7,0) [below] {$3\pi$};
        \node at (11,0) [below] {$5 \pi$};
        \draw[thick] (-1,-0.1) -- (-1,0.1);
        \draw[thick] (11,-0.1) -- (11,0.1);
        \node at (-1,0) [below] {$-\pi$};
        \node at (15,0) [below] {$7\pi$};
        \draw[thick] (3,-0.1) -- (3,0.1);
    \end{tikzpicture}
    \caption{There are four different phases in the mixed $Spin(4k+2)$
 theory at zero temperature. The periodicity of $\theta$
 is $\theta\sim \theta + 8\pi$. This periodicity is changed to be
 $\theta \sim \theta + 4\pi$ when the mixed gauge theory is reduced to the
 $SO(4k+2)_\pm$ Yang-Mills theory, as will be discussed in Sec.~\ref{subsubsec:YM_4k+2}.}
    \label{fig:Spin4k+2_phases}
\end{figure}

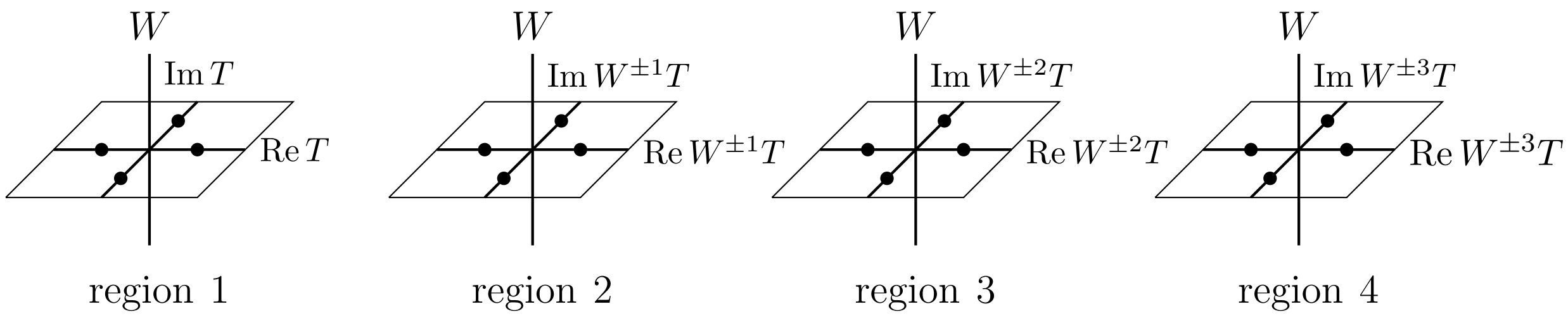
\begin{figure}[h]
  \centering
  \begin{tikzpicture}

      \draw[thick] (1,2.5) -- (3,2.5) node [right] {\footnotesize Re\,$T$};
      \draw[thick] (2,1.5) -- (2,3.5) node [above] {$W$};
      \draw (.5,2) -- (1.5,3) -- (3.5,3) -- (2.5,2) -- (.5,2);
      \draw[thick] (1.5,2) -- (2.5,3) node[above=3mm,right=-5mm]{\footnotesize Im\,$T$};

      \fill (1.5,2.5) circle (2pt);
      \fill (2.5,2.5) circle (2pt);
      \fill (2.3,2.8) circle (2pt);
      \fill (1.7,2.2) circle (2pt);

      \draw[thick] (5,2.5) -- (7,2.5) node [right] {\footnotesize
   Re\,$W^{\pm 1}T$};
      \draw[thick] (6,1.5) -- (6,3.5) node [above] {$W$};
      \draw (4.5,2) -- (5.5,3) -- (7.5,3) -- (6.5,2) -- (4.5,2);
      \draw[thick] (5.5,2) -- (6.5,3) node[above=3mm,
   right=-5mm]{\footnotesize Im\,$W^{\pm 1}T$};
      \fill (5.5,2.5) circle (2pt);
      \fill (6.5,2.5) circle (2pt);
      \fill (6.3,2.8) circle (2pt);
      \fill (5.7,2.2) circle (2pt);

      \draw[thick] (9,2.5) -- (11,2.5) node [right] {\footnotesize
   Re\,$W^{\pm 2}T$};
      \draw[thick] (10,1.5) -- (10,3.5) node [above] {$W$};
      \draw (8.5,2) -- (9.5,3) -- (11.5,3) -- (10.5,2) -- (8.5,2);
      \draw[thick] (9.5,2) -- (10.5,3) node[above =3mm,
   right=-5mm]{\footnotesize Im\,$W^{\pm 2}T$};
      \fill (9.5,2.5) circle (2pt);
      \fill (10.5,2.5) circle (2pt);
      \fill (10.3,2.8) circle (2pt);
      \fill (9.7,2.2) circle (2pt);

      \draw[thick] (13,2.5) -- (15,2.5) node [right] {\small Re\,$W^{\pm
   3}T$};
      \draw[thick] (14,1.5) -- (14,3.5) node [above] {$W$};
      \draw (12.5,2) -- (13.5,3) -- (15.5,3) -- (14.5,2) -- (12.5,2);
      \draw[thick] (13.5,2) -- (14.5,3) node[above =3mm,
   right=-5mm]{\footnotesize Im\,$W^{\pm 3}T$};
      \fill (13.5,2.5) circle (2pt);
      \fill (14.5,2.5) circle (2pt);
      \fill (14.3,2.8) circle (2pt);
      \fill (13.7,2.2) circle (2pt);

      \node at (2.1,1) {region 1};
      \node at (6.1,1) {region 2};
      \node at (10.1,1) {region 3};
      \node at (14.1,1) {region 4};

  \end{tikzpicture}
\bigskip
  \caption{The vacuum structure of {\it mixed} $Spin(4k+2)$ theory for a
 positive integer $k$ for each region in Fig.~\ref{fig:Spin4k+2_phases}. In all these regions, the VEV of $W$
 (corresponding to the Wilson line in the spinor representation of
 $Spin(4k+2)$) vanishes, while a magnetically charged operator has a
 non-vanishing VEV.}
  \label{fig:vacua_mixed_Spin4k+2}
\end{figure}

We now identify a vacuum structure of the mixed $Spin(4k+2)$ gauge theory
for various values of $\theta$. For that, we first need to find the
order parameters of the phases. Recall that our primitive line operators in
four dimensions are the one corresponding to $(z;w) = (0;1)$ and the one
corresponding to $(z;w)=(1;0)$. We denote the former line
operator wrapping on $S^1$ by $T$, and the latter wrapping on $S^1$ by
$W$. These are local operators in the 3d mixed gauge theory, and their
VEVs characterize its phases. Under
$c$ and $h$, these operators transform as
\begin{align}
 c&: T\to T~,\quad W \to \pm i W~,
\label{eq:c_4k+2}
\\
 h&: T \to \pm i T~,\quad W\to W~,
\label{eq:h_4k+2}
\end{align}
where the signs in front of $iW$ and $iT$ are both taken to be plus for odd $k$ and minus for even $k$.
Similarly, under the $CP$, they transform as 
\begin{align}
 CP: T \to W^{\pm \ell} T~,\quad W\to W^{-1}~,
\label{eq:CP_4k+2}
\end{align}
when $\theta = \ell \pi + 4\pi n$ for $\ell=0,1,2,3$ and
$n\in\mathbb{Z}$. The transformation rule of $W$ under the $CP$ can be understood
by noting that the $CP$ flips the sign of the electric charge. 
Note that the above actions of $c,h$ and $CP$ correctly realize a representation of
\eqref{eq:CP1}--\eqref{eq:CP4}. 

Under the same assumption as before, we now identify a vacuum structure of the mixed $Spin(4k+2)$ theory 
that is consistent with the
global symmetry group shown in
Table~\ref{table:globalsym_Spin4k+2}.
 The result is shown in 
in Fig.~\ref{fig:Spin4k+2_phases} and
Fig.~\ref{fig:vacua_mixed_Spin4k+2}. For generic values of $\theta$, the mixed $Spin(4k+2)$ theory has four
vacua, while it has eight vacua when $\theta/\pi$ is an odd integer. We see that these
vacua correctly give representations of the global symmetry groups generated by
\eqref{eq:c_4k+2}, \eqref{eq:h_4k+2} and \eqref{eq:CP_4k+2}. Furthermore, they are
also consistent with the Witten effect
\begin{equation}
 T \to W^{\pm 1}T~,
\end{equation}
under $\theta \to \theta + 2\pi$.
These are strong evidence for our vacuum
structure and phase diagram of the mixed $Spin(4k+2)$ theory.

\subsubsection{Phase diagram of $Spin(4k+2)/SO(4k+2)_\pm$ Yang-Mills theories}
\label{subsubsec:YM_4k+2}

We now read off the phase diagrams of the $Spin(4k+2)/SO(4k+2)_\pm$
Yang-Mills theories from that of the mixed $Spin(4k+2)$ theory identified
above. These Yang-Mills theories are obtained from the mixed $Spin(4k+2)$ theory by
the following discrete gaugings:
\begin{align}
 Spin(4k+2)&= \big(\text{mixed }Spin(4k+2)\big)\,\big/\, \mathbb{Z}_4^{(h)}~,
\label{eq:gauging1}
\\
SO(4k+2)_+&=  \big(\text{mixed }Spin(4k+2)\big)\,\big/\, \big(\mathbb{Z}_2^{(h^2)}\times \mathbb{Z}_2^{(c^2)}\big)~,
\\
SO(4k+2)_-&= \big(\text{mixed }Spin(4k+2)\big)\,\big/\,\mathbb{Z}_4^{(c^2h)}\big)~,
\end{align}
where $(\mathbb{Z}_2)^{(g)}$ is a $\mathbb{Z}_2$-subgroup generated by
$g$. Therefore, the vacuum structures of the $Spin(4k+2)/SO(4k+2)_\pm$
Yang-Mills theories are obtained by taking the corresponding discrete
quotients of the vacuum structure shown in Fig.~\ref{fig:vacua_mixed_Spin4k+2}.

Let us start with the $Spin(4k+2)$ Yang-Mills theory. Since $h$ acts on $T$ and
$W$ as in \eqref{eq:h_4k+2}, the genuine line operators of the
$Spin(4k+2)$ Yang-Mills 
are $W,W^2$ and $W^3$. In any of regions 1 -- 4 of
Fig.~\ref{fig:Spin4k+2_phases}, the discrete gauging \eqref{eq:gauging1}
leaves only
a single vacuum with $\langle W\rangle = 0$. These are confining and oblique
confining vacua. In the convention that region 1 is in the confining
phase, the other regions are in SPT phases protected by the
$\mathbb{Z}_4$ center symmetry of $Spin(4k+2)$. This result is
summarized in the first line of Table~\ref{table:Spin4k+2_phases}. When crossing a boundary
of the regions, one observes an SPT phase transition. At the phase boundary, the CP is
spontaneously broken. The periodicity of $\theta$ is $\theta \sim \theta
+ 8\pi$
in the presence of a generic background gauge field of the
$\mathbb{Z}_4$ center symmetry. 

\begin{table}
 \centering
\begin{tabular}{c||c|c|c|c} 
gauge group $\backslash$ regions in Fig.~\ref{fig:Spin4k+2_phases} &
     region 1  & region 2 & region 3 & region 4 \\
\hline\hline
$Spin(4k+2)$ & trivial & SPT$_1$ & SPT$_2$ & SPT$_3$   \\
\hline
$SO(4k+2)_+$ & $\mathbb{Z}_2$ TQFT & $\mathbb{Z}_2$ TQFT &
	     $\mathbb{Z}_2$ TQFT & $\mathbb{Z}_2$ TQFT \\
\hline
$SO(4k+2)_-$ & trivial & SPT$_1$ & SPT$_2$ & SPT$_3$\\
\end{tabular}
\caption{The IR TQFT present in the four regions in
 Fig.~\ref{fig:Spin4k+2_phases} for 4d
 $Spin(4k+2)/SO(4k+2)_\pm$ Yang-Mills theory. The SPT phases for $Spin(4k+2)$ are protected by the
 electric $\mathbb{Z}_4$ 1-form symmetry (i.e., the center symmetry), while those for $SO(4k+2)_-$ are
 protected by electro-magnetic $\mathbb{Z}_4$ 1-form symmetry.
 }
\label{table:Spin4k+2_phases}
\end{table}

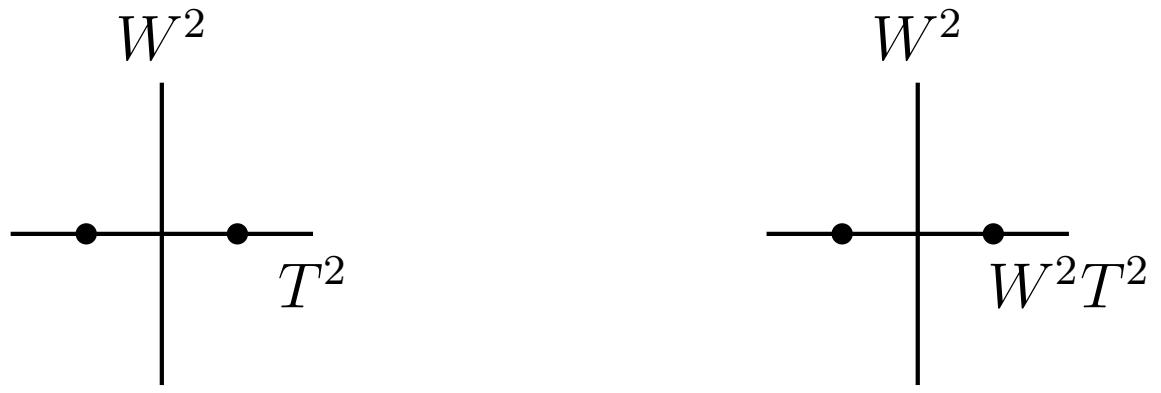
\begin{figure}
  \centering
  \begin{tikzpicture}

      \draw[thick] (1,2.5) -- (3,2.5) node [below] {$T^2$};
      \draw[thick] (2,1.5) -- (2,3.5) node [above] {$W^2$};

      \fill (1.5,2.5) circle (2pt);
      \fill (2.5,2.5) circle (2pt);

      \draw[thick] (6,2.5) -- (8,2.5) node [below] {$W^2T^2$};
      \draw[thick] (7,1.5) -- (7,3.5) node [above] {$W^2$};
      \fill (6.5,2.5) circle (2pt);
      \fill (7.5,2.5) circle (2pt);

      \node at (2.1,1) {regions 1 and 3};
      \node at (7.1,1) {regions 2 and 4};
  \end{tikzpicture}
\bigskip
  \caption{The vacuum structure of $SO(4k+2)_+$ Yang-Mills theory}
  \label{fig:vacua_SO(4k+2)+}
\end{figure}

Let us next consider the $SO(4k+2)_+$ Yang-Mills theory, which is obtained by gauging
the $\mathbb{Z}_2^{(h^2)}\times \mathbb{Z}_2^{(c^2)}$ symmetry of the mixed gauge
theory.
Since $h^2$ and $c^2$ act on $T$ and $W$ as
\begin{equation}
 h^2 : \;\, T \to -T~,\;\; W\to W~,\qquad\quad  c^2 : \;\, T\to T~,\;\; W\to -W~,
\end{equation}
this gauging maps the vacuum structure of 
Fig.~\ref{fig:vacua_mixed_Spin4k+2} to the one shown in
Fig.~\ref{fig:vacua_SO(4k+2)+}. There, the gauge-invariant operators are
taken to be $T^2, W^2$ and $W^2T^2$, corresponding to the genuine line
operators of the $SO(4k+2)_+$ Yang-Mills theory. Every vacuum shown in
Fig.~\ref{fig:vacua_SO(4k+2)+} is accompanied by a $\mathbb{Z}_2$
TQFT.\footnote{The intrinsic TQFTs in region 1 and 3 are identical, and those in
region 2 and 4 are also identical.}
This result is summarized in the second line of
Table~\ref{table:Spin4k+2_phases}. There is a residual
$\mathbb{Z}_2\times \mathbb{Z}_2$ 0-form global symmetry
generated by $c$ and $h$, which is a 3d $0$-form symmetry arising from
the 4d electric and magnetic $\mathbb{Z}_2$ 1-form symmetries of the
$SO(4k+2)_+$ Yang-Mills theory.

Finally, we consider the $SO(4k+2)_-$ Yang-Mills theory obtained by gauging
the $\mathbb{Z}_4^{(c^2h)}$ symmetry of the mixed
gauge theory. Noting that $c^2h$ acts on $T$ and $W$ as
\begin{equation}
c^2h :\;\, T\to \pm i T~,\;\; W\to -W~,
\end{equation}
we find that the gauge-invariant operators are now $W^2,\,WT^2$ and
$W^3T^2$. One can check that they indeed correspond to the genuine line operators of the
$SO(4k+2)_-$ Yang-Mills theory. We see that in each of regions 1 -- 4 there exists only
one vacuum with $\langle W^2\rangle = \langle WT^2\rangle =
\langle W^3T^2\rangle = 0$. The residual 0-form global symmetry is now
$\mathbb{Z}_4$ generated by
\begin{equation}
 c:\quad T \to T~,\quad W \to \pm iW~,
\label{eq:c^2}
\end{equation}
which is a 3d $0$-form symmetry arising from the 4d $\mathbb{Z}_4$
1-form symmetry of $SO(4k+2)_-$ Yang-Mills.\footnote{While the 1-form
symmetry of the 4d $SO(4k+2)_+$ Yang-Mills theory is $\mathbb{Z}_2\times \mathbb{Z}_2$,
that of the 4d $SO(4k+2)_-$ Yang-Mills theory is $\mathbb{Z}_4$. This
can be seen by looking at the charge lattice of genuine line
operators. When focusing on the sub-lattice generated by the charge of
the Wilson line, the $\mathbb{Z}_4$ acts through $\mathbb{Z}_2$ corresponding
to the center of the gauge group $SO(4k+2)$.} 
The vacua in regions 1--4 are SPT phases protected by $\mathbb{Z}_4^{(c)}$.
 This result is summarized in the third line of
 Table~\ref{table:Spin4k+2_phases}.

\subsection{Comparison to soft SUSY breaking}
\label{subsec:Spin4k+2_SUSY}

Here we show that the above phase diagrams read off from the center/CP anomaly
of the non-supersymmetric theories are in perfect agreement with those of the $\mathcal{N}=1$ $Spin(4k+2)/SO(4k+2)_\pm$
SYM with a small gaugino mass turned on.

Let us start with the $Spin(4k+2)$ gauge group. The $\mathcal{N}=1$ $Spin(4k+2)$ SYM has $4k$ SUSY vacua
\cite{Intriligator:1995id, Intriligator:1995au, Aharony:2013hda}, labeled by the 
\begin{equation}
 \langle (\text{Tr}\lambda_\alpha\lambda^\alpha)\rangle = \kappa \Lambda^3
  e^{-\frac{2\pi i }{4k}n}~,\qquad n=0,1,2,\cdots, 4k-1~,
\end{equation}
where $\text{Tr}(\lambda_\alpha\lambda^\alpha)$ is the gaugino bilinear
operator, and $\kappa$ is a constant.
 The dynamical scale
$\Lambda$ is related to the $\theta$-angle by $\Lambda^{12k} =
|\Lambda^{12k}|e^{i\theta}$.
The
line operators are labeled by $(z;w)\in \mathbb{Z}_4\times
\mathbb{Z}_4$, and those with $(z;w) = (1,0), (2,0)$ and $(3,0)$ are
genuine line operators of the $Spin(4k+2)$ SYM.
At the $n$-th vacuum, the condensed
particle has a vanishing Dirac pairing with the line operator of $(z;w) = (\pm n,1)$, where
the sign depends on whether $k$ is even or odd. One can distinguish
different values of $n$ only modulo $4$, corresponding to the
SPT phases protected by the $\mathbb{Z}_4$ center symmetry of
$Spin(4k+2)$.
When turning on the soft SUSY breaking mass term $\delta \mathcal{L} =
m\,\text{Tr}(\lambda_\alpha\lambda^\alpha) + \text{c.c.}$, we find the effective
potential $V = -2|m\kappa\Lambda^3|\cos (\frac{\theta-2\pi n}{4k})$.\footnote{
Here, we choose the origin of $\theta$ so that $m\kappa$ is a positive
real number.} Repeating
exactly the same argument as in Sec.~\ref{subsec:SUSY_Spin4k}, we find
that the same situation as in the first line of Table~\ref{table:Spin4k+2_phases}.

Let us next consider the $\mathcal{N}=1$ $SO(4k+2)_+$ SYM. While the
number of supersymmetric vacua is the same as above, the IR theory existing at each
vacuum is different. The reason for this is that the genuine line
operators of the $SO(4k+2)_+$ theory are those with $(z;w) = (2;0),\,
(0;2)$ and $(2;2)$. At the $n$-th vacuum, the line of $(z;w) = (0;2)$
has a perimeter law if $n$ is even, and the one with $(z;w)=(2;2)$ has a
perimeter law if $n$ is odd, with the other lines having an area law. This
means that at every vacuum the IR theory is a $\mathbb{Z}_2$ TQFT.  
When the soft SUSY breaking term is added, only one vacuum has the
minimum energy for generic values of $\theta$. 
This result coincides with the second line of
Table~\ref{table:Spin4k+2_phases}.

Finally, we consider the $\mathcal{N}=1$ $SO(4k+2)_-$ SYM. The genuine
lines are now those with $(z;w) = (2,0), (1,2)$ and $(3,2)$. Since the
charge lattice of these lines is generated by $(1,2)$, the electric
and magnetic 1-form
symmetries are now combined into $\mathbb{Z}_4$. 
Since every genuine line is in an area
low, these phases are the trivial and SPT phases protected by
this $\mathbb{Z}_4$. When the soft SUSY breaking mass is
turned on, only a single vacuum is selected as the true vacuum,
depending on the value of $\theta$. This result is in perfect
agreement with the third line of Table~\ref{table:Spin4k+2_phases}.

\section{$Spin(2k+1)/SO(2k+1)$ Yang-Mills theory}
\label{sec:Spin2k+1}

In the case of $Spin(2k+1)/SO(2k+1)_\pm$ Yang-Mills theory for $k\geq 2$, there is no mixed
center/CP anomaly  \cite{Aharony:2013hda,
Witten:2000nv, Cordova:2019uob, Chen:2020syd} on spin manifolds. Therefore, one cannot apply the technique of
 mixed gauge theory to these theories. Below, we simply provide the
 phase diagram of the soft broken $Spin(2k+1)/SO(2k+1)_\pm$ SYM for
 completeness. Its comparison to the phase diagram of non-supersymmetric
 $Spin(2k+1)/SO(2k+1)_\pm$ Yang-Mills theories is left for future work.

We start with the $\mathcal{N}=1$ $Spin(2k+1)$ SYM for a positive
integer $k\geq 2$. Line operators are labeled by $(z;w)\in
\mathbb{Z}_2\times \mathbb{Z}_2$, and the genuine line operator of the
$Spin(2k+1)$ Yang-Mills theory is the Wilson line $W_\text{4d}$ in the spinor representation
corresponding to $(z;w) = (1;0)$.
 The theory  has $(2k-1)$ SUSY vacua \cite{Intriligator:1995id, Intriligator:1995au, Aharony:2013hda}
characterized by 
\begin{equation}
 \langle \text{Tr}(\lambda_\alpha\lambda^\alpha)\rangle = \kappa\Lambda^3
  e^{-\frac{2\pi i}{2k-1}n}~,\qquad n=0,1,2,\cdots,2k-2~,
\end{equation}
where $\kappa$ is a constant.
At the $n$-th vacuum, the electro-magnetic charge of the condensed
particles is proportional to that of the line operator of $(z;w) =
(2n,1)=(0,1)$ up to a shift by the root lattice. Therefore the genuine line operator always has an area law,
meaning that every vacuum is in the confining phase. A big difference
from the $Spin(\text{even})/SO(\text{even})_\pm$ theories is that, even
though the theory has a non-trivial center symmetry, one
cannot distinguish vacua with different values of $n$ by it. Indeed, the
center $\mathbb{Z}_2$ symmetry maps the Wilson line in the spinor
representation $W_\text{4d}$ as
\begin{equation}
 W_\text{4d} \to -W_\text{4d}~,
\end{equation}
but preserves the lines with $(z;w) = (2n;1)$ invariant. Therefore, the
above $(2k-1)$ vacua are all in the trivially gapped phase, i.e., the
confining phase. 

\begin{table}
 \centering
\begin{tabular}{c||c|c} 
&
    $\theta/\pi$ is not an odd integer & $\theta =(2n+1)\pi$ for $n\in\mathbb{Z}$ \\
\hline\hline
$Spin(2k+1)$ & trivial & two vacua \\
\hline
$SO(2k+1)_+$ & $\mathbb{Z}_2$ TQFT  & two
	     vacua supporting a $\mathbb{Z}_2$ TQFT\\
\hline
$SO(2k+1)_-$ & trivial  & two trivial vacua\\
\end{tabular}
\caption{The phase structure of the $\mathcal{N}=1$
 $Spin(2k+1)/SO(2k+1)_\pm$ SYM with a small gaugino mass
 $m$ turned on. Since there is no center/CP mixed anomaly, the two-fold
 degeneracy at $\theta = \pi$ modulo $2\pi$ might be resolved when $|m|$ is large.}
\label{table:Spin2k+1_phases}
\end{table}

When turning on the soft SUSY breaking term $\delta \mathcal{L} =
m\,\text{Tr}(\lambda_\alpha\lambda^\alpha) + \text{c.c.}$, one finds the effective
potential $V=-2|m\kappa\Lambda^3|\cos(\frac{\theta - 2\pi n}{2k-1})$ at the
$n$-th vacuum, which means a single true vacuum for generic values of
$\theta$. Since all the $(2k-1)$ vacua are in the trivially gapped
phase, we find that the vacuum structure has the periodicity $\theta
\sim \theta + 2\pi$.
This result is summarized in the first line of Table~\ref{table:Spin2k+1_phases}.

Let us now move on to the $\mathcal{N}=1$ $SO(2k+1)_+$ SYM. Its genuine line operator is the one
with $(z;w) = (0;1)$, which has a perimeter law at all vacua since
whose electro-magnetic charge has a vanishing pairing with the charges of
the condensed particles. Therefore, at every vacuum, we find a $\mathbb{Z}_2$ TQFT in the
infrared.
Thus we find the phase diagram of the softly broken $SO(2k+1)_+$ SYM  as in the second line of
Table~\ref{table:Spin2k+1_phases}.

In the case of $SO(2k+1)_-$ SYM, the genuine line
operator is the one for $(z;w) = (1,1)$. Since this line has a
non-vanishing paring with the electro-magnetic charge of condensed
particles at every vacuum, all vacua of this SYM are in the
confining phase.\footnote{Note that, while the $SO(2k+1)$ Yang-Mills theory has a magnetic
$\mathbb{Z}_2$ 1-form symmetry, one cannot distinguish these
$(2k-1)$ vacua by this magnetic $\mathbb{Z}_2$, since condensed particles
have the same magnetic charge at all these vacua.} This implies the
phase diagram shown in the
third line of Table~\ref{table:Spin2k+1_phases}.

For all $Spin(2k+1)/SO(2k+1)_\pm$ SYM, an interesting situation occurs
when $\theta/\pi$ is an odd integer. In this case, the
$Spin(2k+1)/SO(2k+1)_\pm$ SYM with a small gaugino mass has two degenerate vacua. In contrast to the $Spin(\text{even})/SO(\text{even})$ cases, there is no
mixed center/CP anomaly on spin manifolds that prevents this degeneracy
from being resolved. 
Note that, since the phase diagram is periodic under
$\theta \to \theta + 2\pi$, the absence of phase
transition at $\theta = (2n+1)\pi$ for $n\in \mathbb{Z}$ does not immediately lead to a
contradiction. A possible way to study this degeneracy is to consider the mixed gauge theory on
non-spin manifolds, on which there is a mixed anomaly between $\theta
\to \theta + 2\pi$ and the center symmetry \cite{Cordova:2019uob}. We leave a careful study of this subtlety for future
work.

 \section{Summary and discussions}
 \label{sec:summary}

In this paper, we have studied the low-temperature phase diagrams of $Spin(N)/SO(N)_\pm$ Yang-Mills
theories as functions of the $\theta$-parameter. In particular, for
every even $N\geq 4$, we
have identified a phase diagram consistent with the center/CP mixed
anomaly via the technique of mixed gauge theory. For that, we have made 
an assumption that the $Spin(N)$ Yang-Mills theory is gapped for all values of the $\theta$
parameter and has a trivial non-degenerate vacuum at $\theta = 0$, as in
most sections of \cite{Gaiotto:2017yup}.

Among other cases, the $N=4$ case is
special in the sense that the theory
has two independent $\theta$-parameters. Therefore we have started our
analysis with the $N=4$ case in Sec.~\ref{sec:Spin4}. The resulting
phase diagram is shown in Fig.~\ref{fig:phasetheta1and2-0} and
Table~\ref{table:phase_Spin4}. We have particularly argued that, for the $Spin(4)$
Yang-Mills theory, there are four degenerate vacua
 at intersections of dashed lines in
Fig.~\ref{fig:phasetheta1and2-0}.

 As discussed in Sec.~\ref{sec:Spin4k}, its generalization to the $Spin(4k)/SO(4k)_\pm$ for $k\in \mathbb{N}$
 is straightforward, except that now the theory has only a single
 $\theta$-parameter and the center/CP anomaly for even $k$ is slightly
 different from that for odd $k$. The phase diagrams of $Spin(4k)/SO(4k)_\pm$ Yang-Mills
 theories are shown in Fig.~\ref{fig:Spin4k_phases} and Table~\ref{table:Spin4k_phases}.

The $Spin(4k+2)/SO(4k+2)_\pm$ Yang-Mills theories for $k\in\mathbb{N}$
are essentially different from the above examples, since the center
symmetry is now $\mathbb{Z}_4$ instead of $\mathbb{Z}_2\times
\mathbb{Z}_2$. As a result, in the $Spin(4k+2)$ case, the phase diagram we have identified has
periodicity $\theta \sim \theta +8\pi$, as shown in Fig.~\ref{fig:Spin4k+2_phases} and
Table~\ref{table:Spin4k+2_phases}. 

For all the above gauge groups, we have also analyzed the
phase diagrams of SYM with a small gaugino mass $m$ turned on. The
resulting phase diagrams turn out to be in perfect agreement with
those of the non-supersymmetric Yang-Mills theories read off from the
center/CP mixed anomaly. This strongly suggests that the phase diagram
is independent of $m$. The small-$|m|$ analysis for the $Spin(\text{odd})/SO(\text{odd})_\pm$
Yang-Mills theories is also given in Sec.~\ref{sec:Spin2k+1} for completeness.

The absence of a center/CP mixed anomaly for
$Spin(\text{odd})$ might suggest that the degeneracy
of two
vacua at $\theta = \pi$ in the corresponding softly broken SYM can be resolved by
quantum corrections. However, on non-spin manifolds, there is a mixed
anomaly between the center symmetry and $\theta \to \theta + 2\pi$ for
$Spin(\text{odd})$ \cite{Cordova:2019uob}. It would
therefore be
interesting to study these theories on non-spin manifolds and see this
point further.

Another future direction is to test the phase diagrams obtained in
this paper through numerical simulations. In the conventional Monte Carlo simulation the presence of the $\theta$-term gives rise to a sign
problem, but there are approaches to overcome it 
\cite{Panagopoulos:2011rb, DElia:2013uaf, Bonati:2015sqt,
Bonati:2016tvi, Kitano:2020mfk, Kitano:2021jho, Matsumoto:2023vbf,
Bonanno:2023hhp, Hirasawa:2024vns,Bonanno:2024ggk, Yamada:2024vsk,
Yamada:2024pjy, Hirasawa:2024fjt,
Hirasawa:2025mot, Matsumoto:2026puv, Watanabe:2026ymb}. It would also be interesting to
apply the tensor network and quantum computing in the
Hamiltonian formalism as they are free from the sign problem.
 Indeed, various numerical studies in the Hamiltonian formalism have been conducted on 
the two-dimensional Schwinger model with a non-vanishing
$\theta$-parameter \cite{Funcke:2019zna, Chakraborty:2020uhf, Honda:2021aum, Honda:2021ovk, Honda:2022edn, Itou:2024psm, Matsumoto:2025fjb}.

\section*{Declaration of generative AI and AI-assisted technologies in
the manuscript preparation process}

During the preparation of this work, the authors used ChatGPT (OpenAI)
for language editing and to help identify typos and potential 
inconsistencies in the manuscript. 
After using this tool, the authors critically evaluated the suggestions
and independently verified the calculations and arguments underlying any
revisions. The authors take full responsibility for the content of the
article.

\bigskip
\bigskip

\begin{center}
\noindent{\bf Acknowledgements}
\end{center}
We are grateful to E.~Itou,  A.~Matsumoto, K.~Maruyoshi and S.~Moriyama for illuminating
discussions.
The authors' research is partially supported by JSPS
KAKENHI Grant Number JP21H04993.
In addition, T.~N's research is partially supported by
JSPS KAKENHI Grant Numbers JP18K13547, JP23K03393, JP23K03394 and JP26H01997.

\bigskip
\bigskip

\begin{appendices}
 
\section{The spectrum
of
genuine line 
operators}
\label{app:line}
In this
appendix, we review the 
spectrum of genuine line operators in a four-dimensional
Yang-Mills theory of gauge group $G$ associated 
with a finite-dimensional real Lie algebra $\mathfrak{g}$, following \cite{Aharony:2013hda}. 
We assume that $G$ is connected, which means that there is a sub-group
$H$ of the center $C$ of the simply-connected Lie group $\tilde G$
integrating $\mathfrak{g}$ such that
\begin{equation} 
  G = \widetilde G/H~.
\end{equation} 

A 
genuine line 
operator is generally labeled by
\begin{equation} 
  (\lambda_e,\lambda_m) \in \Lambda_w(\mathfrak{g}) \times
\Lambda_w(\mathfrak{g}^\vee), 
\end{equation} 
where $\Lambda_w(\mathfrak{g})$ is the weight lattice of $\mathfrak{g}$,
and 
$\mathfrak{g}^\vee$ is the Langlands-dual 
of $\mathfrak{g}$. When $\mathfrak{g}$ is
 a simply-laced simple Lie algebra, these two lattices are identical.

The spectrum of genuine line operators in the Yang-Mills theory depends not only on
$\mathfrak{g}$ but $G$. In particular, purely-electric genuine lines
(i.e., Wilson lines) are
those labeled by $(\lambda_e,0)$ invariant under the action of $H$. This
means that, if the gauge group is $\tilde {G}$ then the line with
$(\lambda_e,0)$ for every $\lambda_e\in\Lambda_w(\mathfrak{g})$ is a
genuine Wilson line, but if $G\not \simeq  \tilde G$ not all
representations of $\mathfrak{g}$ give genuine Wilson lines. The
complete set $L$ of genuine line operators, which also  includes lines with
non-vanishing magnetic charges, is then determined so that, for every
pair $\mathcal{L}_1,\,\mathcal{L}_2\in L$, their Dirac pairing $\langle
\mathcal{L}_1, \mathcal{L}_2\rangle$ vanishes.

Note that if $L$ contains a line with
$(\lambda_e^{(1)},\,\lambda_m^{(1)})$ and another line with
$(\lambda_e^{(2)},\,\lambda_m^{(2)})$ then $L$ also contains a line with
$(\lambda_e^{(1)} + \lambda_e^{(2)},\,\lambda_m^{(1)} +
\lambda_m^{(2)})$. In addition, the charge conjugation implies that the presence of a line with
$(\lambda_e,\, \lambda_m)$ gives rise to another line with
$(-\lambda_e,-\lambda_m)$. We also see that the Wilson in the adjoint
representation of $\mathfrak{g}$ and the  't
Hooft line in
the adjoint representation of $\mathfrak{g}^\vee$ always exist. Combining
all these facts, we see that the spectrum of genuine line operators is
essentially characterized by a subset of
$(\Lambda_w(\mathfrak{g})/\Lambda_r(\mathfrak{g}))\times
(\Lambda_w(\mathfrak{g}^\vee)/\Lambda_r(\mathfrak{g}^\vee)) \simeq C
\times C$, where
$\Lambda_r(\mathfrak{g})$ is the root lattice of $\mathfrak{g}$ and we
used the fact that the simply-connected Lie group integrating
$\mathfrak{g}^\vee$ has the same center as $\tilde G$. 
We
therefore label genuine line operators in the theory by
\begin{equation} 
  (z_e, z_m) \in C \times C~.
\end{equation} 

In the rest of this appendix, we focus on the cases of $G = Spin(N)$ and
$G=SO(N)$. Then the center $C$ is $\mathbb{Z}_2$ when $N$ is
odd, $\mathbb{Z}_2\times \mathbb{Z}_2$ when $N/2$ is an even integer,
and $\mathbb{Z}_4$ when $N/2$ is an odd integer. 
 In the case of $G=Spin(N)$, genuine lines are those of
$(z_e,0)$ for all $z_e\in C$. In the case of $G= SO(N)$, the
spectrum of genuine lines is identified by noting that
\begin{equation}
    SO(N) = 
Spin(N)/\mathbb{Z}_2~,
\end{equation}
where $\mathbb{Z}_2$ is a particular sub-group of $C$.
At the level of the charges of genuine line operators, this quotient has two effects:
\begin{enumerate}
    \item The allowed values of $z_e$ 
are restricted to those that are
    invariant under 
$\mathbb{Z}_2$.

    \item Extra lines with non-vanishing $z_m$ can be included in the
	  spectrum.
\end{enumerate}
There are two different options for the set of ``extra lines'' with
non-vanishing $z_m$, corresponding to $SO(N)_+$ and $SO(N)_-$ Yang-Mills theories.

\subsection{$Spin(2k+1)/SO(2k+1)$ case}

Let us
describe the cases of $G=Spin(2k+1)$ and $G=SO(2k+1)$ for a positive
integer $k$ in more detail.
In this case, since 
$C= \mathbb{Z}_2$, 
genuine line operators are labeled by
\begin{equation}
  (z_e, z_m) \in \mathbb{Z}_2 \times \mathbb{Z}_2.~
\end{equation}
For a line with $(z_e^{(1)},z_m^{(1)})$ and another line with
$(z_e^{(2)},z_m^{(2)})$, the vanishing of their Dirac paring is
equivalent to
\begin{equation}
 z_e^{(1)}z_m^{(2)} - z_m^{(1)}z_e^{(2)}= 0 \;\;\text{mod}\;\; 2~.
\label{eq:Dirac_condition}
\end{equation}

When $G=Spin(2k+1)$, the only non-trivial genuine line is the one for
$(z_e,z_m) = (1,0)$, i.e., the Wilson line in the fundamental
representation of $Spin(2k+1)$. When $G=SO(2k+1) \simeq Spin(2k+1)/\mathbb{Z}_2$, this line is no
longer genuine since it is not invariant under
$\mathbb{Z}_2$. Therefore, there is no purely-electric non-trivial
genuine line operator in the $SO(2k+1)$ Yang-Mills theory. Instead, one
can include an extra line with non-vanishing $z_m$. We call the theory
the $SO(2k+1)_+$ Yang-Mills theory when the line with $(z_e,z_m) =
(0,1)$ is included in the spectrum, and the $SO(2k+1)_-$ Yang-Mills theory when the one
with $(z_e,z_m) = (1,1)$ is included. Note that one cannot include
further extra lines without breaking the condition
\eqref{eq:Dirac_condition}.

\begin{figure}[h]
  \centering
  \begin{tabular}{ccc}
    \begin{minipage}[t]{0.32\linewidth} 
      \begin{tikzpicture}[scale=0.8]
        \draw[->,thick]
          (-2.6,0) -- (2.6,0)
          node[right] {$\lambda_e$};

        \draw[->,thick]
          (0,-2.6) -- (0,2.6)
          node[above] {$\lambda_m$};

        \foreach \x in {-2,-1,1,2}
        {
          \draw (\x,0.08) -- (\x,-0.08);
          \node[below=3pt] at (\x,0) {\small $\x$};
        }

        \foreach \y in {-2,-1,1,2}
        {
          \draw (0.08,\y) -- (-0.08,\y);
          \node[left=3pt] at (0,\y) {\small $\y$};
        }
        \foreach \x in {-2,-1,0,1,2}
        {
        \foreach \y in {-2,0,2}
        {
            \fill[black]
              (\x,\y) circle (4pt);
        }
        }
        \node[below left=3pt] at (0,0) {$0$};
      \end{tikzpicture}
      \subcaption{$Spin(3)$}
    \end{minipage} &
    \begin{minipage}[t]{0.32\linewidth}
        \begin{tikzpicture}[scale=0.8]

          \draw[->,thick]
            (-2.6,0) -- (2.6,0)
            node[right] {$\lambda_e$};
          \draw[->,thick]
            (0,-2.6) -- (0,2.6)
            node[above] {$\lambda_m$};

          \foreach \x in {-2,-1,1,2}
          {
            \draw (\x,0.08) -- (\x,-0.08);
            \node[below=3pt] at (\x,0) {\small $\x$};
          }

          \foreach \y in {-2,-1,1,2}
          {
            \draw (0.08,\y) -- (-0.08,\y);
            \node[left=3pt] at (0,\y) {\small $\y$};
          }
          \foreach \x in {-2,0,2}
          {
          \foreach \y in {-2,-1,0,1,2}
          { 
            \fill[black]
              (\x,\y) circle (4pt);
          }
          }
          \node[below left=3pt] at (0,0) {$0$};
        \end{tikzpicture}
        \subcaption{$SO(3)_+$}
    \end{minipage} &
    \begin{minipage}[t]{0.32\linewidth}
      \begin{tikzpicture}[scale=0.8]

        \draw[->,thick]
          (-2.6,0) -- (2.6,0)
          node[right] {$\lambda_e$};
        \draw[->,thick]
          (0,-2.6) -- (0,2.6)
          node[above] {$\lambda_m$};

        \foreach \x in {-2,-1,1,2}
        {
          \draw (\x,0.08) -- (\x,-0.08);
          \node[below=3pt] at (\x,0) {\small $\x$};
        }

        \foreach \y in {-2,-1,1,2}
        {
          \draw (0.08,\y) -- (-0.08,\y);
          \node[left=3pt] at (0,\y) {\small $\y$};
        }

        \foreach \x in {-2,0,2}
        {
        \foreach \y in {-2,0,2}
        {
            \fill[black]
              (\x,\y) circle (4pt);
        }
        }

        \foreach \x in {-1,1}
        {
        \foreach \y in {-1,1}
        {
            \fill[black]
              (\x,\y) circle (4pt);
        }
        }
        \node[below left=3pt] at (0,0) {$0$};
      \end{tikzpicture}
      \subcaption{$SO(3)_-$}
    \end{minipage}
  \end{tabular}
  \caption{The electro-magnetic charges of genuine line operators in
$Spin(3)\simeq SU(2)$ and $SO(3)_{\pm}$ Yang-Mills theories.
}
  \label{fig:weightlattice_Spin(5)}
\end{figure}
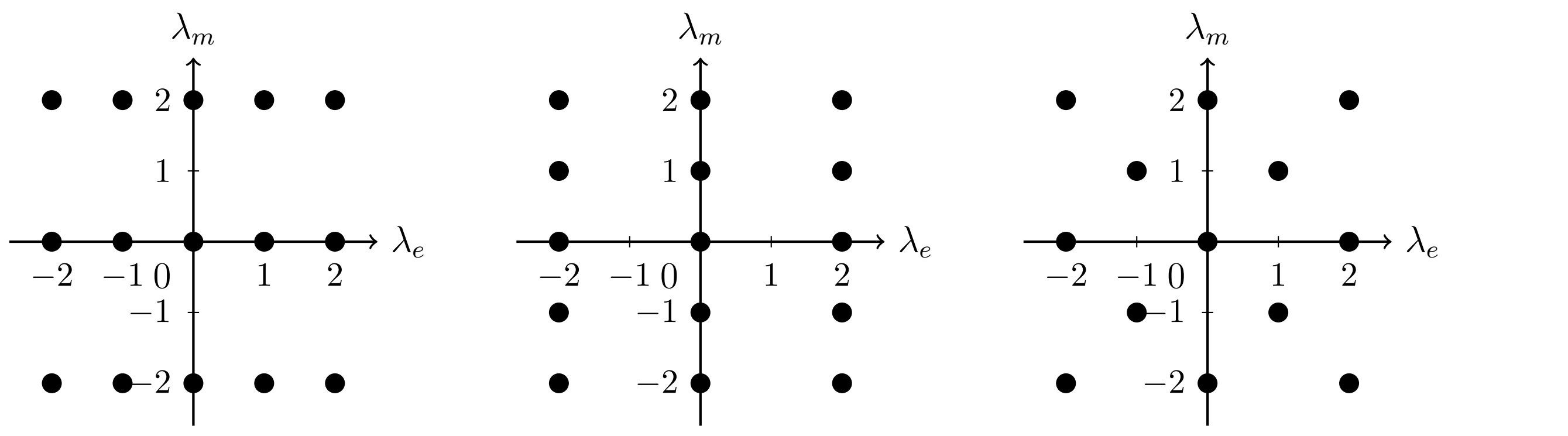

The above discussion tells us about the allowed electro-magnetic charges $(\lambda_e,\lambda_m)$
of genuine line operators.
For example, when  $G=Spin(3)\simeq SU(2)$, the allowed charges are shown 
in the
left picture of
Fig.~\ref{fig:weightlattice_Spin(5)}. 
We see from this picture that the $Spin(3)$ Yang-Mills theory has an
electric $\mathbb{Z}_2$ 1-form symmetry whose generator assigns the phase factor
$e^{\pi i \lambda_e} = e^{\pi i z_e}$ to the line operator associated with
$(\lambda_e,\lambda_m)$.
When $G=SO(3)_+$, the allowed values of $(\lambda_e,\lambda_m)$ are now
as shown in the middle picture of
Fig.~\ref{fig:weightlattice_Spin(5)}. The 1-form symmetry is now a
magnetic $\mathbb{Z}_2$ symmetry whose generator assigns $e^{\pi i
\lambda_m} = e^{\pi i z_m}$ to
line operators.
The spectrum of genuine lines of $SO(3)_-$ Yang-Mills theory is shown in
the right picture of Fig.~\ref{fig:weightlattice_Spin(5)}. We see that
there is now a $\mathbb{Z}_2$ 1-form symmetry whose generator  assigns $e^{\pi i
\lambda_e}=e^{\pi i z_e}$ 
to the lines.\footnote{Here one can take $e^{\pi i \lambda_m}=e^{\pi i z_m}$ instead
of $e^{\pi i \lambda_e}=e^{\pi i z_e}$, since the non-trivial genuine line
operators have $z_e = z_m$.}
We summarize these  lines that transform
non-trivially under the 1-form symmetry in Table~\ref{table:Spin2k+1_charges}. 

\begin{table}
  \begin{center}
    \begin{tabular}{c|c|c}
      gauge group & 
1-form sym  & $(z_e, z_m)$ of non-trivial genuine lines\\ \hline
      $Spin(2k+1)$ & $\mathbb{Z}_2$ & $(1, 0)$ \\
      $SO(2k+1)_+$ & $\mathbb{Z}_2$ & $(0, 1)$ \\
      $SO(2k+1)_-$ & $\mathbb{Z}_2$ & $(1, 1)$ \\
    \end{tabular}
    \caption{The list of non-trivial genuine lines and the 1-form
   symmetry in the case of$\mathfrak{g} = \mathfrak{so}(2k+1)$ $(k \geq 2)$}
    \label{table:Spin2k+1_charges}
  \end{center}
\end{table}

\subsection{$Spin(N)/SO(N)$ for even $N$}

Here, we briefly describe the genuine line operators in $Spin(N)$ and
$SO(N)_\pm$ Yang-Mills theories for even $N$, generalizing the above
discussions. 

We start with the case of  $N=4k+2$ for a positive integer $k$, in which
case $C=\mathbb{Z}_4$ and therefore line operators are classified by $(z_e, z_m) \in \mathbb{Z}_4 \times \mathbb{Z}_4$. 
The vanishing condition of the Dirac pairing is now 
\begin{equation}
  z_e^{(1)} z_m^{(2)} - z_m^{(1)} z_e^{(2)}= 0 \;\;\text{mod}\;\;4~.
\end{equation} 
When $G=Spin(4k+2)$, non-trivial genuine lines are those of $(z_e,z_m) = (1,0),
(2,0)$ and $(3,0)$. There is a 1-form $\mathbb{Z}_4$ symmetry whose generator
assigns 
$e^{\pi i z_e/2}$ to the lines. When
$G=SO(4k+2)_+$, non-trivial genuine lines are those of $(z_e,z_m) =
(2,0),(0,2)$ and $(2,2)$. The 1-form symmetry is now $\mathbb{Z}_2\times
\mathbb{Z}_2$ whose generators assign $e^{\pi i z_e/2}$ and $e^{\pi i
z_m/2}$ to the lines. When $G=SO(4k+2)_-$, non-trivial genuine lines are
those of $(z_e,z_m)=(2,0),(1,2)$ and $(3,2)$. The 1-form symmetry is
$\mathbb{Z}_4$ whose generator assigns $e^{\pi i (z_m-z_e)/2}$ to the
line associated with $(z_e,z_m)$.\footnote{This $\mathbb{Z}_4 = \{1,\omega,\omega^2,\omega^3\}$ contains
a magnetic $\mathbb{Z}_2$ sub-group $\mathbb{Z}_2^\text{mag} =
\{1,\omega^2\}$. The quotient
$\mathbb{Z}_4/\mathbb{Z}_2^\text{mag}\simeq \mathbb{Z}_2^\text{ele}$ can
be regarded as the electric 1-form symmetry corresponding to the center
of $SO(4k+2)$.} These results are summarized in
Table~\ref{table:Spin4k+2_charges}.
\begin{table}
  \begin{center}
    \begin{tabular}{c|c|c}
      gauge group & 1-form sym & $(z_e, z_m)$ of non-trivial genuine lines \\ \hline
      $Spin(4k+2)$ &  $\mathbb{Z}_4$ & $(1,0)$, $(2, 0)$, $(3, 0)$ \\
      $SO(4k+2)_+$ & $\mathbb{Z}_2$ $\times$ 
	 $\mathbb{Z}_2$ & $(2,0)$, $(0, 2)$, $(2, 2)$ \\
      $SO(4k+2)_-$ & $\mathbb{Z}_4$ & $(2,0)$, $(1, 2)$, $(3, 2)$ \\
    \end{tabular}
    \caption{The spectrum of non-trivial genuine line operators for $\mathfrak{g} =
   \mathfrak{so}(4k+2)$ $(k \geq 1)$ and the corresponding 1-form
   symmetry.}
    \label{table:Spin4k+2_charges}
  \end{center}
\end{table}

In the case of $N=4k$ for a positive integer $k$, we have $C =
\mathbb{Z}_2^S\times \mathbb{Z}_2^C$ and therefore line operators are
characterized by 
\begin{equation}
(z_{e, S}, z_{e, C}; z_{m, S}, z_{m, C}) \in (\mathbb{Z}_2)^4~.
\end{equation}
The vanishing condition for the Dirac pairing is
\begin{equation}
  z_{e,S}^{(1)} z_{m,S}^{(2)} - z_{m,S}^{(1)} z_{e,S}^{(2)} + z_{e,C}^{(1)} z_{m,C}^{(2)} - z_{m,C}^{(1)}
   z_{e,C}^{(2)} = 0\;\;\text{mod}\;\;2
  \label{eq:condition_Spin8k+4}
\end{equation}
for any integer $k$. Note that this notation is slightly different from
the one used in \cite{Aharony:2013hda} when $k$ is even. As a result,
the Witten effect under $\theta \to \theta + 2\pi$ is written as
\begin{equation}
 (z_{e,S},z_{e,C};z_{m,S},z_{m,C}) \to  (z_{e,S} + z_{m,S},z_{e,C}+z_{m,C};z_{m,S},z_{m,C})
\end{equation}
when $k$ is odd, while it is expressed as
\begin{equation}
 (z_{e,S},z_{e,C};z_{m,S},z_{m,C}) \to  (z_{e,S} + z_{m,C},z_{e,C}+z_{m,S};z_{m,S},z_{m,C})
\end{equation}
when $k$ is even.

\begin{table}
  \begin{center}
    \begin{tabular}{c|c|c}
      gauge group & 
1-form sym &  $(z_{e, S}, z_{e, C}; z_{m, S}, z_{m, C})$ of non-trivial
	     genuine lines\\ \hline
      $Spin(4k)$ & $\mathbb{Z}_2 \times \mathbb{Z}_2$ & $(1, 0; 0, 0 )$, $(0, 1; 0, 0)$, $(1, 1; 0, 0)$ \\
      $SO(4k)_+$ & $\mathbb{Z}_2\times \mathbb{Z}_2$ & $(1, 1; 0, 0)$, $(0, 0; 1, 1)$, $(1, 1; 1, 1)$ \\
      $SO(4k)_-$ & $\mathbb{Z}_2\times \mathbb{Z}_2$ & $(1, 1; 0, 0 )$, $(1, 0; 1, 1)$, $(0, 1; 1, 1)$ \\
    \end{tabular}
    \caption{The spectrum of genuine line operators for $\mathfrak{g} =
   \mathfrak{so}(4k)$ $(k \geq 1)$, and the corresponding 1-form symmetry.}
    \label{table:Spin4k_charges}
  \end{center}
\end{table}
When $G=Spin(4k)$, non-trivial genuine
lines are those for $(z_{e,S},z_{e,C};z_{m,S},z_{m,C}) =(1,0;0,0),
(0,1;0,0)$ and $(1,1;0,0)$. The 1-form symmetry is $\mathbb{Z}_2\times
\mathbb{Z}_2$ whose generators assign $e^{\pi i z_{e,S}}$ and $e^{\pi i
z_{e,C}}$ to the lines. When $G=SO(4k)_+$, genuine lines are those for
$(z_{e,S},z_{e,C};z_{m,S},z_{m,C})=(1,1;0,0),(0,0;1,1)$ and
$(1,1;1,1)$. The 1-form symmetry in this case is $\mathbb{Z}_2\times
\mathbb{Z}_2$ whose generators assign $e^{\pi i z_{e,S}}$ and $e^{\pi i
z_{m,S}}$ to the lines. Finally, when $G=SO(4k)_-$, genuine
lines are those for $(z_{e,S},z_{e,C};z_{m,S},z_{m,C}) =
(1,1;0,0),(1,0;1,1)$ and $(0,1;1,1)$. The 1-form symmetry is again
$\mathbb{Z}_2\times \mathbb{Z}_2$ whose generators assign $e^{\pi i
z_{e,S}}$ and $e^{\pi i z_{e,C}}$ to the lines. These results are
summarized in Table~\ref{table:Spin4k_charges}.

\end{appendices}

\bibliography{YM}

%bibliography generated by nb.bst v1.01 (C) 2003-2010 Niklas Beisert
\begin{thebibliography}{10}
\ifx\href\asklfhas\newcommand{\href}[2]{#2}\fi
\ifx\arxivref\asklfhas\newcommand{\arxivref}[2]{\href{http://arxiv.org/abs/#1}{#2}}\fi
\ifx\doiref\asklfhas\newcommand{\doiref}[2]{\href{http://dx.doi.org/#1}{#2}}\fi
\parskip 0pt
\normalsize

\bibitem{Gaiotto:2017yup}
D.~Gaiotto, A.~Kapustin, Z.~Komargodski \& N.~Seiberg,
\textit{``{Theta, Time Reversal, and Temperature}''},
\doiref{10.1007/JHEP05(2017)091}{JHEP \textbf{1705}, 091
  (2017)\ignorespaces}\ignorespaces,
\normalsize{\texttt{\arxivref{1703.00501}{arXiv:1703.00501
  \![hep-th]}}}\ignorespaces
\bibitem{Tachikawa:2017gyf}
Y.~Tachikawa,
\textit{``{On gauging finite subgroups}''},
\doiref{10.21468/SciPostPhys.8.1.015}{SciPost~Phys. \textbf{8}, 015
  (2020)\ignorespaces}\ignorespaces,
\normalsize{\texttt{\arxivref{1712.09542}{arXiv:1712.09542
  \![hep-th]}}}\ignorespaces
\bibitem{Tanizaki:2017bam}
Y.~Tanizaki \& Y.~Kikuchi,
\textit{``{Vacuum structure of bifundamental gauge theories at finite
  topological angles}''},
\doiref{10.1007/JHEP06(2017)102}{JHEP \textbf{1706}, 102
  (2017)\ignorespaces}\ignorespaces,
\normalsize{\texttt{\arxivref{1705.01949}{arXiv:1705.01949
  \![hep-th]}}}\ignorespaces
\bibitem{Shimizu:2017asf}
H.~Shimizu \& K.~Yonekura,
\textit{``{Anomaly constraints on deconfinement and chiral phase
  transition}''},
\doiref{10.1103/PhysRevD.97.105011}{Phys.~Rev.~D \textbf{97}, 105011
  (2018)\ignorespaces}\ignorespaces,
\normalsize{\texttt{\arxivref{1706.06104}{arXiv:1706.06104
  \![hep-th]}}}\ignorespaces
\bibitem{Gaiotto:2017tne}
D.~Gaiotto, Z.~Komargodski \& N.~Seiberg,
\textit{``{Time-reversal breaking in QCD$_{4}$, walls, and dualities in 2 + 1
  dimensions}''},
\doiref{10.1007/JHEP01(2018)110}{JHEP \textbf{1801}, 110
  (2018)\ignorespaces}\ignorespaces,
\normalsize{\texttt{\arxivref{1708.06806}{arXiv:1708.06806
  \![hep-th]}}}\ignorespaces
\bibitem{Tanizaki:2017mtm}
Y.~Tanizaki, Y.~Kikuchi, T.~Misumi \& N.~Sakai,
\textit{``{Anomaly matching for the phase diagram of massless
  $\mathbb{Z}_N$-QCD}''},
\doiref{10.1103/PhysRevD.97.054012}{Phys.~Rev.~D \textbf{97}, 054012
  (2018)\ignorespaces}\ignorespaces,
\normalsize{\texttt{\arxivref{1711.10487}{arXiv:1711.10487
  \![hep-th]}}}\ignorespaces
\bibitem{Aitken:2018kky}
K.~Aitken, A.~Cherman \& M.~{\"U}nsal,
\textit{``{Dihedral symmetry in $SU(N)$ Yang-Mills theory}''},
\doiref{10.1103/PhysRevD.100.085004}{Phys.~Rev.~D \textbf{100}, 085004
  (2019)\ignorespaces}\ignorespaces,
\normalsize{\texttt{\arxivref{1804.05845}{arXiv:1804.05845
  \![hep-th]}}}\ignorespaces
\bibitem{Cordova:2018acb}
C.~C{\'o}rdova \& T.~T. Dumitrescu,
\textit{``{Candidate phases for SU(2) adjoint QCD$_4$ with two flavors from
  $\mathcal{N}=2$ supersymmetric Yang-Mills theory}''},
\doiref{10.21468/SciPostPhys.16.5.139}{SciPost~Phys. \textbf{16}, 139
  (2024)\ignorespaces}\ignorespaces,
\normalsize{\texttt{\arxivref{1806.09592}{arXiv:1806.09592
  \![hep-th]}}}\ignorespaces
\bibitem{Tanizaki:2018wtg}
Y.~Tanizaki,
\textit{``{Anomaly constraint on massless QCD and the role of Skyrmions in
  chiral symmetry breaking}''},
\doiref{10.1007/JHEP08(2018)171}{JHEP \textbf{1808}, 171
  (2018)\ignorespaces}\ignorespaces,
\normalsize{\texttt{\arxivref{1807.07666}{arXiv:1807.07666
  \![hep-th]}}}\ignorespaces
\bibitem{Yamaguchi:2018xse}
S.~Yamaguchi,
\textit{``{{\textquoteright}t Hooft anomaly matching condition and chiral
  symmetry breaking without bilinear condensate}''},
\doiref{10.1007/JHEP01(2019)014}{JHEP \textbf{1901}, 014
  (2019)\ignorespaces}\ignorespaces,
\normalsize{\texttt{\arxivref{1811.09390}{arXiv:1811.09390
  \![hep-th]}}}\ignorespaces
\bibitem{Anber:2018xek}
M.~M. Anber \& E.~Poppitz,
\textit{``{Domain walls in high-T SU(N) super Yang-Mills theory and
  QCD(adj)}''},
\doiref{10.1007/JHEP05(2019)151}{JHEP \textbf{1905}, 151
  (2019)\ignorespaces}\ignorespaces,
\normalsize{\texttt{\arxivref{1811.10642}{arXiv:1811.10642
  \![hep-th]}}}\ignorespaces
\bibitem{Wan:2018djl}
Z.~Wan \& J.~Wang,
\textit{``{Adjoint QCD$_4$, Deconfined Critical Phenomena, Symmetry-Enriched
  Topological Quantum Field Theory, and Higher Symmetry-Extension}''},
\doiref{10.1103/PhysRevD.99.065013}{Phys.~Rev.~D \textbf{99}, 065013
  (2019)\ignorespaces}\ignorespaces,
\normalsize{\texttt{\arxivref{1812.11955}{arXiv:1812.11955
  \![hep-th]}}}\ignorespaces
\bibitem{Yonekura:2019vyz}
K.~Yonekura,
\textit{``{Anomaly matching in QCD thermal phase transition}''},
\doiref{10.1007/JHEP05(2019)062}{JHEP \textbf{1905}, 062
  (2019)\ignorespaces}\ignorespaces,
\normalsize{\texttt{\arxivref{1901.08188}{arXiv:1901.08188
  \![hep-th]}}}\ignorespaces
\bibitem{Karasik:2019bxn}
A.~Karasik \& Z.~Komargodski,
\textit{``{The Bi-Fundamental Gauge Theory in 3+1 Dimensions: The Vacuum
  Structure and a Cascade}''},
\doiref{10.1007/JHEP05(2019)144}{JHEP \textbf{1905}, 144
  (2019)\ignorespaces}\ignorespaces,
\normalsize{\texttt{\arxivref{1904.09551}{arXiv:1904.09551
  \![hep-th]}}}\ignorespaces
\bibitem{Poppitz:2019fnp}
E.~Poppitz \& T.~A. Ryttov,
\textit{``{Possible new phase for adjoint QCD}''},
\doiref{10.1103/PhysRevD.100.091901}{Phys.~Rev.~D \textbf{100}, 091901
  (2019)\ignorespaces}\ignorespaces,
\normalsize{\texttt{\arxivref{1904.11640}{arXiv:1904.11640
  \![hep-th]}}}\ignorespaces
\bibitem{Cordova:2019bsd}
C.~C{\'o}rdova \& K.~Ohmori,
\textit{``{Anomaly Obstructions to Symmetry Preserving Gapped Phases}''},
\normalsize{\texttt{\arxivref{1910.04962}{arXiv:1910.04962
  \![hep-th]}}}\ignorespaces
\bibitem{Wang:2019obe}
J.~Wang, Y.-Z. You \& Y.~Zheng,
\textit{``{Gauge enhanced quantum criticality and time reversal deconfined
  domain wall: SU(2) Yang-Mills dynamics with topological terms}''},
\doiref{10.1103/PhysRevResearch.2.013189}{Phys.~Rev.~Res. \textbf{2}, 013189
  (2020)\ignorespaces}\ignorespaces,
\normalsize{\texttt{\arxivref{1910.14664}{arXiv:1910.14664
  \![cond-mat.str-el]}}}\ignorespaces
\bibitem{Chen:2020syd}
S.~Chen, K.~Fukushima, H.~Nishimura \& Y.~Tanizaki,
\textit{``{Deconfinement and $\mathcal {CP}$ breaking at $\theta=\pi$ in
  Yang-Mills theories and a novel phase for SU(2)}''},
\doiref{10.1103/PhysRevD.102.034020}{Phys.~Rev.~D \textbf{102}, 034020
  (2020)\ignorespaces}\ignorespaces,
\normalsize{\texttt{\arxivref{2006.01487}{arXiv:2006.01487
  \![hep-th]}}}\ignorespaces
\bibitem{Honda:2020txe}
M.~Honda \& Y.~Tanizaki,
\textit{``{Topological aspects of 4D Abelian lattice gauge theories with the
  $\theta$ parameter}''},
\doiref{10.1007/JHEP12(2020)154}{JHEP \textbf{2012}, 154
  (2020)\ignorespaces}\ignorespaces,
\normalsize{\texttt{\arxivref{2009.10183}{arXiv:2009.10183
  \![hep-th]}}}\ignorespaces
\bibitem{Bolognesi:2020mpe}
S.~Bolognesi, K.~Konishi \& A.~Luzio,
\textit{``{Dynamics from symmetries in chiral $SU(N)$ gauge theories}''},
\doiref{10.1007/JHEP09(2020)001}{JHEP \textbf{2009}, 001
  (2020)\ignorespaces}\ignorespaces,
\normalsize{\texttt{\arxivref{2004.06639}{arXiv:2004.06639
  \![hep-th]}}}\ignorespaces
\bibitem{Sulejmanpasic:2020zfs}
T.~Sulejmanpasic, Y.~Tanizaki \& M.~{\"U}nsal,
\textit{``{Universality between vector-like and chiral quiver gauge theories:
  Anomalies and domain walls}''},
\doiref{10.1007/JHEP06(2020)173}{JHEP \textbf{2006}, 173
  (2020)\ignorespaces}\ignorespaces,
\normalsize{\texttt{\arxivref{2004.10328}{arXiv:2004.10328
  \![hep-th]}}}\ignorespaces
\bibitem{Furusawa:2020qdz}
T.~Furusawa, Y.~Tanizaki \& E.~Itou,
\textit{``{Finite-density massless two-color QCD at the isospin Roberge-Weiss
  point and the 't Hooft anomaly}''},
\doiref{10.1103/PhysRevResearch.2.033253}{Phys.~Rev.~Res. \textbf{2}, 033253
  (2020)\ignorespaces}\ignorespaces,
\normalsize{\texttt{\arxivref{2005.13822}{arXiv:2005.13822
  \![hep-th]}}}\ignorespaces
\bibitem{Bolognesi:2021yni}
S.~Bolognesi, K.~Konishi \& A.~Luzio,
\textit{``{Probing the dynamics of chiral $SU(N)$ gauge theories via
  generalized anomalies}''},
\doiref{10.1103/PhysRevD.103.094016}{Phys.~Rev.~D \textbf{103}, 094016
  (2021)\ignorespaces}\ignorespaces,
\normalsize{\texttt{\arxivref{2101.02601}{arXiv:2101.02601
  \![hep-th]}}}\ignorespaces
\bibitem{Kaidi:2021xfk}
J.~Kaidi, K.~Ohmori \& Y.~Zheng,
\textit{``{Kramers-Wannier-like Duality Defects in (3+1)D Gauge Theories}''},
\doiref{10.1103/PhysRevLett.128.111601}{Phys.~Rev.~Lett. \textbf{128}, 111601
  (2022)\ignorespaces}\ignorespaces,
\normalsize{\texttt{\arxivref{2111.01141}{arXiv:2111.01141
  \![hep-th]}}}\ignorespaces
\bibitem{Bolognesi:2022beq}
S.~Bolognesi, K.~Konishi \& A.~Luzio,
\textit{``{Dynamical Abelianization and anomalies in chiral gauge theories}''},
\doiref{10.1007/JHEP12(2022)110}{JHEP \textbf{2212}, 110
  (2022)\ignorespaces}\ignorespaces,
\normalsize{\texttt{\arxivref{2206.00538}{arXiv:2206.00538
  \![hep-th]}}}\ignorespaces
\bibitem{Morikawa:2022liz}
O.~Morikawa, H.~Wada \& S.~Yamaguchi,
\textit{``{Phase structure of linear quiver gauge theories from anomaly
  matching}''},
\doiref{10.1103/PhysRevD.107.045020}{Phys.~Rev.~D \textbf{107}, 045020
  (2023)\ignorespaces}\ignorespaces,
\normalsize{\texttt{\arxivref{2211.12079}{arXiv:2211.12079
  \![hep-th]}}}\ignorespaces
\bibitem{Kobayashi:2023ajk}
S.~K. Kobayashi, T.~Yokokura \& K.~Yonekura,
\textit{``{The QCD phase diagram in the space of imaginary chemical potential
  via {\textquoteright}t Hooft anomalies}''},
\doiref{10.1007/JHEP08(2023)132}{JHEP \textbf{2308}, 132
  (2023)\ignorespaces}\ignorespaces,
\normalsize{\texttt{\arxivref{2305.01217}{arXiv:2305.01217
  \![hep-th]}}}\ignorespaces
\bibitem{Nguyen:2023fun}
M.~Nguyen, Y.~Tanizaki \& M.~{\"U}nsal,
\textit{``{Study of gapped phases of 4d gauge theories using temporal gauging
  of the~$\mathbb{Z}_N$ 1-form symmetry}''},
\doiref{10.1007/JHEP08(2023)013}{JHEP \textbf{2308}, 013
  (2023)\ignorespaces}\ignorespaces,
\normalsize{\texttt{\arxivref{2306.02485}{arXiv:2306.02485
  \![hep-th]}}}\ignorespaces
\bibitem{Hayashi:2023wwi}
Y.~Hayashi, Y.~Tanizaki \& H.~Watanabe,
\textit{``{Semiclassical analysis of the bifundamental QCD
  on~$\mathbb{R}^2\times T^2$ with {\textquoteright}t Hooft flux}''},
\doiref{10.1007/JHEP10(2023)146}{JHEP \textbf{2310}, 146
  (2023)\ignorespaces}\ignorespaces,
\normalsize{\texttt{\arxivref{2307.13954}{arXiv:2307.13954
  \![hep-th]}}}\ignorespaces
\bibitem{Hayashi:2024qkm}
Y.~Hayashi \& Y.~Tanizaki,
\textit{``{Semiclassics for the QCD vacuum structure through
  T$^{2}$-compactification with the baryon-{\textquoteright}t Hooft flux}''},
\doiref{10.1007/JHEP08(2024)001}{JHEP \textbf{2408}, 001
  (2024)\ignorespaces}\ignorespaces,
\normalsize{\texttt{\arxivref{2402.04320}{arXiv:2402.04320
  \![hep-th]}}}\ignorespaces
\bibitem{Bolognesi:2024wje}
S.~Bolognesi, K.~Konishi, A.~Luzio \& M.~Orso,
\textit{``{Chiral gauge theories, generalized anomalies and breakdown of the
  color-flavor-locked center symmetry}''},
\doiref{10.1007/JHEP05(2025)062}{JHEP \textbf{2505}, 062
  (2025)\ignorespaces}\ignorespaces,
\normalsize{\texttt{\arxivref{2410.14598}{arXiv:2410.14598
  \![hep-th]}}}\ignorespaces
\bibitem{Hayashi:2024gxv}
Y.~Hayashi, Y.~Tanizaki \& H.~Watanabe,
\textit{``{Non-supersymmetric duality cascade of QCD(BF) via semiclassics on
  {\ensuremath{\mathbb{R}}}$^{2}${\texttimes} T$^{2}$ with the
  baryon-{\textquoteright}t Hooft flux}''},
\doiref{10.1007/JHEP07(2024)033}{JHEP \textbf{2407}, 033
  (2024)\ignorespaces}\ignorespaces,
\normalsize{\texttt{\arxivref{2404.16803}{arXiv:2404.16803
  \![hep-th]}}}\ignorespaces
\bibitem{Brennan:2026ira}
T.~D. Brennan \& K.~Intriligator,
\textit{``{Generalized families of QFTs}''},
\doiref{10.1007/JHEP08(2026)086}{JHEP \textbf{2608}, 086
  (2026)\ignorespaces}\ignorespaces,
\normalsize{\texttt{\arxivref{2602.09105}{arXiv:2602.09105
  \![hep-th]}}}\ignorespaces
\bibitem{Bolognesi:2026eso}
S.~Bolognesi, K.~Konishi \& M.~Orso,
\textit{``{Infrared spectra of some strongly-coupled chiral gauge theories}''},
\doiref{10.1103/4xhl-2fnw}{Phys.~Rev.~D \textbf{114}, 034007
  (2026)\ignorespaces}\ignorespaces,
\normalsize{\texttt{\arxivref{2605.10416}{arXiv:2605.10416
  \![hep-th]}}}\ignorespaces
\bibitem{Kapustin:2014gua}
A.~Kapustin \& N.~Seiberg,
\textit{``{Coupling a QFT to a TQFT and Duality}''},
\doiref{10.1007/JHEP04(2014)001}{JHEP \textbf{1404}, 001
  (2014)\ignorespaces}\ignorespaces,
\normalsize{\texttt{\arxivref{1401.0740}{arXiv:1401.0740
  \![hep-th]}}}\ignorespaces
\bibitem{Gaiotto:2014kfa}
D.~Gaiotto, A.~Kapustin, N.~Seiberg \& B.~Willett,
\textit{``{Generalized Global Symmetries}''},
\doiref{10.1007/JHEP02(2015)172}{JHEP \textbf{1502}, 172
  (2015)\ignorespaces}\ignorespaces,
\normalsize{\texttt{\arxivref{1412.5148}{arXiv:1412.5148
  \![hep-th]}}}\ignorespaces
\bibitem{Hatcher_Chap5}
A.~Hatcher,
\textit{``Chapter 5. Spectral Sequences''},
Additional chapter to the book ``Algebraic Topology"
  \url{https://pi.math.cornell.edu/~hatcher/AT/SSpage.html}\ignorespaces
\bibitem{Kapustin:2013qsa}
A.~Kapustin \& R.~Thorngren,
\textit{``{Topological Field Theory on a Lattice, Discrete Theta-Angles and
  Confinement}''},
\doiref{10.4310/ATMP.2014.v18.n5.a4}{Adv.~Theor.~Math.~Phys. \textbf{18}, 1233
  (2014)\ignorespaces}\ignorespaces,
\normalsize{\texttt{\arxivref{1308.2926}{arXiv:1308.2926
  \![hep-th]}}}\ignorespaces
\bibitem{Benini:2018reh}
F.~Benini, C.~C{\'o}rdova \& P.-S. Hsin,
\textit{``{On 2-Group Global Symmetries and their Anomalies}''},
\doiref{10.1007/JHEP03(2019)118}{JHEP \textbf{1903}, 118
  (2019)\ignorespaces}\ignorespaces,
\normalsize{\texttt{\arxivref{1803.09336}{arXiv:1803.09336
  \![hep-th]}}}\ignorespaces
\bibitem{Witten:1979ey}
E.~Witten,
\textit{``{Dyons of Charge e theta/2 pi}''},
\doiref{10.1016/0370-2693(79)90838-4}{Phys.~Lett.~B \textbf{86}, 283
  (1979)\ignorespaces}\ignorespaces
\bibitem{Itoyama:1982up}
H.~Itoyama \& A.~H. Mueller,
\textit{``{The Axial Anomaly at Finite Temperature}''},
\doiref{10.1016/0550-3213(83)90370-X}{Nucl.~Phys.~B \textbf{218}, 349
  (1983)\ignorespaces}\ignorespaces
\bibitem{Tachikawa:2016xvs}
Y.~Tachikawa \& K.~Yonekura,
\textit{``{Gauge interactions and topological phases of matter}''},
\doiref{10.1093/ptep/ptw131}{PTEP \textbf{2016}, 093B07
  (2016)\ignorespaces}\ignorespaces,
\normalsize{\texttt{\arxivref{1604.06184}{arXiv:1604.06184
  \![hep-th]}}}\ignorespaces
\bibitem{Intriligator:1995id}
K.~A. Intriligator \& N.~Seiberg,
\textit{``{Duality, monopoles, dyons, confinement and oblique confinement in
  supersymmetric SO(N(c)) gauge theories}''},
\doiref{10.1016/0550-3213(95)00159-P}{Nucl.~Phys.~B \textbf{444}, 125
  (1995)\ignorespaces}\ignorespaces,
\normalsize{\texttt{\arxivref{hep-th/9503179}{hep-th/9503179}}}\ignorespaces
\bibitem{Intriligator:1995au}
K.~A. Intriligator \& N.~Seiberg,
\textit{``{Lectures on supersymmetric gauge theories and electric-magnetic
  duality}''},
\doiref{10.1016/0920-5632(95)00626-5}{Nucl.~Phys.~B~Proc.~Suppl. \textbf{45BC},
  1 (1996)\ignorespaces}\ignorespaces,
\normalsize{\texttt{\arxivref{hep-th/9509066}{hep-th/9509066}}}\ignorespaces
\bibitem{Aharony:2013hda}
O.~Aharony, N.~Seiberg \& Y.~Tachikawa,
\textit{``{Reading between the lines of four-dimensional gauge theories}''},
\doiref{10.1007/JHEP08(2013)115}{JHEP \textbf{1308}, 115
  (2013)\ignorespaces}\ignorespaces,
\normalsize{\texttt{\arxivref{1305.0318}{arXiv:1305.0318
  \![hep-th]}}}\ignorespaces
\bibitem{Witten:2000nv}
E.~Witten,
\textit{``{Supersymmetric index in four-dimensional gauge theories}''},
\doiref{10.4310/ATMP.2001.v5.n5.a1}{Adv.~Theor.~Math.~Phys. \textbf{5}, 841
  (2002)\ignorespaces}\ignorespaces,
\normalsize{\texttt{\arxivref{hep-th/0006010}{hep-th/0006010}}}\ignorespaces
\bibitem{Cordova:2019uob}
C.~C{\'o}rdova, D.~S. Freed, H.~T. Lam \& N.~Seiberg,
\textit{``{Anomalies in the Space of Coupling Constants and Their Dynamical
  Applications II}''},
\doiref{10.21468/SciPostPhys.8.1.002}{SciPost~Phys. \textbf{8}, 002
  (2020)\ignorespaces}\ignorespaces,
\normalsize{\texttt{\arxivref{1905.13361}{arXiv:1905.13361
  \![hep-th]}}}\ignorespaces
\bibitem{Panagopoulos:2011rb}
H.~Panagopoulos \& E.~Vicari,
\textit{``{The 4D SU(3) gauge theory with an imaginary $\theta$ term}''},
\doiref{10.1007/JHEP11(2011)119}{JHEP \textbf{1111}, 119
  (2011)\ignorespaces}\ignorespaces,
\normalsize{\texttt{\arxivref{1109.6815}{arXiv:1109.6815
  \![hep-lat]}}}\ignorespaces
\bibitem{DElia:2013uaf}
M.~D'Elia \& F.~Negro,
\textit{``{Phase diagram of Yang-Mills theories in the presence of a $\theta$
  term}''},
\doiref{10.1103/PhysRevD.88.034503}{Phys.~Rev.~D \textbf{88}, 034503
  (2013)\ignorespaces}\ignorespaces,
\normalsize{\texttt{\arxivref{1306.2919}{arXiv:1306.2919
  \![hep-lat]}}}\ignorespaces
\bibitem{Bonati:2015sqt}
C.~Bonati, M.~D'Elia \& A.~Scapellato,
\textit{``{$\theta$ dependence in $SU(3)$ Yang-Mills theory from analytic
  continuation}''},
\doiref{10.1103/PhysRevD.93.025028}{Phys.~Rev.~D \textbf{93}, 025028
  (2016)\ignorespaces}\ignorespaces,
\normalsize{\texttt{\arxivref{1512.01544}{arXiv:1512.01544
  \![hep-lat]}}}\ignorespaces
\bibitem{Bonati:2016tvi}
C.~Bonati, M.~D'Elia, P.~Rossi \& E.~Vicari,
\textit{``{$\theta$ dependence of 4D $SU(N)$ gauge theories in the large-$N$
  limit}''},
\doiref{10.1103/PhysRevD.94.085017}{Phys.~Rev.~D \textbf{94}, 085017
  (2016)\ignorespaces}\ignorespaces,
\normalsize{\texttt{\arxivref{1607.06360}{arXiv:1607.06360
  \![hep-lat]}}}\ignorespaces
\bibitem{Kitano:2020mfk}
R.~Kitano, N.~Yamada \& M.~Yamazaki,
\textit{``{Is $N = 2$ Large?}''},
\doiref{10.1007/JHEP02(2021)073}{JHEP \textbf{2102}, 073
  (2021)\ignorespaces}\ignorespaces,
\normalsize{\texttt{\arxivref{2010.08810}{arXiv:2010.08810
  \![hep-lat]}}}\ignorespaces
\bibitem{Kitano:2021jho}
R.~Kitano, R.~Matsudo, N.~Yamada \& M.~Yamazaki,
\textit{``{Peeking into the {\ensuremath{\theta}} vacuum}''},
\doiref{10.1016/j.physletb.2021.136657}{Phys.~Lett.~B \textbf{822}, 136657
  (2021)\ignorespaces}\ignorespaces,
\normalsize{\texttt{\arxivref{2102.08784}{arXiv:2102.08784
  \![hep-lat]}}}\ignorespaces
\bibitem{Matsumoto:2023vbf}
A.~Matsumoto, K.~Hatakeyama, M.~Hirasawa, M.~Honda, J.~Nishimura \&
  A.~Yosprakob,
\textit{``{Numerical studies on the finite-temperature CP restoration in 4D
  SU($N$) gauge theory at $\theta=\pi$}''},
\doiref{10.22323/1.430.0378}{PoS \textbf{LATTICE2022}, 378
  (2023)\ignorespaces}\ignorespaces,
\normalsize{\texttt{\arxivref{2301.03879}{arXiv:2301.03879
  \![hep-lat]}}}\ignorespaces
\bibitem{Bonanno:2023hhp}
C.~Bonanno, M.~D'Elia \& L.~Verzichelli,
\textit{``{The $\theta$-dependence of the $SU(N)$ critical temperature at large
  $N$}''},
\doiref{10.1007/JHEP02(2024)156}{JHEP \textbf{2402}, 156
  (2024)\ignorespaces}\ignorespaces,
\normalsize{\texttt{\arxivref{2312.12202}{arXiv:2312.12202
  \![hep-lat]}}}\ignorespaces
\bibitem{Hirasawa:2024vns}
M.~Hirasawa, K.~Hatakeyama, M.~Honda, A.~Matsumoto, J.~Nishimura \&
  A.~Yosprakob,
\textit{``{Determination of the CP restoration temperature at $\theta=\pi$ in
  4D SU(2) Yang-Mills theory through simulations at imaginary $\theta$}''},
\doiref{10.22323/1.453.0193}{PoS \textbf{LATTICE2023}, 193
  (2024)\ignorespaces}\ignorespaces,
\normalsize{\texttt{\arxivref{2401.05726}{arXiv:2401.05726
  \![hep-lat]}}}\ignorespaces
\bibitem{Bonanno:2024ggk}
C.~Bonanno, C.~Bonati, M.~Papace \& D.~Vadacchino,
\textit{``{The {\ensuremath{\theta}}-dependence of the Yang-Mills spectrum from
  analytic continuation}''},
\doiref{10.1007/JHEP05(2024)163}{JHEP \textbf{2405}, 163
  (2024)\ignorespaces}\ignorespaces,
\normalsize{\texttt{\arxivref{2402.03096}{arXiv:2402.03096
  \![hep-lat]}}}\ignorespaces
\bibitem{Yamada:2024vsk}
N.~Yamada, M.~Yamazaki \& R.~Kitano,
\textit{``{Subvolume method for SU(2) Yang-Mills theory at finite temperature:
  topological charge distributions}''},
\doiref{10.1007/JHEP07(2024)198}{JHEP \textbf{2407}, 198
  (2024)\ignorespaces}\ignorespaces,
\normalsize{\texttt{\arxivref{2403.10767}{arXiv:2403.10767
  \![hep-lat]}}}\ignorespaces
\bibitem{Yamada:2024pjy}
N.~Yamada, M.~Yamazaki \& R.~Kitano,
\textit{``{{\ensuremath{\theta}} dependence of T$_{c}$ in SU(2) Yang-Mills
  theory}''},
\doiref{10.1007/JHEP02(2025)211}{JHEP \textbf{2502}, 211
  (2025)\ignorespaces}\ignorespaces,
\normalsize{\texttt{\arxivref{2411.00375}{arXiv:2411.00375
  \![hep-lat]}}}\ignorespaces
\bibitem{Hirasawa:2024fjt}
M.~Hirasawa, M.~Honda, A.~Matsumoto, J.~Nishimura \& A.~Yosprakob,
\textit{``{Evidence of a CP broken deconfined phase in 4D SU(2) Yang-Mills
  theory at {\ensuremath{\theta}} = {\ensuremath{\pi}} from imaginary
  {\ensuremath{\theta}} simulations}''},
\doiref{10.1007/JHEP05(2025)009}{JHEP \textbf{2505}, 009
  (2025)\ignorespaces}\ignorespaces,
\normalsize{\texttt{\arxivref{2412.03683}{arXiv:2412.03683
  \![hep-th]}}}\ignorespaces
\bibitem{Hirasawa:2025mot}
M.~Hirasawa, M.~Honda, A.~Matsumoto, J.~Nishimura \& A.~Yosprakob,
\textit{``{Numerical evidence for a CP broken deconfined phase at $\theta =\pi$
  in 4D SU(2) Yang-Mills theory through simulations at imaginary $\theta$}''},
\doiref{10.22323/1.466.0407}{PoS \textbf{LATTICE2024}, 407
  (2025)\ignorespaces}\ignorespaces,
\normalsize{\texttt{\arxivref{2502.09115}{arXiv:2502.09115
  \![hep-lat]}}}\ignorespaces
\bibitem{Matsumoto:2026puv}
A.~Matsumoto, M.~Hirasawa, J.~Nishimura \& A.~Yosprakob,
\textit{``{Phase diagram of 4D SU(3) Yang-Mills theory at
  {\ensuremath{\theta}}={\ensuremath{\pi}} via imaginary theta simulations}''},
\doiref{10.22323/1.518.0187}{PoS \textbf{LATTICE2025}, 187
  (2026)\ignorespaces}\ignorespaces,
\normalsize{\texttt{\arxivref{2603.09604}{arXiv:2603.09604
  \![hep-lat]}}}\ignorespaces
\bibitem{Watanabe:2026ymb}
H.~Watanabe, I.~Kanamori, O.~Morikawa, Y.~Nagai, Y.~Tanizaki \& A.~Tomiya,
\textit{``{Numerical hints for dyon condensation at {\ensuremath{\theta}} =
  2{\ensuremath{\pi}} via Wilson-'t Hooft loops in SU(2) Yang-Mills theory}''},
\doiref{10.1007/JHEP10(2026)068}{JHEP \textbf{2610}, 068
  (2026)\ignorespaces}\ignorespaces,
\normalsize{\texttt{\arxivref{2606.13428}{arXiv:2606.13428
  \![hep-lat]}}}\ignorespaces
\bibitem{Funcke:2019zna}
L.~Funcke, K.~Jansen \& S.~K{\"u}hn,
\textit{``{Topological vacuum structure of the Schwinger model with matrix
  product states}''},
\doiref{10.1103/PhysRevD.101.054507}{Phys.~Rev.~D \textbf{101}, 054507
  (2020)\ignorespaces}\ignorespaces,
\normalsize{\texttt{\arxivref{1908.00551}{arXiv:1908.00551
  \![hep-lat]}}}\ignorespaces
\bibitem{Chakraborty:2020uhf}
B.~Chakraborty, M.~Honda, T.~Izubuchi, Y.~Kikuchi \& A.~Tomiya,
\textit{``{Classically emulated digital quantum simulation of the Schwinger
  model with a topological term via adiabatic state preparation}''},
\doiref{10.1103/PhysRevD.105.094503}{Phys.~Rev.~D \textbf{105}, 094503
  (2022)\ignorespaces}\ignorespaces,
\normalsize{\texttt{\arxivref{2001.00485}{arXiv:2001.00485
  \![hep-lat]}}}\ignorespaces
\bibitem{Honda:2021aum}
M.~Honda, E.~Itou, Y.~Kikuchi, L.~Nagano \& T.~Okuda,
\textit{``{Classically emulated digital quantum simulation for screening and
  confinement in the Schwinger model with a topological term}''},
\doiref{10.1103/PhysRevD.105.014504}{Phys.~Rev.~D \textbf{105}, 014504
  (2022)\ignorespaces}\ignorespaces,
\normalsize{\texttt{\arxivref{2105.03276}{arXiv:2105.03276
  \![hep-lat]}}}\ignorespaces
\bibitem{Honda:2021ovk}
M.~Honda, E.~Itou, Y.~Kikuchi \& Y.~Tanizaki,
\textit{``{Negative string tension of a higher-charge Schwinger model via
  digital quantum simulation}''},
\doiref{10.1093/ptep/ptac007}{PTEP \textbf{2022}, 033B01
  (2022)\ignorespaces}\ignorespaces,
\normalsize{\texttt{\arxivref{2110.14105}{arXiv:2110.14105
  \![hep-th]}}}\ignorespaces
\bibitem{Honda:2022edn}
M.~Honda, E.~Itou \& Y.~Tanizaki,
\textit{``{DMRG study of the higher-charge Schwinger model and its
  {\textquoteright}t Hooft anomaly}''},
\doiref{10.1007/JHEP11(2022)141}{JHEP \textbf{2211}, 141
  (2022)\ignorespaces}\ignorespaces,
\normalsize{\texttt{\arxivref{2210.04237}{arXiv:2210.04237
  \![hep-lat]}}}\ignorespaces
\bibitem{Itou:2024psm}
E.~Itou, A.~Matsumoto \& Y.~Tanizaki,
\textit{``{DMRG study of the theta-dependent mass spectrum in the 2-flavor
  Schwinger model}''},
\doiref{10.1007/JHEP09(2024)155}{JHEP \textbf{2409}, 155
  (2024)\ignorespaces}\ignorespaces,
\normalsize{\texttt{\arxivref{2407.11391}{arXiv:2407.11391
  \![hep-lat]}}}\ignorespaces
\bibitem{Matsumoto:2025fjb}
A.~Matsumoto, E.~Itou \& Y.~Tanizaki,
\textit{``{Computing theta-dependent mass spectrum of the 2-flavor Schwinger
  model in the Hamiltonian formalism}''},
\doiref{10.22323/1.466.0210}{PoS \textbf{LATTICE2024}, 210
  (2025)\ignorespaces}\ignorespaces,
\normalsize{\texttt{\arxivref{2501.18960}{arXiv:2501.18960
  \![hep-lat]}}}\ignorespaces
\end{thebibliography}

\end{document}